\documentclass[modern]{aastex631}

\hypersetup{linkcolor=red,citecolor=blue,filecolor=cyan,urlcolor=magenta}
\usepackage{url}
\usepackage{etex}
\usepackage{amsmath,amssymb,latexsym} 
\usepackage{xcolor, fontawesome} 
\usepackage{graphicx}
\usepackage{lipsum}
\usepackage{pgfplots}
\pgfplotsset{compat=1.3}
\usepackage{color,soul}

\begin{document}

\title{Hydrodynamic Simulations of Oxygen - Neon Classical Novae as Galactic $^7$Li Producers and Potential Accretion Induced Collapse  Progenitors\footnote{
We dedicate this paper to the memory of R. Mark Wagner. Friend, collaborator, exemplary observational astronomer, and contributor to studies of classical novae and 
transient objects for more than 40 years. He will be missed.}}

\author[0000-0002-1359-6312]{Sumner Starrfield}
\affiliation{Earth and Space Exploration, Arizona State University, P.O. Box 871404, Tempe, Arizona, 85287-1404, USA
starrfield@asu.edu}

\author[0000-0002-7978-6570]{Maitrayee Bose}
\affiliation{Earth and Space Exploration, Arizona State University, P.O. Box 871404, Tempe, Arizona, 85287-1404, USA}
\affiliation{Center for Isotope Analysis (CIA), Arizona State University, Tempe, Arizona, 85287-1404, USA}

\author[0000-0003-2381-0412]{Christian Iliadis}
\affiliation{Department of Physics \& Astronomy, University of North Carolina, Chapel Hill, NC 27599-3255}
\affiliation{Triangle Universities Nuclear Laboratory, Durham, NC 27708-0308, USA}

\author[0000-0002-9481-9126]{W. Raphael Hix}
\affiliation{Physics Division, Oak Ridge National Laboratory, Oak Ridge TN, 37831-6354}
\affiliation{Department of Physics and Astronomy, University of Tennessee, Knoxville, TN 37996 }

\author[0000-0001-6567-627X]{Charles E. Woodward}
\affiliation{MN Institute for Astrophysics, 116 Church Street, SE University of Minnesota, Minneapolis, MN 55455 }

\author[0000-0003-1892-2751]{R. Mark Wagner}
\affiliation{Large Binocular Telescope Observatory, Tucson, AZ 85721}
\affiliation{Department of Astronomy, Ohio State University, Columbus, OH 43210}
\affiliation{deceased, September 2, 2023}

\newcommand{\kms}{km\,s$^{-1}$}
\def\lesssim{\mathrel{\hbox{\rlap{\hbox{\lower4pt\hbox{$\sim$}}}\hbox{$<$}}}}
\def\gtrsim{\mathrel{\hbox{\rlap{\hbox{\lower4pt\hbox{$\sim$}}}\hbox{$>$}}}}
\def\apj{$Astrophys.\ J.$}
\def\apjl{$Astrophys.\ J.$}
\def\aj{$Astron.\ J.$}
\def\aap{$Astron.\ Astrophys.$}
\def\mnras{$Mon.\ Not.\ R.\ Astron.\ Soc.$}
\def\pasj{$Publ.\ Astron.\ Soc.\ Jpn$}
\def\iaucirc{$IAU Circ.$}


\bigskip

\begin{abstract}
We report on studies of Classical Nova (CN) explosions where we follow the
evolution of thermonuclear runaways (TNRs) on oxygen-neon (ONe) white dwarfs (WDs). 
Using NOVA, a one-dimensional hydrodynamic computer code,  we accrete
Solar matter until the TNR is ongoing and then switch to a mixed composition.  This approach is guided by the results of multi-dimensional studies of TNRs in WDs  which find that sufficient mixing with WD core material occurs after the TNR is well underway, and levels of enrichment of the CNONeMg elements are reached that 
agree with observations of CN ejecta abundances.  Because the amount of accreted material is inversely proportional to the oxygen abundance, 
by first accreting Solar matter, the amount of accreted material is larger than in those simulations with an initially enriched composition. 
We vary the mass of the WD (from 0.6 M$_\odot$ to 1.35 M$_\odot$) and the composition of the mixed materials.   Our results show large enrichments of $^7$Be in the ejected gases implying that ONe CNe and CO CNe  { \citep{starrfield_2020_aa}} may be responsible for a significant fraction ($\sim$ 100 M$_\odot$) of the galactic $^7$Li  ($\sim$1000~M$_\odot$).  The production of $^{22}$Na and  $^{26}$Al in CN explosions and the $\gamma$-ray emission predicted by our simulations is discussed. The WDs in all our simulations eject less material than they accrete and we predict that the WD is growing in mass as a consequence of the CN outburst. ONe CNe, therefore, may be an important channel for accretion induced collapse (AIC) events.  
 \end{abstract}

\keywords{Cataclysmic variable stars (203) -- Novae (1127) -- Recurrent Novae (1366) -- Galaxy chemical evolution (580) -- Galaxy abundances (574) -- Milky Way Galaxy (1054)}

\section{Introduction}

This paper is a continuation of our hydrodynamic studies of the consequences of accretion of Solar composition matter, from a secondary star, onto a white dwarf (WD), which results in a thermonuclear runaway (TNR) and is observed as either a classical nova (CN), recurrent nova (RN), or Symbiotic Nova (SyN) explosion \citep[][and references therein]{starrfield_2020_aa}.  Observations of nova ejecta show that there are basically two composition classes for the WD component. In one class the WD core composition is carbon and oxygen (CO), while in the other
class the WD core is mostly oxygen and neon (ONe)  \citep[][and references therein]{gehrz_1998_aa, starrfield_2012_aa}.  Previously,  we concentrated on simulating TNRs in ONe enriched material \citep{starrfield_1998_aa, starrfield_2000_aa, starrfield_2009_aa, starrfield_2012_aa, starrfield_2016_aa}, but
in \citet{starrfield_2020_aa}; however, we assumed that the core composition was that of a CO WD.  In this paper we perform a complementary set of simulations to \citet{starrfield_2020_aa} but use the core composition of an ONe WD. 

Until recently there was no agreement on when and how matter from the outer edges of the WD core was mixed into the accreted material, and we and others assumed that this mixing happened from the beginning of the simulation.  This would, for example, represent the effects of shear mixing of the accreting material with the outer core layers \citep{sparks_1987_aa, kutter_1987_aa, alexakis_2004_aa}. The   observations of the abundances and energetics in CN explosions demanded that such mixing occur \citep{gehrz_1998_aa, downen_2013_aa}.  Now, we are guided by the multi-dimensional studies that show that material is dredged up from the outer layers of the WD by convectively associated instabilities once the TNR is well underway  \citep{casanova_2010_ab, casanova_2011_ab, casanova_2016_aa, casanova_2018_aa, jose_2014_aa, jose_2020_aa}.

As described in \citet{starrfield_2020_aa}, we assume that mixing does not occur until the late stages of the TNR when the convective region has nearly reached the surface of the WD.   At this time in the evolution, we switch to a mixed composition.  \cite{starrfield_2020_aa} found that the amount of {\it accreted} material is an inverse function of both the initial abundance of $^{12}$C and the mass of the WD.   Assuming material mixed from the beginning, results in a much larger amount of $^{12}$C in the accreted material, and since $^{12}$C is the catalyst
 in the CNO cycle, the temperature rises faster  per unit accreted material and the explosion occurs earlier in the evolution with less total matter accreted.  The result is a less violent outburst \citep{starrfield_2020_aa}.  We will show that the same evolutionary results occurs with enriched $^{16}$O in an ONe CN.
 
 In contrast, for a given WD mass, accreting solar material allows for more matter to be accreted by the time 
 the growing energy generation in the TNR reaches $\sim10^{11}$ erg gm$^{-1}$ s$^{-1}$  compared to those simulations where mixing with accreted material occurs from the beginning of the evolution.  Switching to a mixed composition at this time in the evolution produces a more violent outburst. Moreover, reducing the metallicity of the accreting mixture to values seen in the LMC, SMC, or even lower, also reduces the initial $^{12}$C and $^{16}$O abundances, allowing more material to be accreted before the TNR is well underway \citep{starrfield_1999_ac, jose_2007_aa, chen_2019_aa}.

 Whatever the mechanism, and whenever during the accretion phase that mixing of accreted matter with core matter takes place, the TNR theory of the outburst, in combination with the observations, demands that mixing occur.  Four predictions of the TNR theory have been verified by subsequent observations:

\begin{enumerate}
\item The need for a nuclear burning region enriched  in CNO nuclei to produce fast CNe was initially predicted by \citet{starrfield_1972_aa} and verified by observational studies of novae ejecta which were done in the late 1970's \citep[][and references therein]{starrfield_1989_aa, gehrz_1998_aa}.

\item The prediction that $^7$Li was produced in the CN outburst was made by \citet{starrfield_1978_aa} and confirmed in 2015 using high dispersion spectroscopy \citep{tajitsu_2015_aa}.  Since that time, there have been numerous studies showing enriched $^7$Be and $^7$Li in early CNe spectra \citep[][and references therein]{wagner_2018_aa, dellavalle_2020_aa, molaro_2022_aa, molaro_2023_aa}.  \citet{molaro_2022_aa, molaro_2023_aa} provide a detailed comparison of observations and theory and find that it is likely that CNe produce most of the $^7$Li in the galaxy (see Section \ref{lithium}). 

\item Theoretical studies done in the 1980's, with both accretion and time-dependent convection, showed that shortly after the rising temperature in the nuclear burning region exceeded $10^8$ K, convection carried the $\beta^+$-unstable nuclei to the surface. Their decays heated the surface layers to both high temperatures and high luminosities causing an early ''flash'' of x-ray photons \citep{starrfield_1990_aa}.  This prediction has now been verified by observations with the eRosita x-ray satellite \citep{konig_2022_aa}.

\item An exciting recent  development has been the discovery of VHE $\gamma$-rays from CNe using the Fermi-LAT telescope launched in 2008 \citep[][and references therein]{cheung_2016_aa, chomiuk_2021_aa}.  \citet{tatischeff_2007_aa, tatischeff_2023_aa} predicted the possibility of VHE $\gamma$-rays based on their analysis of the x-rays produced in the outburst of RS Oph.  In addition, they predicted the possibility of cosmic rays from CNe, and those have now been detected by the MAGIC collaboration \citep{acciari_2022_aa}.

\end{enumerate} 

Finally, based on our studies \citep[][and references therein]{starrfield_2020_aa} and those to be reported in this paper, we find that more matter is accreted than ejected during the outburst and, therefore, the WD is growing in mass as a result of the CN phenomenon.  We discussed the implications of these results for CO novae as SN Ia progenitors in \citet{starrfield_2020_aa}.  Here, we discuss the implications of ONe novae evolving to Accretion Induced Collapse (AIC) and neutron stars.  Not all CN or CV systems will become either SN Ia or undergo AIC, however, since it is necessary for the secondary to be sufficiently massive to transfer enough matter to the WD for it to ultimately evolve to ''near'' the Chandrasekhar Limit.

In Section \ref{novacode} we describe NOVA our 1-D hydrodynamic computer code.  In Section \ref{MFB} we present the simulations where we assume mixing from the beginning and show that they do not fit the observations.  In Section \ref{MDTNR} we present the simulations where we do not mix until near the peak of the TNR.  It consists of 2 subsections: in  \ref{Solar},  we present simulations with only a Solar mixture; and in \ref{mixdtnr}  we present the main results of the paper in which we mix during the TNR.  Section \ref{nucleo} gives the nucleosynthesis results.
 We also report the amount of $^7$Li, $^{22}$Na, and $^{26}$Al produced in ONe CNe (Sections \ref{lithium} and \ref{sodium}) and compare the values to both our CO CNe studies and those of \citet{jose_1998_aa} and  \citet{rukeya_2017_aa}. 
 We end the this paper with a discussion (Section \ref{discuss}) and conclusions (Section \ref{conclude}).

\section{NOVA: Our 1-Dimensional Hydrodynamic Code}
\label{novacode}

We use NOVA \citep{kutter_1972_aa,sparks_1972_aa, kutter_1974_aa, kutter_1980_aa, starrfield_2009_aa, 
starrfield_2016_aa, starrfield_2020_aa} in this study.   The most recent descriptions of NOVA can be found in 
\citet[][and references therein]{starrfield_2009_aa, starrfield_2020_aa} and we only briefly describe it here.  NOVA is a  one-dimensional (1-D), fully implicit, hydrodynamic, computer code that has been well tested against standard problems \citep{kutter_1972_aa, sparks_1972_aa}. NOVA includes the OPAL opacities  \citep{iglesias_1996_aa}, a large nuclear reaction network with 187 nuclei (up to $^{64}$Ge), nuclear reaction rates from \citet{longland_2010_aa, iliadis_2010_aa, iliadis_2010_ab, iliadis_2010_ac} and Starlib \citep{sallaska_2013_aa}, the nuclear reaction network solver developed by \citet{hix_1999_aa}, and the Timmes equations of state \citep{timmes_1999_aa, timmes_2000_ab}.  

NOVA also includes the \citet{arnett_2010_aa} algorithm for mixing-length convection and the Potekhin electron degenerate conductivities described in \citet{cassisi_2007_aa}.   These improvements have had the effect of changing the initial structures of the WDs so that they have smaller radii and, thereby, larger surface gravities compared to our previous ONe studies \citep{starrfield_2009_aa}.  We also include the possible effects of a binary companion (an extra source of heating at radii of $\sim 10^{11}$ cm) as described by \citet{macdonald_1980_aa}, which can increase the amount of mass lost during the late stages of the outburst.

In contrast to our previous studies, where we used 95 or 150 mass zones, we now use 300 mass zones with the mass of the zone decreasing from the center to the surface.  The effects of increasing the number of zones on the resulting evolution is described in the appendix of  \citet{starrfield_2020_aa}.  The simulations reported on in that appendix show that they have converged by the time we have increased the number of zones to 150 but given slight differences (and a faster CPU) we chose to increase the number of zones to 300.  The mass of the surface zone is $\sim 2 \times 10^{-9}$ in units of the WD mass.  This is much less than the total accreted mass.  A surface mass this low decreases the maximum time step during the accretion phase (although the code is implicit, the size of the maximum time step is tied to the mass of the outer zone), but allows us to fully resolve the behavior of the evolution once the TNR is underway.

We accrete material at a rate (\.M) of $1.6 \times 10^{-10}$ M$_\odot$yr$^{-1}$ onto complete WDs (the structure extends to the WD center) with masses of 0.6 M$_\odot$, 0.8 M$_\odot$, 1.00 M$_\odot$, 1.15 M$_\odot$, 1.25 M$_\odot$, and 1.35 M$_\odot$.  We choose this value of \.M because it is the value used by \citet{hernanz_1996_aa, jose_1998_aa}, \citet{rukeya_2017_aa}, \citet{starrfield_2009_aa}, and \citet{starrfield_2020_aa} and we compare our results to their results. 

We use two different mixed compositions in this study.  The first is similar to that used in \citet{starrfield_2009_aa} and is 50\% ONe WD matter and 50\% Solar matter \citep{lodders_2003_aa}.   The second composition is 25\% ONe WD matter and 75\% Solar matter which allows a better comparison of our results with \citet{hernanz_1996_aa}, \citet{jose_1998_aa}, and \citet{rukeya_2017_aa},  who also investigated the consequences of 25\% ONe WD matter and 75\% Solar matter.  For both compositions, we use the abundance distributions tabulated in \citet{kelly_2013_aa}.    \citet{kelly_2013_aa}  did post-processing of simulations done with the SHIVA code \citep{jose_1998_aa} and reported that the 25\% WD - 75\% Solar mixture was a better fit to the observations.

We choose an extremely broad range in WD mass even though it is commonly assumed that an ONe WD should not have a mass below $\sim 1.10$ M$_\odot$; a limit derived from the evolution of single (non-binary) stars \citep{iben_1991_aa, ritossa_1996_aa, iben_1997_aa, doherty_2015_aa}.  { This choice allows us to compare our ONe simulations to the CO simulations reported in \citet{starrfield_2020_aa} and the CO and ONe simulations reported in \citet{jose_1998_aa} and \citet{rukeya_2017_aa}.}

\bigskip
\bigskip
\section{Simulations with a mixed composition from the beginning}
\label{MFB}


The principal motivation for this paper is to present the results of a new set of simulations where we assume the composition is not mixed until the TNR is well underway.  In order to compare these new simulations to previous simulations \citep{jose_1998_aa, starrfield_2009_aa}, we first present  the results from ``Mixing From the Beginning'' (hereafter, MFB) and assume some mechanism is instantaneously mixing accreted with core matter from the beginning of accretion.  This was the assumption in our previous CN simulations \citep[][and references therein]{starrfield_2009_aa, starrfield_2016_aa} and it follows the same outline as in \citet{starrfield_2020_aa}.  MFB was used previously \citep{starrfield_2016_aa, dellavalle_2020_aa} because there was no consensus on how or when WD core matter was mixed into the accreted envelope.  A discussion of mixing mechanisms can be found in \cite{jose_2007_aa, jose_2020_aa}, \citet{ dellavalle_2020_aa}, and \citet{chomiuk_2021_aa}. The observations of both fast CO and ONe CNe require that such mixing occur \citep[][and references therein]{gehrz_1998_aa, starrfield_2009_aa, downen_2013_aa, starrfield_2016_aa, starrfield_2020_aa}. Moreover, it is not physically reasonable to assume that the accreted matter is fully mixed with core matter from the beginning of accretion. 


\begin{deluxetable}{@{}lccccccc}
\tabletypesize{\small}
\tablecaption{Initial Parameters and Evolutionary Results for Accretion onto ONe WDs: MFB
\tablenotemark{}\label{evolONeMFB}}
\tablewidth{0pt}
\tablecolumns{7}
 \tablehead{ \colhead{WD Mass:} &
\colhead{0.6M$_\odot$} &
\colhead{0.8M$_\odot$} &
\colhead{1.0M$_\odot$} &
\colhead{1.15M$_\odot$} &
\colhead{1.25M$_\odot$} &
\colhead{1.35M$_\odot$}}
 \startdata
Initial:  L/L$_\odot$($10^{-3}$)&5.0&5.0&5.0&5.0&5.2&5.4\\
Initial:  R($10^3$km)&8.5&6.8&5.2&4.1&3.2&2.2\\
Initial: T$_c$($10^7$K)&1.9&1.7&1.6&1.5&1.5&1.5\\
Initial: $\rho_c$ ($10^7$ gm cm$^{-3}$)&$0.37$&$1.0$&$3.4$&$10.0$&$26.0$&$130.0$\\
Initial: P$_c$(dynes cm$^{-2}$)&$1.8 \times10^{23}$&$8.9 \times10^{23}$&$4.7 \times10^{24}$&$2.1 \times10^{25}$&$7.8 \times10^{25}$&$7.0 \times10^{26}$\\
Initial:  T$_{\rm eff}$($10^4$K)&1.4&1.6&1.8&2.0&2.3&2.8\\
 \hline
{\bf 25\% WD - 75\% Solar: MFB}&&&&&&\\
\hline
$\tau$$_{\rm acc}$($10^5$ yr)&19.0&6.3&3.3&1.8&1.0&0.4\\
M$_{\rm acc}$($10^{-5}$M$_{\odot}$)&30.0&10.0&5.3&2.8&1.5&0.6\\
T$_{\rm peak}$($10^8$K)&1.3&1.4&1.8&2.1&2.3&2.9\\
$\epsilon_{\rm nuc-peak}$(erg gm$^{-1}$s$^{-1}$)&$3.6 \times10^{13}$&$3.7 \times10^{13}$&$7.5\times10^{14}$&$2.7 \times10^{15}$&$4.4 \times10^{15}$&$1.1 \times10^{16}$\\
L$_{\rm peak}$/L$_\odot$ ($10^4$)&2.0&2.1&2.5&3.2&3.7&4.5\\
T$_{\rm eff-peak}$($10^5$K)&0.8&3.1&3.0&5.2&6.6&10.0 \\
M$_{\rm ej}$(M$_{\odot}$)&$8.7 \times10^{-7}$&0.0\tablenotemark{a}&$3.2\times10^{-7}$&$1.6 \times10^{-7}$&$5.7 \times 10^{-8}$&$5.1\times10^{-9}$\\
N($^7$Li/H)$_{\rm ej}$/N($^7$Li/H)$_{\odot}$\tablenotemark{c} &$2.0 \times10^{-2}$&$0.0$&$4.8 \times10^{-1}$&$3.8$&7.4&15.2\\
$^7$Li$_{\rm ej}$($10^{-15}$M$_{\odot}$)&0.1&0.0&1.1&4.2&2.8&0.5\\
$^{22}$Na$_{\rm ej}$($10^{-12}$M$_{\odot}$)&96.0&0.0&13.0&8.3&5.3&1.2\\
$^{26}$Al$_{\rm ej}$($10^{-10}$M$_{\odot}$)&3.3&0.0&6.1&0.4&0.1&0.005\\
RE\tablenotemark{b}: (M$_{\rm acc}$ -  M$_{\rm ej}$)/M$_{\rm acc}$&1.0&1.0&0.99&0.99&1.0&1.0\\
V$_{\rm max}$($10^2$km s$^{-1}$)&4.6&0.0&4.8&5.2&5.1&4.5\\
\hline
{\bf 50\% WD - 50\% Solar: MFB}&&&&&&\\
\hline
$\tau$$_{\rm acc}$($10^5$ yr)&11.0&5.9&3.1&1.7&0.9&0.3\\
M$_{\rm acc}$($10^{-5}$M$_{\odot}$)&18.0&9.4&5.0&2.7&1.4&0.5\\
T$_{\rm peak}$($10^8$K)&1.3&1.6&2.0&2.4&2.7&3.2\\
$\epsilon_{\rm nuc-peak}$(erg gm$^{-1}$s$^{-1}$)&$2.1\times10^{13}$&$3.1\times10^{14}$&$2.8\times10^{15}$&$1.3\times10^{16}$&$3.0\times10^{16}$&$8.9\times10^{16}$\\
L$_{\rm peak}$/L$_\odot$ ($10^4$)&12.6&4.5&2.5&3.2&5.0&7.4\\
T$_{\rm eff-peak}$($10^5$K)&1.5&3.3&3.5&5.6&7.0&11.0 \\
M$_{\rm ej}$(M$_{\odot}$)&$1.0 \times10^{-6}$&$1.7 \times10^{-6}$&$3.4 \times10^{-7}$&$5.1 \times10^{-7}$&$5.6\times10^{-7}$&$1.4\times10^{-6}$\\
N($^7$Li/H)$_{\rm ej}$/N($^7$Li/H)$_{\odot}$ \tablenotemark{c}&$5.7 \times10^{-2}$&$1.7 \times10^{-2}$&$2.2 \times10^{1}$&$9.7 \times 10^{1}$&$1.5 \times10^{2}$&$2.3 \times 10^{2}$\\
$^7$Li$_{\rm ej}$($10^{-13}$M$_{\odot}$)&0.003&0.01&0.3&2.1&3.5&11.0\\
$^{22}$Na$_{\rm ej}$($10^{-10}$M$_{\odot}$)&4.6&2.0&0.3&1.1&4.3&7.4\\
$^{26}$Al$_{\rm ej}$($10^{-10}$M$_{\odot}$)&4.2&31.0&15.0&3.0&3.1&9.0\\
RE\tablenotemark{b}: (M$_{\rm acc}$ -  M$_{\rm ej}$)/M$_{\rm acc}$&0.99&0.98&0.99&0.98&0.96&0.72\\
V$_{\rm max}$($10^2$km s$^{-1}$)&4.1&4.1&4.4&4.9&6.7&14.0\\
\enddata
\tablenotetext{a}{This simulation did not eject any material.}
\tablenotetext{b}{RE: Retention Efficiency see Figure \ref{retention}}
\tablenotetext{c}{This notation follows that of \citet{hernanz_1996_aa}}

\end{deluxetable}

The basic properties of each WD initial model (luminosity, radius, central temperature, central density, central pressure, and effective temperature) are given in the first 6 rows of Table \ref{evolONeMFB}.    Comparing to the data in Table 1 in \citet{starrfield_2020_aa},  where the only difference in the two structures is going from 150 to 300 zones, we see a slight decrease in radius that results in a larger central density for the most massive WDs.  This change, a direct result of the increase in the number of mass zones, is discussed in the appendix in \citet{starrfield_2020_aa}.

We accrete onto the given WD with NOVA at the chosen rate and composition, held constant through the accretion phase, until we reach a peak nuclear reaction rate 
in the envelope of $10^{11}$ erg gm$^{-1}$ s$^{-1}$. 
  {(While this value is arbitrary, it is the value that we have used since we began accretion studies with NOVA.
The time to the peak temperature and energy generation is now sufficiently short that only a small amount of accreted material is neglected.  We find that NOVA runs faster through the TNR with accretion turned off.) }  We now stop accreting and use NOVA to
evolve the simulation through the peak of the TNR and the following decline in the temperature toward quiescence. 

Examining Table \ref{evolONeMFB}, as the initial WD mass increases its radius decreases, which is a well known result of electron degeneracy.  We choose an initial luminosity of $\sim 5 \times 10^{-3}$ L$_\odot$ in order to obtain as large an amount of accreted mass as possible.  Since it is virtually the same initial luminosity for all the WD masses, as the radius decreases the initial T$_{\rm eff}$ must increase.  The decrease in radius, in turn, increases the gravitational potential energy at the surface and the TNR is reached with a smaller amount of accreted mass and, thereby, a smaller accretion time \citep{starrfield_1989_aa}.   The latter effect is evident in the next set of rows which give the evolutionary results for the first mixture which is 25\% WD matter and 75\% Solar matter \citep{lodders_2003_aa}. 
The model composition is noted in bold.  The rows are the accretion time to the beginning of the 
TNR, $\tau_{\rm acc}$,  and M$_{\rm acc}$ is the total accreted mass with the scaling factor in parentheses for all rows.
The next set of rows tabulate, as a function of WD mass, the peak temperature in the simulation (T$_{\rm peak}$),
the peak rate of energy generation ($\epsilon_{\rm nuc-peak}$), the peak surface luminosity in units of the Solar luminosity, (L$_{\rm peak}$/L$_\odot$), the peak effective temperature (T$_{\rm eff-peak}$), the amount of
mass ejected in Solar masses (M$_{\rm ej}$),  the amount of $^7$Li ejected with respect to the Solar value, (N($^7$Li/H)$_{\rm ej}$/N($^7$Li/H)$_{\odot}$),  {followed by the amount of $^7$Li ejected in units of solar mass, $^7$Li$_{\rm ej}$.  (We assume that all the $^7$Be produced in the TNR decays to $^7$Li).}  The 0.8M$_\odot$ simulation does not eject any mass. Although the accreted plus core layers reached to radii exceeding $\sim10^{12}$ cm, their velocities were below escape velocity.

We express the $^7$Li ratio (N($^7$Li/H)$_{\rm ej}$/N($^7$Li/H)$_{\odot}$) in the same form as \citet[][Table 1]{hernanz_1996_aa} in order to provide a direct comparison (see Section \ref{lithium}).  We use the \citet{anders_1989_aa} value for N($^7$Li/H)$_{\odot}$ of $2.04 \times 10^{-9}$ only in this table to assist in the comparison with \citet{hernanz_1996_aa}  because that is the value that they used.  In our simulations, however, we use the value from the \citet{lodders_2003_aa} meteoritic abundances since our isotopic abundance distribution is taken from \citet{kelly_2013_aa}. 
The actual value of the initial $^7$Li does not matter in the simulations, however,  because it is mostly destroyed during the TNR.

The next two rows give the amount of $^{22}$Na and $^{26}$Al ejected in Solar masses.  These two
radioactive elements are predicted to be produced in ONe CNe and here we provide our predictions for these
elements.  The final two rows give the Retention Efficiency (RE): (M$_{\rm acc}$ -  M$_{\rm ej}$)/M$_{\rm acc}$,
and the velocity of the surface zone which is the maximum velocity in each simulation (V$_{\rm max}$).  Clearly,
the RE shows that most of the accreted material remains on the WD surface and is not ejected. Therefore, the
WD, in these simulations,  is growing in mass as a consequence of the CN phenomenon.

In the following rows, we tabulate the same information but for the MFB simulations with 50\% WD and 50\% Solar matter. Because of the increase in the initial $^{16}$O abundance, once the accreting material gets sufficiently hot for CNO burning rather than the initial $p-p$ chain (which now includes the $pep$ reaction: $p + e^{-}
+p \rightarrow d + \nu$, as discussed in Starrfield et al. 2009), the increased energy generation per unit accreted mass reduces the time to the TNR, and thereby the amount of accreted mass, when compared to the 25\% WD - 75\% Solar simulations.

 As shown in Table \ref{evolONeMFB} for both compositions, the most massive WD exhibits the highest peak temperature, which is reached a few hundred seconds after the temperature in the TNR exceeds $10^{8}$K.   The rise in temperature ends when virtually all the light nuclei in the convective region have become 
positron-decay nuclei ($^{13}$N, $^{14}$O, $^{15}$O, and $^{17}$F). No further proton captures can occur on $^{14}$O and $^{15}$O until they have decayed \citep[][]{starrfield_1972_aa, starrfield_2009_aa, starrfield_2016_aa, starrfield_2020_aa}.  The temperature then starts to drop since most of the accreted plus WD layers are now only partially degenerate and expanding.  The total time around peak temperature and decline is about
2000~s for the mass zones adjacent to the WD core.  The simulation for the 0.6 M$_\odot$ WD evolves so slowly, however,  that it takes almost 8000~s for the temperature to reach its peak and the decline is even slower.  In fact, the evolution of the TNR on this low a mass WD is so slow that it rules out this WD mass for all but the slowest CNe (either CO or ONe) such as Nova Velorum 2022 \citep{aydi_2023_aa}.  

We follow each of the simulations through peak temperature and the following decline to values where no further nuclear burning is occurring in the outermost layers and the outer radius has reached to $\sim 10^{12}$ cm.  
NOVA now tests the radii, optical depths, and velocities of the expanding layers to determine if they have been ejected or will soon be ejected.   We tabulate, as the amount of mass that has been ejected, those mass zones that are expanding at speeds above the local escape velocity and are optically thin.  We do not remove any mass zones during the evolution since this reduces the {\it numerical} pressure on the zones just below those escaping, causing them to accelerate outward and possibly reach escape speed.

We find that when the material is ejected, we can only follow the mass zones until they have reached radii of a few times $10^{12}$ cm.  At these radii, the density in the outer layers has fallen to values that are below the lower limit in density ($\rho = 10^{-12}$ gm cm$^{-3}$) of the physics tables (opacity, pressure equation of state, energy equation of state) and we end the evolution.  
This procedure is in contrast to the recent studies by \citet{denissenkov_2014_aa} and \citet{chen_2019_aa} who use the super-Eddington wind prescription in MESA \citep{paxton_2011_aa, paxton_2013_aa, paxton_2015_aa, paxton_2018_aa} or \citet{epelstain_2007_aa} who use the \citet{kovetz_2009_aa} code and remove the mass zones from the simulation as it is evolving. 
Their prescriptions begin mass loss as soon as the luminosity exceeds the Eddington luminosity.  However, in NOVA, the outer layers may exceed the Eddington luminosity it is only by small amounts and the acceleration is sufficiently small to neither drive a wind or mass loss. 
In addition, the mass zones are typically also convective which is expected in a super-Eddington regime \citep{quataert_2016_aa}.  A comparison of the MESA and NOVA mass loss methods can be found in \citet{newsham_2014_aa} and
\citet{starrfield_2017_aa}.  
The criticism of mass loss prescriptions stated in \citet{chomiuk_2021_aa} does not apply to NOVA.  
 {We do examine the velocities of the mass zones just below the escaping layers, and in most cases they have velocities far below escape velocity 
and even when we run the simulation longer they do not reach escape velocity.}

Table \ref{evolONeMFB} shows that peak energy generation is an increasing function of WD mass for both compositions.  In addition, the larger mass fraction of $^{16}$O in the accreted layers (.0261 to 0.133) results in peak energy generation being larger in the 50\% WD -50\% Solar simulations for all WD masses except that at 0.6M$_\odot$.  At this low a WD mass, the temperatures during the TNR never become sufficiently high to reach the $\beta^+$ - decay limited value of the CNO cycle
 ($\sim 2 \times 10^{14}$ erg gm$^{-1}$ s$^{-1}$).  We tabulate the peak luminosity for each WD mass, which occurs as the outermost zones reach radii exceeding $10^{11}$ cm and are expanding close to the escape velocity.  The matter is still optically thick at this time. 

Peak effective temperature is also an increasing function of WD mass but it occurred shortly after the convective region reached the surface early in the evolution, when those layers reached their maximum temperature and began to expand and cool. The increased initial mass fraction of $^{16}$O in the 50\% WD - 50\% Solar simulations results in a higher effective temperature for the same WD mass.  As emphasized in \citet{starrfield_2009_aa, starrfield_2020_aa},
peak effective temperature appears to be a measure of the WD mass and, for the more massive WDs, the peak is sufficiently hot to be detected by all sky X-ray instruments.  \citet{konig_2022_aa} have now reported detection of such emission with eRosita.

Examining the ejected masses for both MFB compositions in Table \ref{evolONeMFB}, we see that the 25\% WD - 75\% Solar simulations eject insufficient material to agree with the observations \citep[][and references therein]{gehrz_1998_aa, dellavalle_2020_aa}.  The simulations for the 50\% WD - 50\% Solar 
at 0.6 M$_\odot$, 0.8 M$_\odot$, and 1.35 M$_\odot$ barely agree with observed ejecta velocites, while the other 3 simulations for this mixture also eject insufficient amounts of material to agree with observations. The most material ejected at the highest velocities occurs for the 50\% WD - 50\% Solar simulation on the 1.35 M$_\odot$ WD.  However, the amount of ejected mass, $1.4 \times 10^{-6}$M$_\odot$, is lower than the typical ejecta mass estimates for CNe \citep{warner_1995_aa, gehrz_1998_aa, bode_evans_2008, starrfield_2012_basi, dellavalle_2020_aa, chomiuk_2021_aa}.  The comparison with the amounts of ejected material is improved for the simulations where we mix during the TNR (see Section \ref{MDTNR}). We also tabulate for both compositions the RE and it is clear that the WD is growing in mass as a consequence of accretion and the resulting  TNR. 


\begin{deluxetable}{@{}lccccccc}
\tabletypesize{\small}
\tablecaption{ MFB comparison with S2009 \citep[][]{starrfield_2009_aa}
\tablenotemark{}\label{ONecomp09}}
\tablewidth{0pt}
\tablecolumns{5}
 \tablehead{ \colhead{WD Mass:} &
\colhead{1.25M$_\odot$} &
\colhead{1.25M$_\odot$} &
\colhead{1.35M$_\odot$} &
\colhead{1.35M$_\odot$}}
 \startdata
 \hline
{\bf 50\% WD- 50\% Solar}&&\\
\hline
 Reference:&S2009&This Work&S2009&This Work\\
Initial:  L/L$_\odot$($10^{-3}$)&3.2&5.2&4.2&5.4\\
Initial:  R($10^3$km)&3.5&3.2&2.5&2.2\\
Initial:  T$_{\rm eff}$($10^4$K)&1.9&2.3&2.5&2.8\\
M$_{\rm acc}$($10^{-5}$M$_{\odot}$)&6.1&1.4&2.8&0.5\\
T$_{\rm peak}$($10^8$K)&3.2&2.7&3.9&3.2\\
$\epsilon_{\rm nuc-peak}$($10^{17}$erg gm$^{-1}$s$^{-1}$)&$1.3$&$0.3$&$4.4$&$0.9$\\
L$_{\rm peak}$($10^5$L$_\odot$)&2.6&0.5&5.9&0.7\\
T$_{\rm eff-peak}$($10^5$ K)&6.6&7.0&8.8&11.0 \\
M$_{\rm ej}$(M$_{\odot}$)&$1.5 \times10^{-5}$&$5.6 \times 10^{-7}$&$1.7\times10^{-5}$&$1.4\times10^{-6}$\\
RE\tablenotemark{a}: (M$_{\rm acc}$ -  M$_{\rm ej}$)/M$_{\rm acc}$&0.75&0.96&0.39&0.72\\
V$_{\rm max}$($10^3$km s$^{-1}$)&3.1&0.7&4.5&1.4\\
\enddata
\tablenotetext{a}{RE: Retention Efficiency see Figure \ref{retention}}

\end{deluxetable}

The 3 rows tabulated after the ejected mass in Table \ref{evolONeMFB} show some of the nucleosynthetic results from these simulations.  We single out $^7$Be (which decays to $^7$Li), $^{22}$Na, and $^{26}$Al because they are all radioactive and produced at some level in CNe.  The results for $^7$Li (produced as $^7$Be) show that it is enriched in ONe ejecta just as we found for
CO CNe \citep{starrfield_2020_aa} although the amount of enrichment (with respect to Solar) is only important for the more massive WDs. The enrichment is higher for CO CNe \citep{starrfield_2020_aa}. We will return to the $^7$Li ($^7$Be) results in Section \ref{lithium} and the $^{22}$Na results in Section \ref{sodium}.

$^{22}$Na, and $^{26}$Al are both predicted to be produced in ONe CNe because of the enriched $^{20}$Ne in an ONe WD, 
which is dredged up into the accreted matter during the TNR.  
As the temperatures increase in the TNR, repeated proton captures fuse $^{20}$Ne to more massive nuclei.  We find that while neither of these nuclei are produced in large amounts in the 25\% WD - 75\% Solar simulations, there is some enrichment in the 50\% WD - 50\% Solar simulations.  However, as shown in the next section (see Table \ref{evolONeMDTNR}), we find larger enrichments in the simulations based on ''Mixing During the TNR'' (MDTNR).

Finally, we compare our latest results for ONe WDs to those we published in \citet{starrfield_2009_aa}. 
In \citet{starrfield_2009_aa}, we also assumed MFB, used only one composition (50\% WD and 50\% Solar material),
and followed the TNR for only 1.25 M$_\sun$ and 1.35 M$_\sun$ WDs.  The purpose of that study was to analyze the effects of changes and improvements in the nuclear reaction rate library, the use of a new reaction rate solver
\citep{hix_1999_aa}, and the inclusion of the $pep$ reaction ($p + e^{-} +p \rightarrow d + \nu$),  which provided additional nuclear energy because of the high densities (compared to the center of the Sun) in the WD envelope.  Since \citet{starrfield_2009_aa} was published, as described in \citet{starrfield_2020_aa}, we have updated the reaction rate library to that of the Starlib nuclear reaction rates  \citep{sallaska_2013_aa}, the microphysics (opacities and equations of state), the initial composition, and increased the number of mass zones from 95 to 300.   We use the same value of \.M. 

The broad properties of the comparison are given in Table \ref{ONecomp09}, which has the same defined variables as listed in Table \ref{evolONeMFB}.  The results quoted from  \citet{starrfield_2009_aa} are only for the I2005A nuclear reaction rate library in that paper. As can be seen in Table \ref{ONecomp09}, the two sets of simulations are different.  We have run other simulations to determine the cause of the differences and, primarily, 
it is the change in the electron degenerate conductivities as discussed in \citet{starrfield_2020_aa}.  Implementing the new conductivities has caused the radius to shrink at each value of the WD mass, which, in turn, caused the initial model to have a larger central density and pressure.  We also found that it was necessary to initiate the simulation with a slightly higher luminosity and higher effective temperature.  
In addition, as described in the appendix in \citet{starrfield_2020_aa}, increasing the number of mass zones from 95 to 150 also changed the results.
 Because the radius is smaller and, thereby, the gravity is larger for both masses, it now takes less mass to reach the conditions to initiate the TNR.  With less mass involved in the TNR, the peak conditions are less violent and less material is ejected at lower velocities.    

\clearpage
\section{Simulations with the composition mixed during the Thermonuclear Runaway}
\label{MDTNR}

We began the work being reported here using MFB, two different ONe compositions, and a broad range of 
WD mass and found that the ejected mass and ejecta velocities were generally too small to agree with the observations.  In order to increase the amount of accreted material, therefore, we now use the results of multi-dimensional simulations as guides.  These studies show that sufficient material is dredged-up into the accreted envelope, from the outer layers of the WD, by convectively associated instabilities when the TNR is well underway \citep{casanova_2010_ab, casanova_2011_ab, casanova_2016_aa, casanova_2018_aa, jose_2014_aa, jose_2020_aa}.  We simulate their calculations by first accreting a Solar mixture \citep{lodders_2003_aa} until the peak energy generation in the nuclear burning region exceeds $10^{11}$ erg gm$^{-1}$ s$^{-1}$  so that $\sim$ 96\% of the accreted material is convective.  At this time, we switch the composition of the accreted layers to a mixed composition (both abundances and the associated equations of state and opacities) and subsequently evolve the simulation through peak temperature and decline.  This technique was discussed in detail in the appendix in \citet{starrfield_2020_aa}.

A similar technique was used by \citet{jose_2007_aa} in their study of the ``First Nova Explosions.'' They
explored a variety of time scales for mixing the WD material into the accreted layers, once convection was
underway, and found that short mixing time scales (seconds or less) was warranted.   More recently, \citet{jose_2020_aa} have
expanded their earlier studies by using both 1-D and multi-D codes to explore this technique and
its effect on CN explosions.  They choose a temperature of $10^8$K to make the switch from 1-D to 
multi-D based on the multi-D study of \citet{glasner_2007_aa}, who explored the consequences of mixing times for TNR  temperatures ranging from 
$5 \times 10^7$ K to $9 \times 10^7$ K.  Our chosen energy generation at which we mix (for the various WD masses in this study) is equivalent to temperatures of $\sim 6 \times 10^7$ K, which results in a convective region that spans almost the entire accreted envelope. It typically 
takes NOVA less than $\sim$2 s of ``star'' time (but many time steps) to converge to the new composition.

\subsection{Solar Accretion}
\label{Solar}

In this subsection, we present the evolution of just the Solar accretion phase of the study.  A similar section
was provided in \citet{starrfield_2020_aa} and the small differences are a slightly higher initial
luminosity and that all the simulations now have 300 mass zones instead of the 150 mass zones used in that paper.
The next subsection describes the simulations assuming the mixed compositions. The initial conditions and evolutionary results are presented in Table \ref{evolONeMDTNR}.  The variables in the tables are the same as already described for Table \ref{evolONeMFB}.  The initial conditions for each of the 6 ONe WD masses are given in the first 3 rows.  The values in these rows are identical to the first 3 rows in Table \ref{evolONeMFB},
 and are repeated here for clarity. The next two rows give the accretion time to when the peak energy generation reaches $10^{11}$ erg gm$^{-1}$ s$^{-1}$ and the accreted matter is nearly completely convective,  $\tau$$_{\rm acc}$,  and M$_{\rm acc}$ is the total accreted mass at that time.  These values are those used both for the Solar accretion simulations and, later, for the two mixed composition simulations for each of the listed WD masses (Section \ref{mixdtnr}).  We begin each of the sets of simulations with the composition listed in bold. We use a mass accretion rate of $1.6 \times 10^{-10}$M$_\odot$ yr$^{-1}$.  

\begin{figure}[htb!]
\includegraphics[width=1.0\textwidth]{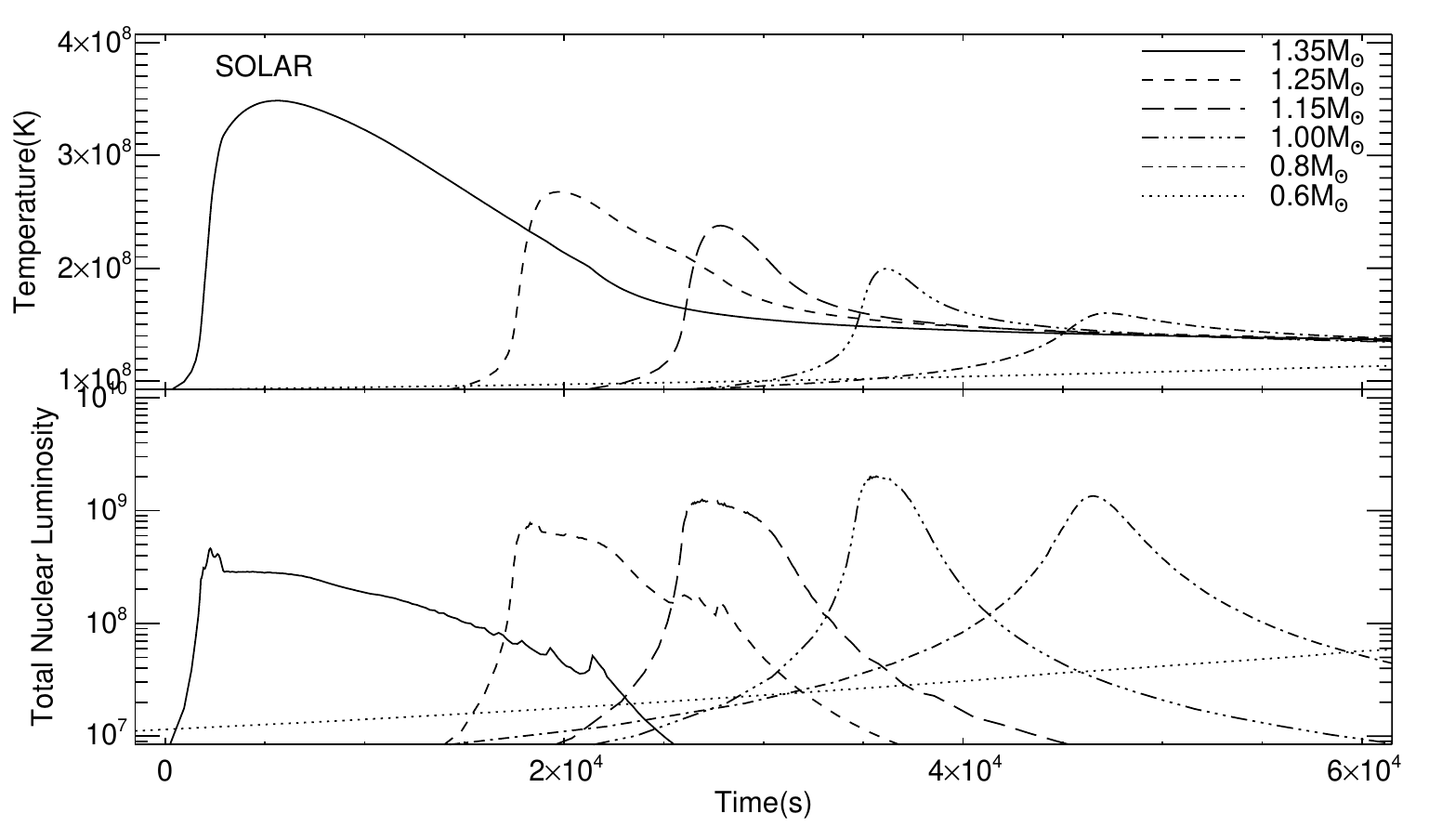}
\caption{Top panel:  the variation with time of the temperature in those mass zones near the interface
between the outer layers of the WD and the accreted Solar matter. The results for all six simulations are shown with the WD mass identified in the legend.  The curve for each sequence has been shifted in time to improve its visibility. As expected, the peak temperature achieved in each simulation is an increasing function of WD mass.
Bottom panel: The variation with time of the total nuclear luminosity in Solar units (L$_\odot$) around the
time of peak temperature during the TNR. We integrate over all zones with ongoing nuclear fusion to obtain the plotted numbers. The identification with each WD mass is given on the plot and the evolution time has again been shifted to improve visibility.  The irregularities seen in the more massive WD evolution are caused by the convective region changing its spatial extent, with respect to the mass zones,  as the material expands. Unlike all the other sets of simulations, the total energy increases as the WD mass decreases. This is because the accreted mass
has declined with increasing WD mass and there is less material involved in the evolution. In contrast,
both peak temperature and peak energy generation do increase with increasing WD mass as is
shown in Table \ref{evolONeMDTNR}.}
\label{ONESolarmulti}
\end{figure}

We immediately see the effects of accreting a Solar composition instead of a mixed composition.
Comparing the results given in Table \ref{evolONeMDTNR} to those in Table \ref{evolONeMFB}, the reduced amount of $^{16}$O in the Solar accretion simulations (a mass fraction of $6.6 \times 10^{-3}$ compared to either 0.133 or 0.261, respectively) increases the amount of mass accreted before the TNR.  For example, comparing the Solar accretion simulation to the 25\% WD  -75\% Solar (MFB) simulation, we find that slightly more mass is accreted at 0.6 M$_\odot$ and about 66\% more mass at 1.35 M$_\odot$.   The difference is more significant in the 50\% WD - 50\% Solar simulations.  Peak temperature is higher for all WD masses (except 0.6 M$_\sun$) in the Solar accretion simulation when compared to the MFB simulations.  The increased mass and degeneracy at the bottom of the accreted material clearly compensates for the larger amount of $^{16}$O in the MFB simulations.

\begin{figure}[htb!]
\includegraphics[width=1.0\textwidth]{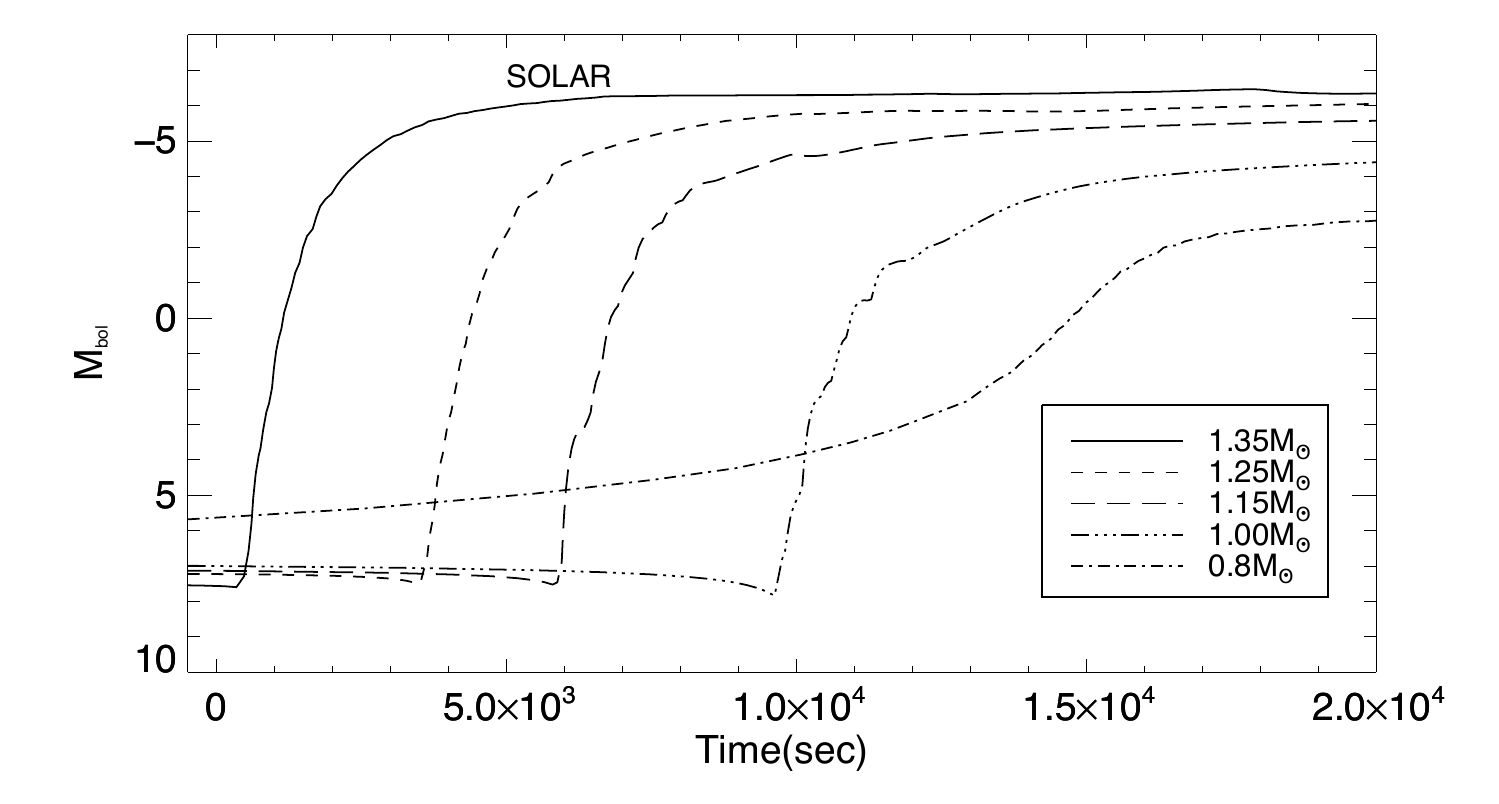}
\caption{The variation with time of the absolute bolometric magnitude for the Solar composition
simulations.  Only the simulations on the most massive WDs reach peak values close to
those that are observed.  These simulations evolve extremely slowly  compared to the mixed compositions (MDTNR) as can be seen on the time axis.  We do not plot the evolution of the 0.6 M$_\odot$ simulation since it is still rising after $2 \times 10^4$s. }
\label{ONESolarmbolt}
\end{figure}

These simulations are used to determine the amount of mass accreted before switching to a mixed composition.  Nevertheless, we follow them through the explosion. The time axis is much longer than for any of the other 
simulations because of the slow evolution of these sequences.  The 0.6 M$_\odot$ simulation does not
reach peak temperature until after $6 \times 10^4$ s of evolution although it started from the same beginning
temperature as the simulations for other WD masses.

Figure \ref{ONESolarmulti} (top panel) shows the evolution of the temperature with time for the zones where peak conditions in the TNR occur for all the WD masses 
accreting just a Solar composition.  Although there is more accreted mass in each of the simulations, the temperature evolution is extremely slow, as shown by the time axis.  
While the 1.35~M$_\odot$ simulation takes  $\sim 2 \times10^4$ s to evolve through the peak and decline of the TNR,  the 0.6 M$_\odot$ simulation is still on the rise after $\sim 6 \times 10^4$ s.  In contrast, the equivalent  MFB and MDTNR simulations take far shorter times to evolve through the peak.

Figure \ref{ONESolarmulti} (bottom panel) shows the variation in total nuclear energy generation around the peak of the TNR. 
The rise to peak nuclear energy generation is extremely slow and the decline is also slow.  
The ``glitches'' seen in the more massive WD evolution are caused by the convective region
changing its spatial extent, with respect to the mass zones, as the material expands. 
The total energy as a function of time increases as the WD mass decreases. This is because the total accreted mass
has declined with increasing WD mass and there is less material involved in the evolution. In contrast,
both peak temperature and peak energy generation do increase with increasing  WD mass as is
shown in Table \ref{evolONeMDTNR}.

Figure \ref{ONESolarmbolt} shows the time evolution of the bolometric magnitude for the Solar accretion simulations.  Peak M$_{\rm{bol}}$ is an increasing function of WD mass but even the simulation on the most massive WD does not reach a value that is observed in a typical CN outburst.  The simulations on low mass WDs might fit some of the slowest and faintest CNe shown in \citet{kasliwal_2011_aa} or the evolution of Nova Velorum 2022 as described by \citet{aydi_2023_aa}.  Obtaining the element abundances for the Nova Velorum 2022 ejecta could test these simulations.  In addition, the existence of such slow novae might provide information about the outer layers of the WD.  
For example, if there is a helium layer, the remnant of previous outbursts, it is possible that mixing during the TNR might not 
reach down to the CO or ONe outer layers of the WD.



\subsection{The Simulations using Compositions Mixed During the TNR (MDTNR)}
\label{mixdtnr}

\begin{deluxetable}{@{}lccccccc}
\centerwidetable 
\tabletypesize{\footnotesize}
\tablecaption{Initial Parameters and Evolutionary Results for Accretion onto ONe WDs Mixing after the TNR (MDTNR)
\tablenotemark{}\label{evolONeMDTNR}}
\tablewidth{8pt}
\tablecolumns{7}
 \tablehead{ \colhead{WD Mass:} &
 \colhead{0.6M$_\odot$} &
 \colhead{0.8M$_\odot$} &
\colhead{1.0M$_\odot$} &
\colhead{1.15M$_\odot$} &
\colhead{1.25M$_\odot$} &
\colhead{1.35M$_\odot$}}
 \startdata
 Initial:  L/L$_\odot$($10^{-3}$)&5.0&5.0&5.0&5.0&5.2&5.4\\
Initial: R($10^3$km)&8.5&6.8&5.2&4.1&3.2&2.2\\
Initial: T$_{\rm eff}$($10^4$K)&1.4&1.6&1.8&2.0&2.3&2.8\\
\hline
 $\tau$$_{\rm acc}$($10^5$ yr)&18.0&9.6&4.7&2.4&1.3&0.6\\
M$_{\rm acc}$($10^{-5}$M$_{\odot}$)&29.0&15.0&7.5&3.8&2.0&1.0\\
\hline
{\bf Solar mixture}&&&&&&\\
\hline
T$_{\rm peak}$($10^8$K)&1.2&1.6&2.0&2.4&2.6&3.2\\
$\epsilon_{\rm nuc-peak}$($10^{14}$erg gm$^{-1}$s$^{-1}$)&$0.03$&$0.84$&$1.8$&$1.7$&$1.8$&$2.1$\\
L$_{\rm peak}$/L$_\odot$ ($10^4$)&3.4&6.1&2.4&2.7&3.2&4.2\\
T$_{\rm eff-peak}$($10^5$K)&0.9&1.8&1.9&3.6&5.2&8.8 \\
M$_{\rm ej}$(M$_{\odot}$)&$5.8\times 10^{-7}$&$1.4\times10^{-6}$&$2.0\times10^{-8}$&$5.6\times10^{-8}$&$3.7\times10^{-8}$&$7.8\times10^{-8}$\\
N($^7$Li/H)$_{\rm ej}$/N($^7$Li/H)$_{\odot}$ &$1.0 \times10^{-5}$&$1.6 \times10^{-5}$&$4.4 \times10^{-3}$&$3.4 \times10^{-4}$&$2.8 \times10^{-3}$&$2.5 \times10^{-2}$\\
$^7$Li$_{\rm ej}$($10^{-19}$M$_{\odot}$)&0.6&2.1&8.6&1.7&9.2&160.0\\
$^{22}$Na$_{\rm ej}$($10^{-14}$M$_{\odot}$)&116.0&7.0&2.8&17.9&3.4&0.002\\
$^{26}$Al$_{\rm ej}$($10^{-13}$M$_{\odot}$)&170.0&270.0&0.4&1.7&0.3&0.0005\\
RE\tablenotemark{a}: (M$_{\rm acc}$ -  M$_{\rm ej}$)/M$_{\rm acc}$&1.0&0.99&1.0&0.98&0.98&0.92\\
V$_{\rm max}$($10^2$km s$^{-1}$)&4.3&5.0&3.5&4.5&5.7&3.8\\
\hline
{\bf 25\% WD - 75\% Solar: MDTNR}&&&&&&\\
\hline
T$_{\rm peak}$($10^8$K)&1.3&1.7&2.1&2.5&2.8&3.7\\
$\epsilon_{\rm nuc-peak}$(erg gm$^{-1}$s$^{-1}$)&$2.1\times10^{13}$&$6.0 \times10^{14}$&$3.0 \times10^{15}$&$7.2 \times10^{15}$&$1.2 \times10^{16}$&$6.8 \times10^{16}$\\
L$_{\rm peak}$/L$_\odot$ ($10^4$)&1.3&3.2&2.4&3.9&5.7&12.0\\
T$_{\rm eff-peak}$($10^5$K)&0.9&2.2&2.6&4.2&6.1&9.2 \\
M$_{\rm ej}$(M$_{\odot}$)&$7.9 \times10^{-7}$&$1.1 \times10^{-6}$&$3.6 \times10^{-9}$&$1.3 \times10^{-8}$&$6.7 \times10^{-8}$&$2.6 \times10^{-7}$\\
N($^7$Li/H)$_{\rm ej}$/N($^7$Li/H)$_{\odot}$ &$3.0 \times10^{-1}$&$3.4$&$6.7 \times10^{1}$&$1.6 \times 10^{2}$&$2.0\times10^{2}$&$2.4 \times10^{2}$\\
$^7$Li$_{\rm ej}$($10^{-14}$M$_{\odot}$)&0.2&2.6&0.2&1.6&9.4&39.0\\
$^{22}$Na$_{\rm ej}$($10^{-11}$M$_{\odot}$)&8.7&5.9&0.04&0.86&10.7&102.0\\
$^{26}$Al$_{\rm ej}$($10^{-11}$M$_{\odot}$)&29.0&180.0&0.5&1.6&1.7&2.1\\
RE\tablenotemark{a}: (M$_{\rm acc}$ -  M$_{\rm ej}$)/M$_{\rm acc}$&1.0&0.99&1.0&1.0&1.0&0.97\\
V$_{\rm max}$($10^2$km s$^{-1}$)&4.2&4.8&3.2&4.1&4.8&6.4\\
\hline
{\bf 50\% WD - 50\% Solar: MDTNR}&&&&&&\\
\hline
T$_{\rm peak}$($10^8$K)&1.4&1.8&2.2&2.6&2.9&3.8\\
$\epsilon_{\rm nuc-peak}$(erg gm$^{-1}$s$^{-1}$)&$3.2 \times10^{13}$&$1.2 \times10^{15}$&$8.2  \times10^{15}$&$2.7 \times10^{16}$&$4.9 \times10^{16}$&$1.6 \times10^{17}$\\
L$_{\rm peak}$/L$_\odot$ ($10^5$)&0.2&0.8&0.6&1.0&1.6&4.1\\
T$_{\rm eff-peak}$($10^5$K)&1.0&2.8&3.0&5.2&6.7&10.0 \\
M$_{\rm ej}$(M$_{\odot}$)&$7.9 \times10^{-7}$&$1.4 \times10^{-6}$&$3.9 \times10^{-8}$&$3.6 \times10^{-6}$&$7.5 \times10^{-6}$&$5.5 \times10^{-6}$\\
N($^7$Li/H)$_{\rm ej}$/N($^7$Li/H)$_{\odot}$ &$4.4$&$1.2 \times10^2$&$6.9 \times10^{2}$&$8.5 \times10^{2}$&$2.6 \times 10^{3}$&$3.3 \times 10^{3}$\\
$^7$Li$_{\rm ej}$($10^{-11}$M$_{\odot}$)&0.002&0.08&0.01&1.3&7.5&6.0\\
$^{22}$Na$_{\rm ej}$($10^{-10}$M$_{\odot}$)&1.3&1.3&0.01&22.0&38.0&143.0\\
$^{26}$Al$_{\rm ej}$($10^{-10}$M$_{\odot}$)&5.1&63.0&1.4&20.5&82.5&187.0\\
RE\tablenotemark{a}: (M$_{\rm acc}$ -  M$_{\rm ej}$)/M$_{\rm acc}$&1.0&0.99&1.0&0.90&0.62&0.45\\
V$_{\rm max}$($10^2$km s$^{-1}$)&4.2&4.7&4.3&13.0&26.0&38.0\\
\enddata
\tablenotetext{a}{RE: Retention Efficiency see Figure \ref{retention}}

\end{deluxetable}

We now take the results for each WD mass from the simulations reported in the last subsection and
switch to a mixed composition when the peak energy generation in the simulation reaches  $10^{11}$ erg gm$^{-1}$ s$^{-1}$.  
The TNR and convection are well underway and the convective region extends almost to the surface.  
Once we have switched the composition, we continue the evolution through peak temperature of the TNR
and follow the decline in temperature until there is no further nuclear burning in the
outer layers.  We use the same two mixtures (either 25\% WD and 75\% Solar material or 50\% WD and 50\% Solar material) as used in the MFB simulations.   These two sets of simulations are identified on the plots as either 25\_75 Mixing During the TNR (25\_75 MDTNR) or 50\_50 Mixing During the TNR (50\_50 MDTNR).

\begin{figure}[htb!]
\includegraphics[width=1.0\textwidth]{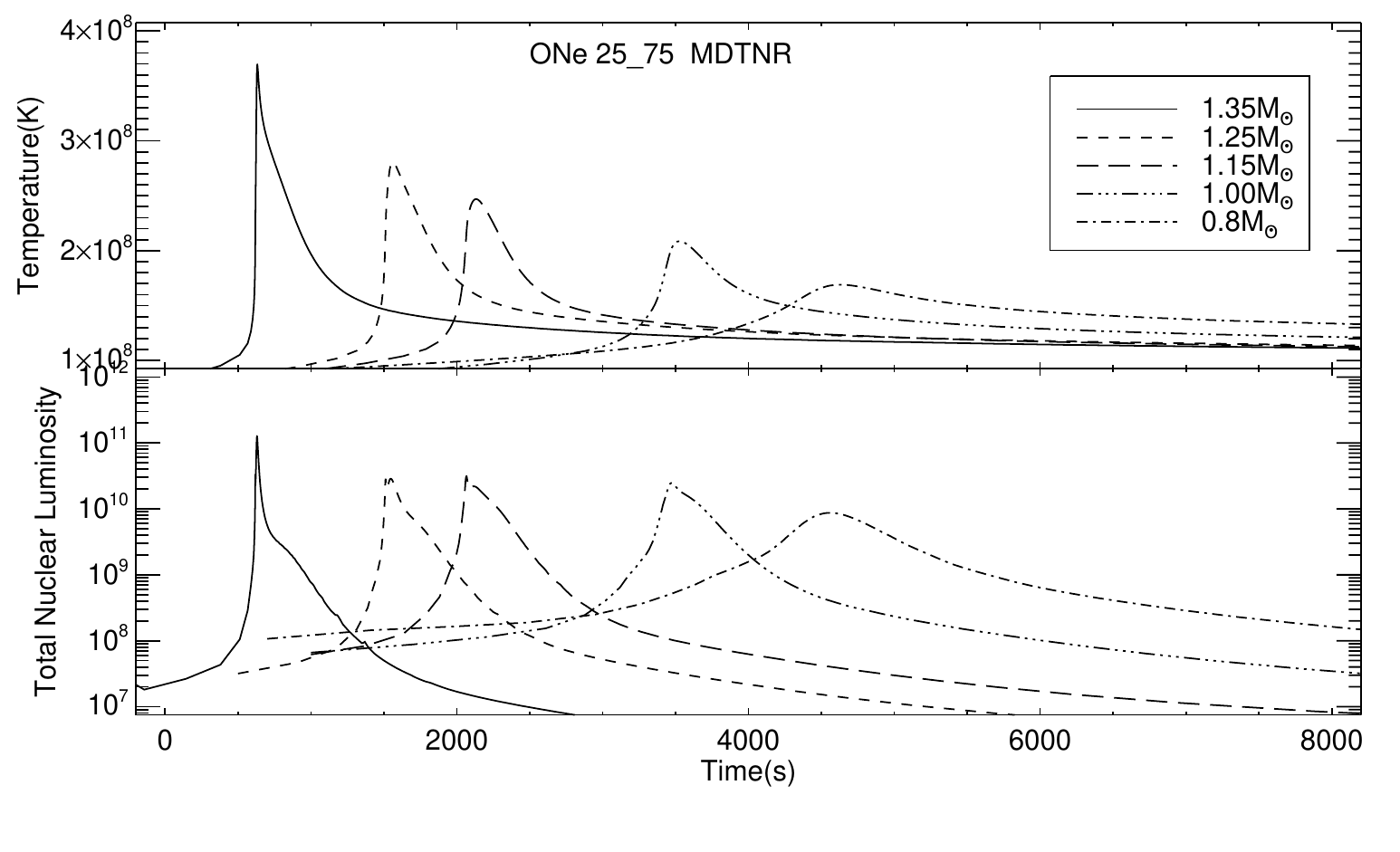}
\caption{Top panel: the variation with time of the temperature in those mass zones near the interface
between the outer layers of the WD and the accreted plus WD matter 
for the MDTNR simulations with 25\% WD material and 75\% Solar. 
The results for five of the six simulations are shown (the WD mass is identified in
the inset).  We do not show that for the 0.6 M$_\sun$ WD because it evolves too slowly to show in this plot.  The curve for each sequence has
been shifted in time to improve its visibility. As expected, the peak temperature
achieved in each simulation is an increasing function of WD mass. 
Bottom panel: the variation with time of the total nuclear
luminosity in Solar units (L$_\odot$) around the
time of peak temperature during the TNR. We
integrate over all zones with ongoing nuclear fusion to obtain the plotted numbers. The
identification with each WD mass is given on the plot and the evolution time has again been
shifted by a few 100 s to improve visibility.}
\label{figuremultiaa}
\end{figure}

These are the same {\it initial} models (thermal structure, spatial structure, and amount of accreted mass distributed through the same number of mass zones) that are used for the following sets of simulations where we use either 25\% WD and 75\% Solar material or 50\% WD and 50\% Solar material.  The properties of the resulting evolution now depend on the WD mass and the amount of $^{16}$O in the mixed layers (Solar plus WD) {\it after} the switch in composition.  Since, the pressure and density are unchanged by the switch in composition,  increasing the $^{16}$O abundance  changes the mean molecular weight, $\mu$,
which increases the temperature and, thereby, the rate of energy generation.  
Peak temperature is reached about one hundred seconds after the temperature exceeds $10^{8}$ K.  The rise in
temperature ends because virtually all the CNO nuclei in the convective region have been converted to radioactive isotopes that will decay 
subsequently by positron emission.

As discussed in the last subsection, the first set of evolutionary results in Table \ref{evolONeMDTNR} shows the consequences of 
following each of the Solar accretion simulations through the TNR and its return to quiescence.  
The next two sets of rows provide exactly the same information but for the mixed composition simulations with 25\% WD and 75\% Solar matter, and followed
below by the results for the 50\% WD and 50\% Solar matter simulations.   Comparing the 25\% WD - 75\% Solar MDTNR simulations to the
50\% WD - 50\% Solar MDTNR simulations, the enrichment of $^{16}$O in the simulations with more WD matter causes a more extreme set of evolutionary results.

\begin{figure}[htb!]
\includegraphics[width=1.0\textwidth]{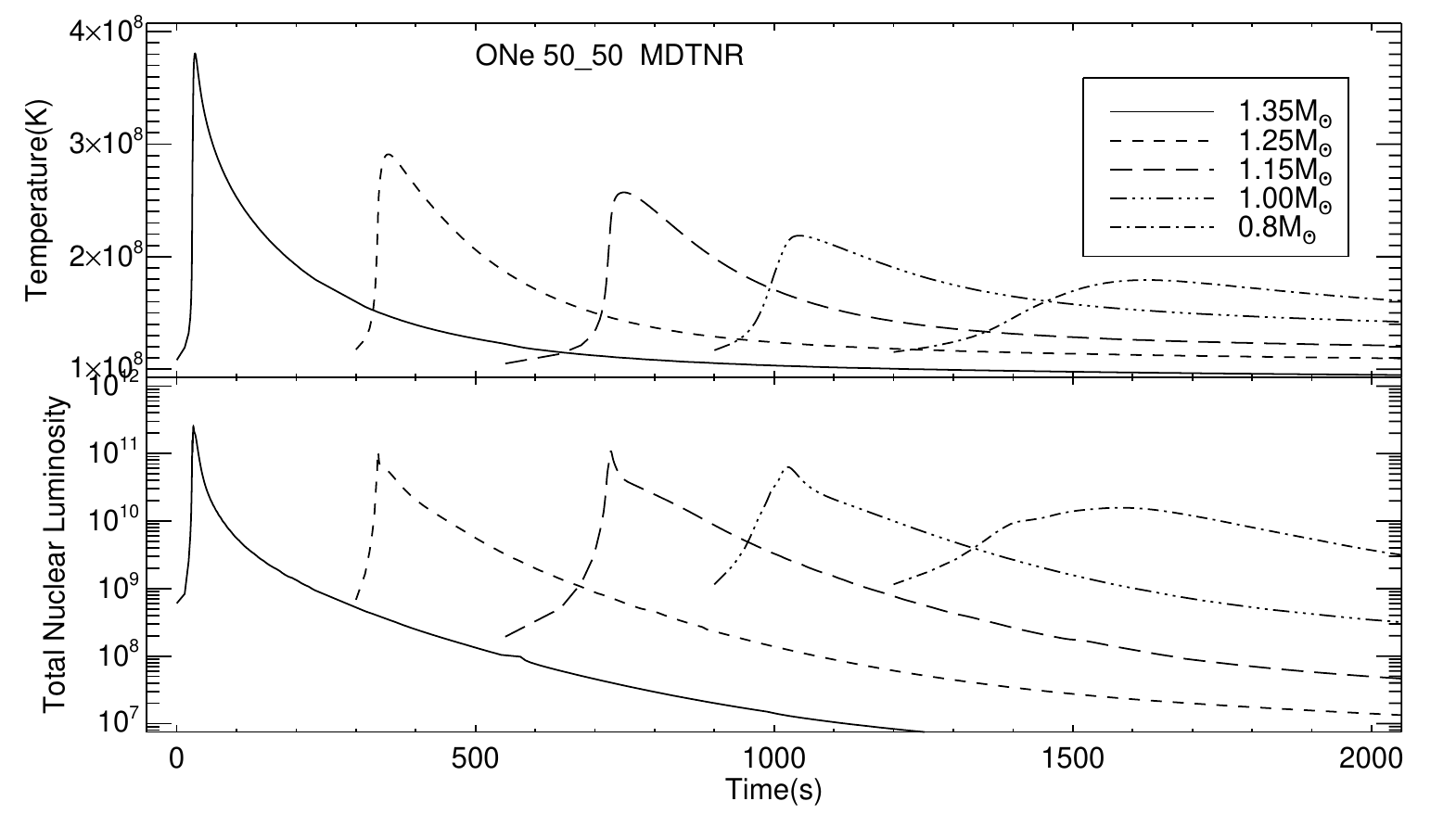}
\caption{Top panel: same as Figure \ref{figuremultiaa} but for the simulations with 
50\% WD matter and 50\% accreted matter.  Peak temperature is an increasing function of WD mass.  
In addition, because of the increased $^{16}$O abundance, the rate of energy generation is larger for a given temperature
and density, and these simulations evolve much more rapidly than the simulations with a lower $^{16}$O
abundance.  
Interestingly, in the CO simulations, where there was a much higher abundance of $^{12}$C, the
extremely rapid increase in temperature formed a shock wave in the
zone where peak temperature occurred \citep{starrfield_2020_aa}. This did not occur in these simulations. 
Bottom panel: the same plot as in Figure \ref{figuremultiaa} but for the simulation with
50\% WD matter and 50\% accreted matter.  Because of the larger abundance of $^{16}$O in these simulations, 
the evolution is more extreme and faster than for those simulations with a smaller amount
of $^{16}$O.}
\label{figuremultiab}
\end{figure}

The plots of temperature versus time for these two sets of simulations are given in Figures \ref{figuremultiaa} and \ref{figuremultiab}.  The horizontal scales in these two figures are different since the simulations with the more enriched nuclei evolve faster.   As already noted, the peak temperature during the TNR is an increasing function of WD mass.  For example, as shown in Figure \ref{figuremultiaa}, the simulation with 25\% WD - 75\% Solar MDTNR involving a WD with a mass of 1.35 M$_\odot$ reaches the highest temperature of $3.7 \times 10^8$ K, while the simulation on the 0.6 M$_\odot$ WD reaches only a peak temperature of $1.3 \times 10^8$ K.  Comparing the results of the solar and two mixed (MDTNR) compositions, however, we find that peak temperature depends more on WD mass than composition.

Figures \ref{figuremultiaa} (for the 25\% WD - 75\% Solar MDTNR simulations) and \ref{figuremultiab} (for the 50\% WD - 50\% Solar MDTNR simulations) also show (bottom panels) the variation with time of the  total nuclear
luminosity in solar units (L$_\odot$) around the time of peak temperature.   Both sets of simulations show a rapid rise to maximum followed by a slower decline.   The rise to maximum nuclear luminosity occurs as the convective region encompasses all the accreted layers, thus
carrying the  $\beta^+$- unstable nuclei to the surface and unprocessed nuclei down to the nuclear burning region.

As in the MFB evolutionary sequences, the rise time for the most massive WD MDTNR simulations is
shorter than for the lower mass simulations.  In contrast to the Solar simulations, as shown in
Figure \ref{figuremultiaa}  and Figure \ref{figuremultiab}, the total nuclear luminosity also increases as the WD mass increases. The peak is highest for the  50\% WD - 50\% Solar MDTNR simulations.  
The nuclear energy in the 50\% WD - 50\% Solar MDTNR simulations declines faster than in the 25\% WD - 75\% Solar simulations because the expansion velocities are larger and the temperatures in the envelope are decreasing more rapidly. 

\begin{figure}[htb!]
\includegraphics[width=1.0\textwidth]{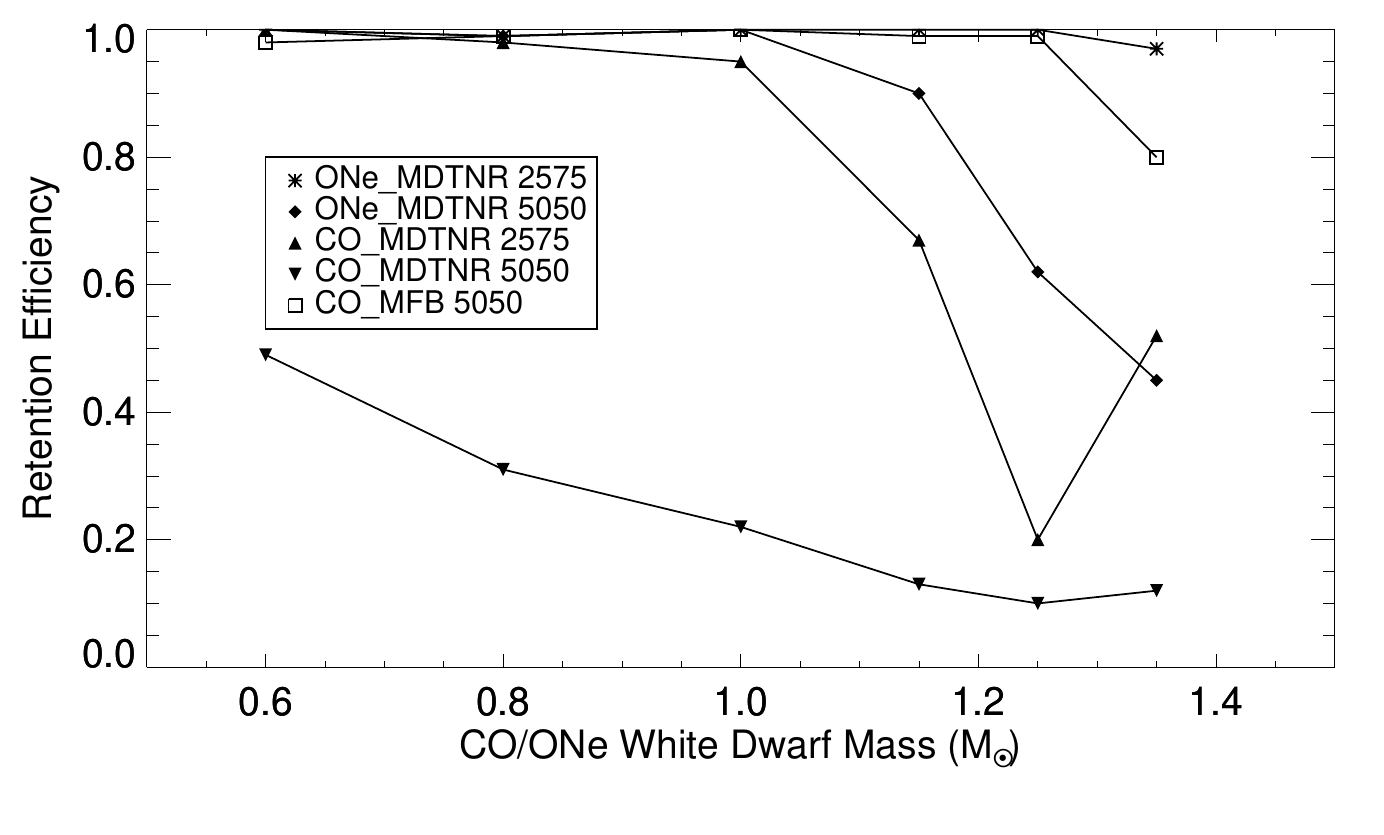}
\caption{The retention efficiency (RE), (M$_{\rm acc}$ -  M$_{\rm ej}$)/M$_{\rm acc}$, as a function
of WD mass (either CO or ONe) for some of our 
simulations (using different symbols) from both this paper and \citet{starrfield_2020_aa}.  Those sets of simulations that are not plotted but discussed in both papers have a RE of nearly 1.0. This plot shows clearly that most of the matter
accreted is not ejected during the explosion and the WD is growing in mass.}
\label{retention}
\end{figure}

Table \ref{evolONeMDTNR} shows that the peak luminosities for the 25\% WD - 75\% Solar MDTNR simulations range from $1.3 \times 10^4$ L$_\odot$ for the 0.6 M$_\odot$ WD to $1.2 \times 10^5$ L$_\odot$  for the 1.35 M$_\odot$ WD.  The peak luminosities for the 50\% WD - 50\% Solar MDTNR mixture range from $2.0 \times 10^4$ L$_\odot$ to $4.1 \times 10^5$ L$_\odot$. These luminosities fall in the range of observed luminosities for fast CNe in outburst and, 
 in combination with the predicted effective temperatures, which range from  $1.0 \times 10^5$ K for the 0.6 M$_\odot$ simulation to $1.0 \times 10^6$ K for the 1.35 M$_\odot$ simulation, should trigger responses in some of the X-ray detectors currently in orbit as has now been realized \citep{konig_2022_aa}.  The cause of these high luminosities and
effective temperatures is that convection has transported large amounts of the $\beta^+$-unstable nuclei to the
surface and their decays have resulted in the energy generation in those layers exceeding  $10^{12}$ erg gm$^{-1}$ s$^{-1}$.

We follow each of the simulations through peak nuclear burning, peak temperature, and decline.  We end the simulation when the outer layers have reached radii of a few $\times 10^{12}$ cm.  At these radii the outer mass zones are become optically thin and, in most of the simulations, some of the expanding material has exceeded the escape velocity and we tabulate it as escaping. We continue the simulation until the densities in those mass zones have declined to below $\sim 10^{-12}$ gm cm$^{-3}$, are off the physics tables, and we normally end the evolution.  
 {In some of the simulations, however, we linearly extrapolate below the EOS tables to lower densities and evolve the simulation until the outer zones reach radii of  $\sim 10^{13}$ cm.  In none of these cases is significantly more material ejected (one or two mass zones at a maximum) before we end the evolution. We then examine the mass zones just below those that are escaping and find velocities that are far less than the escape velocity. }

\begin{figure}[htb!]
\includegraphics[width=0.9\textwidth]{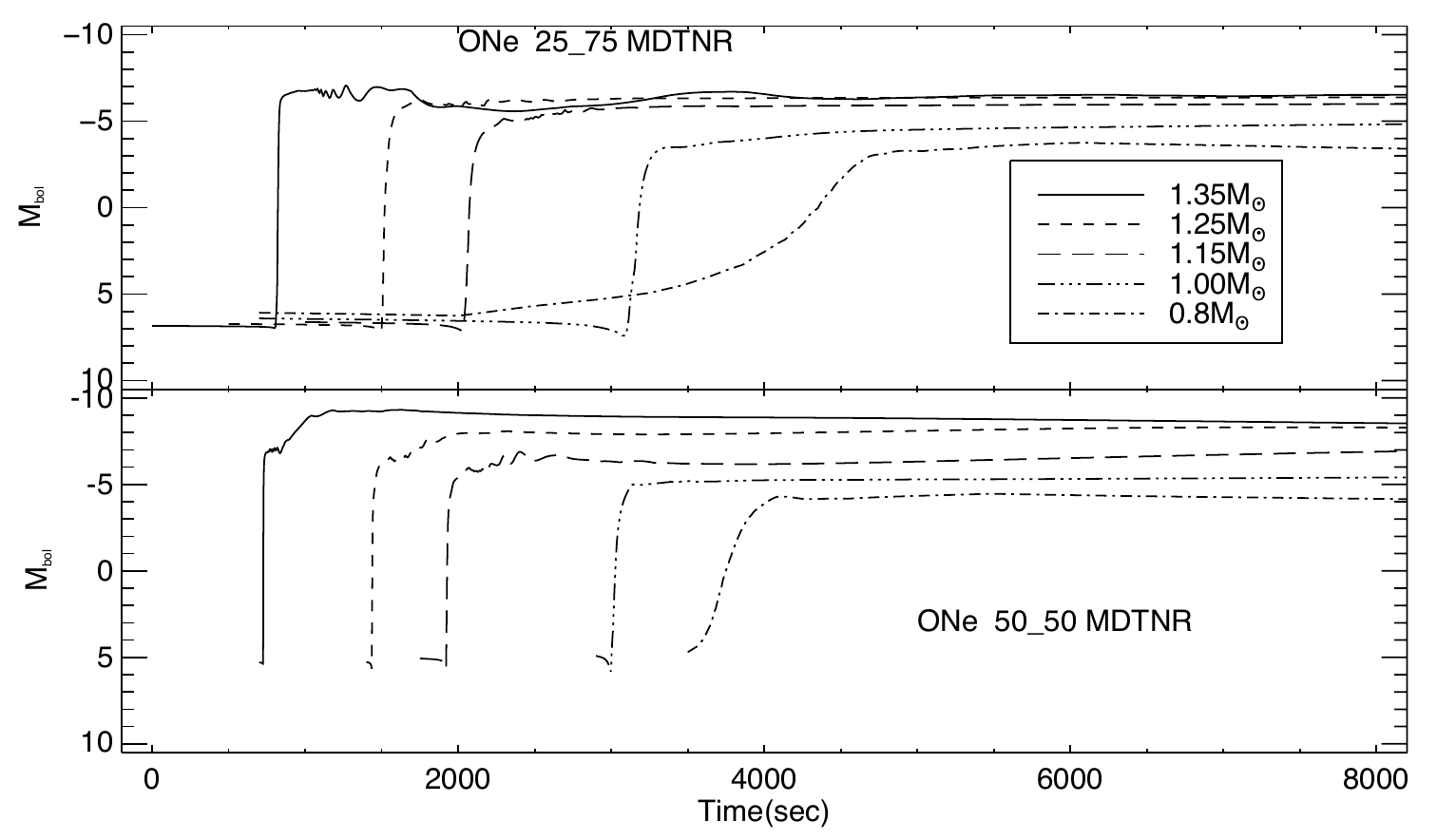}
\caption{Top panel: the variation with time of the absolute bolometric magnitude for
the simulations where we used a composition of 25\% WD matter and
75\% Solar after the TNR was well underway (MDTNR).  After the initial
few hundred seconds they show a range in peak bolometric 
magnitude as a function of WD mass with the most massive WD exhibiting
the brightest bolometric magnitude.
Bottom panel: The same plot as in the top panel but for a
composition of 50\% WD and 50\% Solar matter.  The larger amount of
$^{16}$O in these simulations results in bolometric magnitudes that are in
much better agreement with observations.  The peak M$_{\rm{bol}}$
around -8 is approximately that for a typical CN outburst. }
\label{MDTNRLC}
\end{figure}

 Tables \ref{evolONeMFB} and \ref{evolONeMDTNR} give the mass ejected by each of the simulations
along with the RE as a function of WD mass.  For most cases, the RE is so close to 1.0 that it is clear that almost
no mass is ejected during the outburst and the WD must be gaining in mass.  This can be seen more clearly
in Figure \ref{retention} which plots the RE as a function of WD mass and composition. 
The only sequences that eject a significant amount  of material are the 50\% WD - 50\% Solar MDTNR sequences on the most massive CO WDs.  However, for these simulations only 50\% of the ejecta is WD material.  
Moreover, the amount of ejected material will be reduced by increasing the mass accretion rate, or the 
initial WD luminosity, or both \citep{yaron_2005_aa, starrfield_2016_aa, hillman_2015_aa, hillman_2015_ac, hillman_2016_aa, 
hillman_2022_aa, chen_2019_aa}.  
We assert, therefore, that for most observed ONe CNe less than about half of the material in the accreted region comes from the WD and it is gaining 
in mass as a result of accretion, TNR, and ejection. These tables also give the ejecta abundances for $^7$Li, $^{22}$Na, and $^{26}$Al.  We discuss these results in the next
section (Section \ref{nucleo}).

We end this subsection with plots of the evolution of the bolometric magnitude (M$_ {\rm{bol}}$) with time.  Figure \ref{MDTNRLC} shows the first hours of the evolution of both sets of simulations (top panel: 25\_75 MDTNR and bottom panel: 50\_50 MDTNR).  Both panels show the rapid rise in M$_ {\rm{bol}}$ as the energy produced in the nuclear burning region reaches the surface. 
Subsequently,  M$_ {\rm{bol}}$ becomes roughly constant with time up to the end of the simulations.  
The light curves for the most massive WDs on the 25\% WD - 75\% Solar appear a bit low for a typical CNe in outburst.
Nevertheless, M$_{\rm{bol}}$ for the 50\% WD - 50\% Solar MDTNR simulations, on the more massive WDs, 
are in the observed range for CNe and are brighter than those shown for the other compositions.  
Since we are plotting absolute bolometric magnitudes, we do not attempt to fit them to observations. 
Finally, we end our simulations before these CNe would have been discovered.  Our predicted absolute visual magnitudes rise slowly
and reach values close to those plotted for peak M$_{\rm{bol}}$ after a few hours when the T$_{\rm eff}$ 
of the simulation has fallen below $10^4$ K.

\section{Nucleosynthesis}
\label{nucleo}

In this section we discuss the nucleosynthesis results from our simulations.  We provide these results both as tables 
of the ejecta abundances and as production plots.  In the next subsection we discuss the results for $^7$Li, $^{22}$Na, and $^{26}$Al  and end with a 
subsection on the other isotopes.  The abundances for all the nuclei in these tables (Table \ref{ONe2575MFBabund}, Table \ref{ONe5050MFBabund}, Table \ref{solarabund}, Table \ref {ONe2575MDTNRabund}, and Table \ref{ONe5050MDTNRabund}) are given in mass fraction while
Tables \ref{evolONeMFB} and \ref{evolONeMDTNR} give the predicted ejecta abundances in Solar masses for 
$^{22}$Na and $^{26}$Al, and for $^7$Li as a ratio to the Solar abundance and also the amount ejected in Solar masses.


\begin{deluxetable}{@{}lcccccc}
\tablecaption{Ejecta Abundances for 25-75  mixture in ONe WDs Mixing from Beginning (MFB)\tablenotemark{a} \label{ONe2575MFBabund}}
\tablewidth{0pt}
\tablecolumns{7}
 \tablehead{ \colhead{WD Mass}&
\colhead{0.6 M$_\odot$ \tablenotemark{}} & 
\colhead{0.8 M$_\odot$\tablenotemark{b}} &
\colhead{1.0 M $_\odot${}} &
\colhead{1.15 M$_\odot${}} &
\colhead{1.25 M$_\odot${}} &
\colhead{1.35 M$_\odot${}}} 
\startdata
$^1$H&$5.2 \times10^{-1}$&$5.1 \times10^{-1}$&$5.0 \times10^{-1}$&$4.9 \times10^{-1}$&$4.7 \times10^{-1}$&$4.4 \times10^{-1}$ \\
$^{3}$He&$9.9\times10^{-9}$  &$4.7 \times 10^{-10}$&$3.2 \times 10^{-9}$& $9.3 \times 10^{-10}$&$4.9\times 10^{-10}$ &$3.1 \times 10^{-12}$\\
$^4$He&$2.3 \times10^{-1}$&$2.3 \times10^{-1}$&$2.4 \times10^{-1}$&$2.6 \times10^{-1}$&$2.8 \times10^{-1}$&$2.9 \times10^{-1}$\\
$^{7}$Be&$1.6 \times10^{-10}$  &$1.7\times 10^{-11}$&$3.4 \times 10^{-9}$& $2.6 \times 10^{-8}$&$4.9 \times 10^{-8}$ &$9.4 \times 10^{-8}$\\
$^{7}$Li\tablenotemark{c}  &$0.0$&$0.0$&$1.4 \times 10^{-15}$& $2.3 \times 10^{-15}$&$2.2 \times 10^{-15}$&$0.0$\\
$^{12}$C &$8.8 \times10^{-4}$&$1.3\times10^{-3}$&$3.7 \times 10^{-3}$& $9.1 \times 10^{-3}$&$1.1 \times 10^{-2}$ &$6.0\times 10^{-3}$\\
$^{13}$C &$3.8 \times10^{-4}$&$7.6 \times10^{-4}$&$6.6 \times10^{-3}$ &$1.7 \times 10^{-2}$&$1.9 \times 10^{-2}$&$1.1\times 10^{-2}$\\
$^{14}$N&$3.4 \times10^{-2}$ &$3.9 \times 10^{-2}$&$3.3 \times10^{-2}$&$2.8 \times10^{-2}$&$2.9 \times10^{-2}$&$3.4 \times10^{-2}$\\
$^{15}$N &$1.7 \times10^{-5}$ &$8.1\times10^{-5}$&$4.7\times10^{-3}$& $2.4\times 10^{-2}$&$4.3 \times10^{-2}$&$6.9\times 10^{-2}$ \\
$^{16}$O &$9.9 \times10^{-2}$ &$9.2 \times10^{-2}$&$8.2 \times10^{-2}$&$4.2 \times10^{-2}$ &$1.3 \times10^{-2}$&$9.2 \times10^{-5}$\\
$^{17}$O &$6.4 \times10^{-4}$ &$5.6 \times 10^{-4}$&$1.4 \times 10^{-3}$& $6.4 \times 10^{-3}$&$8.6 \times 10^{-3}$ &$4.5 \times 10^{-3}$\\
$^{18}$O &$2.6 \times10^{-7}$ &$2.5 \times 10^{-7}$&$2.5 \times 10^{-7}$& $2.0 \times 10^{-6}$&$6.9 \times 10^{-6}$ &$2.1 \times 10^{-5}$\\
$^{18}$F &$4.3 \times10^{-9}$ &$8.3\times 10^{-9}$&$1.4 \times 10^{-8}$& $1.0 \times 10^{-7}$&$3.9 \times 10^{-7}$ &$1.3\times 10^{-6}$\\
$^{19}$F &$4.6 \times10^{-9}$ &$1.0 \times 10^{-9}$&$5.2 \times 10^{-10}$& $5.4 \times 10^{-9}$&$2.4 \times 10^{-8}$ &$1.6\times 10^{-7}$\\
$^{20}$Ne &$9.1\times10^{-2}$ &$9.3\times 10^{-2}$&$9.3 \times 10^{-2}$& $9.1\times 10^{-2}$&$8.8\times 10^{-2}$ &$6.4\times 10^{-2}$\\
$^{21}$Ne &$1.7 \times10^{-6}$ &$2.2 \times 10^{-6}$&$1.1 \times 10^{-5}$& $2.3 \times 10^{-5}$&$2.8 \times 10^{-5}$ &$3.0 \times 10^{-5}$\\
$^{22}$Ne&$7.4 \times10^{-4}$ &$1.0\times 10^{-3}$&$1.0 \times 10^{-3}$& $8.2 \times 10^{-4}$&$4.1 \times 10^{-4}$ &$6.9 \times 10^{-7}$\\
$^{22}$Na &$1.1 \times10^{-4}$ &$7.4 \times 10^{-5}$&$4.1 \times 10^{-5}$& $5.2 \times 10^{-5}$&$9.3 \times 10^{-5}$ &$2.4 \times 10^{-4}$\\
$^{23}$Na&$3.8 \times10^{-3}$ &$1.1 \times 10^{-3}$&$2.2 \times 10^{-4}$& $1.7 \times 10^{-4}$&$2.8\times 10^{-4}$ &$8.2\times 10^{-4}$\\
$^{24}$Mg &$4.0 \times10^{-5}$ &$1.2 \times 10^{-5}$&$9.4 \times 10^{-6}$& $4.5 \times 10^{-6}$&$3.8 \times 10^{-6}$ &$6.6 \times 10^{-6}$\\
$^{25}$Mg &$1.9\times10^{-2}$ &$1.9 \times 10^{-2}$&$1.3\times 10^{-2}$& $8.1 \times 10^{-4}$&$5.6 \times 10^{-4}$ &$3.9\times 10^{-4}$\\
$^{26}$Mg&$2.4\times10^{-3}$ &$2.3\times 10^{-3}$&$1.0 \times 10^{-3}$& $5.8 \times 10^{-5}$&$3.5 \times 10^{-5}$ &$2.2 \times 10^{-5}$\\
$^{26}$Al &$3.8\times10^{-4}$&$5.3 \times 10^{-4}$&$1.9 \times 10^{-3}$& $2.3 \times 10^{-4}$&$1.8 \times 10^{-4}$&$1.0 \times 10^{-4}$ \\
$^{27}$Al &$2.8\times10^{-3}$&$2.8 \times 10^{-3}$&$6.3 \times 10^{-3}$& $1.2 \times 10^{-3}$&$7.5 \times 10^{-4}$&$5.7 \times 10^{-4}$ \\
$^{28}$Si &$9.1 \times10^{-4}$&$1.3 \times 10^{-3}$&$6.3 \times 10^{-3}$& $2.9 \times 10^{-2}$&$3.1 \times 10^{-2}$ &$2.0 \times 10^{-2}$\\
$^{29}$Si &$1.8\times10^{-5}$&$1.8 \times 10^{-5}$&$1.9 \times 10^{-5}$& $3.1 \times 10^{-4}$&$3.3 \times 10^{-3}$ &$6.0 \times 10^{-4}$\\
$^{30}$Si &$1.9\times10^{-5}$&$1.9 \times 10^{-5}$&$2.1 \times 10^{-5}$& $3.1 \times 10^{-4}$&$3.3 \times 10^{-3}$&$9.7 \times 10^{-3}$ \\
$^{31}$P &$5.2\times10^{-6}$&$5.2 \times 10^{-6}$&$5.2 \times 10^{-6}$& $1.8 \times 10^{-5}$&$6.2 \times 10^{-4}$ &$4.0 \times 10^{-3}$\\
$^{32}$S &$2.6\times10^{-4}$&$2.6 \times 10^{-4}$&$2.6 \times 10^{-4}$& $2.6 \times 10^{-4}$&$4.6 \times 10^{-4}$&$4.1 \times 10^{-2}$ \\
$^{33}$S &$2.1\times10^{-6}$&$2.1\times 10^{-6}$&$2.1 \times 10^{-6}$& $2.0 \times 10^{-6}$&$1.8\times 10^{-6}$&$3.2 \times 10^{-4}$ \\
$^{34}$S &$1.4\times10^{-5}$&$1.4 \times 10^{-5}$&$1.4 \times 10^{-5}$& $1.3 \times 10^{-5}$&$1.0\times 10^{-5}$&$1.5 \times 10^{-4}$ \\
$^{35}$Cl &$3.1\times10^{-6}$&$3.1 \times 10^{-6}$&$3.2 \times 10^{-6}$& $4.2 \times 10^{-6}$&$7.2 \times 10^{-6}$ &$1.5 \times 10^{-4}$\\
$^{36}$Ar &$5.8\times10^{-5}$&$5.8 \times 10^{-5}$&$5.7 \times 10^{-5}$& $5.2 \times 10^{-5}$&$3.3 \times 10^{-5}$&$1.4 \times 10^{-5}$ \\
$^{40}$Ca &$4.8\times10^{-5}$&$4.8 \times 10^{-5}$&$4.8 \times 10^{-5}$& $4.8 \times 10^{-5}$&$4.8 \times 10^{-5}$&$4.8 \times 10^{-5}$ \\
\enddata

\tablenotetext{a}{All abundances are Mass Fraction.}

\tablenotetext{b} {No material was ejected, these abundances are just for the surface zone.}

\tablenotetext{c}{The ``0.0'' in this row means that the abundance of $^7$Li is less than $10^{-15}$.}




\end{deluxetable}

\clearpage


\begin{deluxetable}{@{}lcccccc}
\tablecaption{Ejecta Abundances for 50-50  mixture in ONe WDs: Mixing from Beginning (MFB)\tablenotemark{a} \label{ONe5050MFBabund}}
\tablewidth{0pt}
\tablecolumns{7}
 \tablehead{ \colhead{WD Mass}&
\colhead{0.6 M$_\odot$ \tablenotemark{}} & 
\colhead{0.8 M$_\odot$\tablenotemark{}} &
\colhead{1.0 M $_\odot${}} &
\colhead{1.15 M$_\odot${}} &
\colhead{1.25 M$_\odot${}} &
\colhead{1.35 M$_\odot${}}} 
\startdata
H&$3.4 \times10^{-1}$&$3.3 \times10^{-1}$&$3.2 \times10^{-1}$&$3.1 \times10^{-1}$&$3.0 \times10^{-1}$&$2.4 \times10^{-1}$ \\
$^{3}$He&$1.2 \times10^{-8}$  &$1.3 \times 10^{-8}$&$4.0 \times 10^{-8}$& $1.5 \times 10^{-8}$&$5.9 \times 10^{-9}$ &$1.1 \times 10^{-10}$\\
$^4$He&$1.6\times10^{-1}$&$1.7 \times10^{-1}$&$1.8 \times10^{-1}$&$1.8 \times10^{-1}$&$1.9 \times10^{-1}$&$2.4 \times10^{-1}$\\
$^{7}$Be&$2.7 \times10^{-10}$  &$7.8 \times 10^{-10}$&$9.8 \times 10^{-8}$& $4.2 \times 10^{-7}$&$6.3\times 10^{-7}$ &$7.8\times 10^{-7}$\\
$^{7}$Li\tablenotemark{b}  &$0.0$&$0.0$&$4.6\times 10^{-13}$& $8.3\times 10^{-14}$&$1.5\times 10^{-14}$&$0.0$\\
$^{12}$C &$1.1 \times10^{-3}$&$2.1\times10^{-3}$&$8.1 \times 10^{-3}$& $7.3 \times 10^{-3}$&$4.2 \times 10^{-3}$ &$1.7\times 10^{-2}$\\
$^{13}$C &$4.6 \times10^{-4}$&$1.1 \times10^{-3}$&$8.6\times10^{-3}$ &$1.0 \times 10^{-2}$&$6.3 \times 10^{-3}$&$2.3 \times 10^{-2}$\\
$^{14}$N&$4.0\times10^{-2}$ &$5.1 \times 10^{-2}$&$4.2 \times10^{-2}$&$2.1 \times10^{-2}$&$2.1 \times10^{-2}$&$4.1 \times10^{-2}$\\
$^{15}$N &$3.2 \times10^{-5}$ &$8.6 \times10^{-5}$&$9.9 \times10^{-3}$& $6.0 \times 10^{-2}$&$9.5 \times10^{-2}$&$1.2 \times 10^{-1}$ \\
$^{16}$O &$2.2 \times10^{-1}$ &$2.1 \times10^{-1}$&$1.8 \times10^{-1}$&$1.2 \times10^{-1}$ &$4.2\times10^{-2}$&$1.1 \times10^{-3}$\\
$^{17}$O &$1.4\times10^{-3}$ &$1.2 \times 10^{-3}$&$7.5 \times 10^{-3}$& $4.6 \times 10^{-2}$&$9.3 \times 10^{-2}$ &$4.1 \times 10^{-2}$\\
$^{18}$O &$6.2 \times10^{-7}$ &$4.7 \times 10^{-7}$&$1.4 \times 10^{-6}$& $2.0 \times 10^{-5}$&$1.2 \times 10^{-4}$ &$1.0 \times 10^{-4}$\\
$^{18}$F &$3.2\times10^{-8}$ &$1.4 \times 10^{-8}$&$7.8 \times 10^{-8}$& $1.1 \times 10^{-6}$&$9.5 \times 10^{-6}$ &$7.6 \times 10^{-6}$\\
$^{19}$F &$6.1 \times10^{-9}$ &$3.7 \times 10^{-9}$&$5.0 \times 10^{-9}$& $6.9 \times 10^{-8}$&$6.9 \times 10^{-7}$ &$1.0 \times 10^{-6}$\\
$^{20}$Ne &$1.7 \times10^{-1}$ &$1.8 \times 10^{-1}$&$1.8 \times 10^{-1}$& $1.8 \times 10^{-1}$&$1.7\times 10^{-1}$ &$1.4\times 10^{-1}$\\
$^{21}$Ne &$3.3 \times10^{-6}$ &$4.2 \times 10^{-6}$&$2.9 \times 10^{-5}$& $7.3 \times 10^{-5}$&$9.0 \times 10^{-5}$ &$6.1 \times 10^{-5}$\\
$^{22}$Ne&$2.2\times10^{-3}$ &$2.0\times 10^{-3}$&$2.0 \times 10^{-3}$& $1.6 \times 10^{-3}$&$3.6 \times 10^{-4}$ &$1.4 \times 10^{-6}$\\
$^{22}$Na &$4.6 \times10^{-4}$ &$1.2 \times 10^{-4}$&$9.3 \times 10^{-5}$& $2.2 \times 10^{-4}$&$7.7 \times 10^{-4}$ &$5.3 \times 10^{-4}$\\
$^{23}$Na&$1.6 \times10^{-2}$ &$1.2 \times 10^{-3}$&$6.0 \times 10^{-4}$& $7.4 \times 10^{-4}$&$2.8 \times 10^{-3}$ &$2.1 \times 10^{-3}$\\
$^{24}$Mg &$4.7 \times10^{-4}$ &$1.6 \times 10^{-5}$&$4.8 \times 10^{-5}$& $1.4 \times 10^{-5}$&$1.7 \times 10^{-5}$ &$3.4 \times 10^{-5}$\\
$^{25}$Mg &$3.7\times10^{-2}$ &$3.8 \times 10^{-2}$&$1.3 \times 10^{-2}$& $1.5 \times 10^{-3}$&$2.0 \times 10^{-3}$ &$1.4 \times 10^{-3}$\\
$^{26}$Mg&$4.9 \times10^{-3}$ &$4.1\times 10^{-3}$&$1.0 \times 10^{-3}$& $1.0 \times 10^{-4}$&$9.6 \times 10^{-5}$ &$9.8 \times 10^{-5}$\\
$^{26}$Al &$4.2 \times10^{-4}$&$1.8 \times 10^{-3}$&$4.3 \times 10^{-3}$& $5.9 \times 10^{-4}$&$5.6 \times 10^{-4}$&$6.4 \times 10^{-4}$ \\
$^{27}$Al &$5.4\times10^{-3}$&$6.3\times 10^{-3}$&$1.8 \times 10^{-2}$& $3.4 \times 10^{-3}$&$2.9 \times 10^{-3}$&$3.7 \times 10^{-3}$ \\
$^{28}$Si &$6.6 \times10^{-4}$&$2.5 \times 10^{-3}$&$2.3 \times 10^{-2}$& $5.8 \times 10^{-2}$&$5.3 \times 10^{-2}$ &$4.6 \times 10^{-2}$\\
$^{29}$Si &$1.7\times10^{-5}$&$1.7 \times 10^{-5}$&$4.3 \times 10^{-5}$& $8.4 \times 10^{-4}$&$1.7 \times 10^{-3}$ &$1.5 \times 10^{-3}$\\
$^{30}$Si &$1.2\times10^{-5}$&$1.2 \times 10^{-5}$&$2.0 \times 10^{-5}$& $1.4 \times 10^{-3}$&$1.1 \times 10^{-2}$&$1.8 \times 10^{-2}$ \\
$^{31}$P &$4.1 \times10^{-6}$&$4.0 \times 10^{-6}$&$4.0 \times 10^{-6}$& $9.7 \times 10^{-5}$&$3.1 \times 10^{-3}$ &$8.7 \times 10^{-3}$\\
$^{32}$S &$2.0 \times10^{-4}$&$2.0 \times 10^{-4}$&$2.0 \times 10^{-4}$& $2.1 \times 10^{-4}$&$1.7 \times 10^{-3}$&$5.0 \times 10^{-2}$ \\
$^{33}$S &$1.6\times10^{-6}$&$1.6\times 10^{-6}$&$1.6 \times 10^{-6}$& $1.5 \times 10^{-6}$&$2.4 \times 10^{-6}$&$7.1 \times 10^{-4}$ \\
$^{34}$S &$9.5 \times10^{-6}$&$9.5 \times 10^{-6}$&$9.4 \times 10^{-6}$& $8.5 \times 10^{-6}$&$5.9 \times 10^{-6}$&$2.3 \times 10^{-4}$ \\
$^{35}$Cl &$1.9 \times10^{-6}$&$1.9 \times 10^{-6}$&$2.0 \times 10^{-6}$& $3.0 \times 10^{-6}$&$5.7 \times 10^{-6}$ &$1.4 \times 10^{-4}$\\
$^{36}$Ar &$3.9 \times10^{-5}$&$3.9 \times 10^{-5}$&$3.8 \times 10^{-5}$& $3.2 \times 10^{-5}$&$1.5 \times 10^{-5}$&$1.5 \times 10^{-5}$ \\
$^{40}$Ca &$3.0 \times10^{-5}$&$3.0 \times 10^{-5}$&$3.0 \times 10^{-5}$& $3.0 \times 10^{-5}$&$3.0 \times 10^{-5}$&$3.0 \times 10^{-5}$ \\
\enddata
\tablenotetext{a}{All abundances are Mass Fraction.}
\tablenotetext{b}{The ``0.0'' in this row means that the abundance of $^7$Li is less than $10^{-15}$.}




\end{deluxetable}

\clearpage

 
\begin{deluxetable}{@{}lcccccc}
\tablecaption{Ejecta and/or  Surface Abundances for Solar Accretion and No Mixing with Core Material\tablenotemark{a} \label{solarabund}}
\tablewidth{0pt}
\tablecolumns{7}
\tablehead{ \colhead{WD Mass (M$_\odot$): }&
\colhead{0.6\tablenotemark{}}&
\colhead{0.8\tablenotemark{}} &
\colhead{1.0\tablenotemark{}} &
\colhead{1.15\tablenotemark{}} &
\colhead{1.25\tablenotemark{}} &
\colhead{1.35\tablenotemark{}}} 

\startdata
$^1$H&$7.0 \times10^{-1}$&$6.9 \times10^{-1}$&$6.9 \times10^{-1}$&$6.6 \times10^{-1}$&$6.4 \times10^{-1}$&$6.0 \times10^{-1}$ \\
$^{3}$He&$7.7 \times10^{-12}$  &$3.0 \times 10^{-12}$&$4.9 \times 10^{-9}$& $3.2 \times 10^{-11}$&$1.8 \times 10^{-10}$ &$9.4 \times 10^{-12}$\\
$^4$He&$2.8 \times10^{-1}$&$3.0 \times10^{-1}$&$3.0 \times10^{-1}$&$3.3 \times10^{-1}$&$3.4 \times10^{-1}$&$3.9 \times10^{-1}$ \\
$^{7}$Be&$1.0\times10^{-13}$  &$1.5 \times 10^{-13}$&$4.3 \times 10^{-11}$& $3.1 \times 10^{-12}$&$2.5 \times 10^{-11}$ &$2.1 \times 10^{-10}$\\
$^{7}$Li\tablenotemark{b} &0.0 &0.0&$1.4 \times 10^{-12}$& $5.7 \times 10^{-15}$&$2.5 \times 10^{-14}$&0.0\\
$^{12}$C&$1.8\times 10^{-4}$  &$3.6 \times 10^{-4}$&$6.6 \times 10^{-4}$& $8.8 \times 10^{-4}$&$1.3 \times 10^{-3}$ &$1.4 \times 10^{-3}$\\
$^{13}$C &$1.1\times 10^{-4}$ &$3.0 \times 10^{-4}$&$1.0 \times 10^{-3}$& $1.0 \times 10^{-3}$&$2.3 \times 10^{-3}$&$2.5 \times 10^{-3}$\\
$^{14}$N &$7.1\times10^{-3}$ &$7.8 \times 10^{-3}$&$3.4 \times 10^{-3}$& $6.9 \times 10^{-3}$&$4.1 \times 10^{-3}$ &$4.0 \times 10^{-3}$\\
$^{15}$N &$6.7 \times10^{-6}$ &$4.0 \times10^{-5}$&$4.4 \times10^{-3}$& $4.7 \times 10^{-4}$&$1.5 \times 10^{-3}$ &$1.2 \times10^{-3}$\\
$^{16}$O &$2.4 \times10^{-3}$ &$1.1 \times 10^{-3}$&$1.0 \times 10^{-4}$& $2.2 \times 10^{-5}$&$2.3 \times 10^{-5}$ &$2.6 \times 10^{-5}$\\
$^{17}$O&$1.5\times10^{-5}$  &$7.6 \times 10^{-6}$&$3.1 \times 10^{-5}$& $4.0 \times 10^{-7}$&$5.2 \times 10^{-7}$ &$4.1 \times 10^{-7}$\\
$^{18}$O &$7.1 \times10^{-9}$ &$3.0 \times 10^{-9}$&$1.4 \times 10^{-8}$& $3.2 \times 10^{-10}$&$3.7 \times 10^{-10}$ &$8.4\times 10^{-11}$\\
$^{18}$F &$6.3 \times10^{-11}$ &$1.2 \times 10^{-10}$&$8.2 \times 10^{-10}$& $2.3 \times 10^{-11}$&$2.1 \times 10^{-11}$ &$1.0 \times 10^{-12}$\\
$^{19}$F &$9.5 \times10^{-11}$ &$1.5 \times 10^{-11}$&$1.0 \times 10^{-10}$& $2.0 \times 10^{-12}$&$5.0 \times 10^{-12}$ &$1.6 \times 10^{-13}$\\
$^{20}$Ne &$1.2\times10^{-3}$ &$1.2\times 10^{-3}$&$1.2 \times 10^{-3}$& $5.0\times 10^{-4}$&$9.2 \times 10^{-5}$ &$7.1 \times 10^{-8}$\\
$^{21}$Ne &$2.3\times10^{-8}$ &$3.9 \times 10^{-8}$&$4.9 \times 10^{-7}$& $4.6 \times 10^{-8}$&$1.8 \times 10^{-8}$ &$1.2 \times 10^{-11}$\\
$^{22}$Ne&$8.8 \times10^{-5}$ &$3.6 \times 10^{-5}$&$1.5 \times 10^{-5}$& $1.4 \times 10^{-8}$&$3.9 \times 10^{-9}$ &$4.0 \times 10^{-16}$\\
$^{22}$Na &$2.0 \times10^{-6}$ &$9.5 \times 10^{-7}$&$1.4 \times 10^{-6}$& $3.2 \times 10^{-7}$&$9.3 \times 10^{-8}$ &$2.6 \times 10^{-11}$\\
$^{23}$Na&$6.4 \times10^{-6}$ &$3.1 \times 10^{-6}$&$4.0 \times 10^{-6}$& $9.8 \times 10^{-7}$&$2.8 \times 10^{-7}$ &$7.0 \times 10^{-11}$\\
$^{24}$Mg &$5.5 \times10^{-8}$ &$3.2\times 10^{-8}$&$1.2 \times 10^{-7}$& $1.3 \times 10^{-8}$&$6.0 \times 10^{-9}$ &$1.8 \times 10^{-12}$\\
$^{25}$Mg &$6.0 \times10^{-4}$ &$3.1 \times 10^{-4}$&$1.1 \times 10^{-5}$& $2.2 \times 10^{-6}$&$4.5 \times 10^{-7}$ &$3.0 \times 10^{-10}$\\
$^{26}$Mg&$7.6 \times10^{-5}$ &$2.3 \times 10^{-5}$&$7.2 \times 10^{-7}$& $8.8 \times 10^{-8}$&$2.0 \times 10^{-8}$ &$1.3 \times 10^{-11}$\\
$^{26}$Al &$2.9 \times10^{-5}$&$1.9 \times 10^{-5}$&$1.8 \times 10^{-6}$& $3.1 \times 10^{-7}$&$8.4 \times 10^{-8}$&$6.4 \times 10^{-11}$ \\
$^{27}$Al &$9.6 \times10^{-5}$&$1.4 \times 10^{-4}$&$8.1 \times 10^{-6}$& $1.6 \times 10^{-6}$&$3.3 \times 10^{-7}$&$2.2 \times 10^{-10}$ \\
$^{28}$Si &$7.8 \times10^{-4}$&$1.1 \times 10^{-3}$&$1.7 \times 10^{-3}$& $3.9 \times 10^{-4}$&$5.8 \times 10^{-5}$ &$6.5 \times 10^{-8}$\\
$^{29}$Si &$4.0\times10^{-5}$&$3.6 \times 10^{-5}$&$1.9 \times 10^{-5}$& $4.9 \times 10^{-6}$&$7.4 \times 10^{-7}$ &$6.6 \times 10^{-10}$\\
$^{30}$Si&$2.7 \times 10^{-5}$&$3.2 \times 10^{-5}$&$1.3 \times 10^{-4}$& $5.0 \times 10^{-4}$&$5.7 \times 10^{-5}$&$8.6 \times 10^{-8}$ \\
$^{31}$P &$7.5 \times10^{-6}$&$7.2 \times 10^{-6}$&$1.3 \times 10^{-5}$& $5.6 \times 10^{-5}$&$6.5 \times 10^{-6}$ &$7.2 \times 10^{-9}$\\
$^{32}$S &$4.0\times10^{-4}$ &$4.0\times 10^{-4}$&$4.0 \times 10^{-4}$& $2.6 \times 10^{-3}$&$3.5 \times 10^{-3}$ &$1.5 \times 10^{-4}$\\
$^{33}$S &$3.2 \times10^{-6}$&$3.2 \times 10^{-6}$&$2.6 \times 10^{-6}$& $4.5 \times 10^{-6}$&$6.6 \times 10^{-6}$&$2.4 \times 10^{-7}$ \\
$^{34}$S &$1.9\times10^{-5}$ &$1.8\times 10^{-5}$&$1.5 \times 10^{-5}$& $3.3 \times 10^{-6}$&$5.8 \times 10^{-6}$ &$2.8\times 10^{-7}$\\
$^{35}$Cl &$4.1\times10^{-6}$&$4.4\times 10^{-6}$&$8.8 \times 10^{-6}$& $1.8 \times 10^{-5}$&$5.0 \times 10^{-5}$ &$1.2 \times 10^{-5}$\\
$^{36}$Ar &$9.1\times10^{-5}$&$9.1 \times 10^{-5}$&$6.8 \times 10^{-5}$& $1.5 \times 10^{-6}$&$4.1 \times 10^{-6}$&$1.0 \times 10^{-6}$ \\
$^{40}$Ca &$7.1\times10^{-5}$&$7.1 \times 10^{-5}$&$7.1 \times 10^{-5}$& $7.3 \times 10^{-5}$&$6.3 \times 10^{-4}$&$4.7 \times 10^{-3}$ \\
\enddata
\tablenotetext{a}{All abundances are Mass Fraction.}

\tablenotetext{b}{The ``0.0'' in this row means that the abundance of $^7$Li is less than $10^{-15}$.}




\end{deluxetable}

\clearpage

\begin{deluxetable}{@{}lcccccc}
\tablecaption{Ejecta Abundances for 25-75  mixture assuming MDTNR in ONe White Dwarfs\tablenotemark{a} \label{ONe2575MDTNRabund}}
\tablewidth{0pt}
\tablecolumns{7}
 \tablehead{ \colhead{WD Mass (M$_\odot$): }&
\colhead{0.6\tablenotemark{}} & 
\colhead{0.8\tablenotemark{}} &
\colhead{1.0\tablenotemark{b}} &
\colhead{1.15\tablenotemark{}} &
\colhead{1.25\tablenotemark{}} &
\colhead{1.35\tablenotemark{}}} 
\startdata
$^1$H&$5.2 \times10^{-1}$&$5.1 \times10^{-1}$&$5.3 \times10^{-1}$&$5.2 \times10^{-1}$&$4.9  \times10^{-1}$&$4.5 \times10^{-1}$ \\
$^{3}$He&$1.4 \times10^{-7}$  &$1.5 \times 10^{-7}$&$4.5 \times 10^{-6}$& $5.8 \times 10^{-7}$&$1.1 \times 10^{-7}$ &$5.4 \times 10^{-11}$\\
$^4$He&$2.2 \times10^{-1}$&$2.3 \times10^{-1}$&$2.1 \times10^{-1}$&$2.2 \times10^{-1}$&$2.4\times10^{-1}$&$2.5 \times10^{-1}$\\
$^{7}$Be&$2.2\times10^{-9}$  &$2.4 \times 10^{-8}$&$5.0 \times 10^{-7}$& $1.2 \times 10^{-6}$&$1.4 \times 10^{-6}$ &$1.5 \times 10^{-6}$\\
$^{7}$Li\tablenotemark{c}  &$0.0$&$9.0 \times 10^{-15}$&$4.4 \times 10^{-10}$& $4.3 \times 10^{-11}$&$8.3 \times 10^{-12}$&$1.3 \times 10^{-15}$\\
$^{12}$C &$7.6 \times10^{-4}$&$2.1\times10^{-3}$&$5.3 \times 10^{-4}$& $3.5 \times 10^{-4}$&$1.6 \times 10^{-3}$ &$2.5 \times 10^{-3}$\\
$^{13}$C &$3.3 \times10^{-4}$&$2.0 \times10^{-3}$&$9.0 \times10^{-4}$ &$4.1 \times 10^{-4}$&$1.4 \times 10^{-3}$&$1.6 \times 10^{-3}$\\
$^{14}$N&$2.9\times10^{-2}$ &$3.6 \times 10^{-2}$&$6.2 \times10^{-3}$&$5.5 \times10^{-3}$&$1.6 \times10^{-2}$&$2.2 \times10^{-2}$\\
$^{15}$N &$1.6 \times10^{-5}$ &$6.3 \times10^{-4}$&$1.0 \times10^{-2}$& $2.6 \times 10^{-2}$&$6.0 \times10^{-2}$&$8.4 \times 10^{-2}$ \\
$^{16}$O &$1.0 \times10^{-1}$ &$9.2 \times10^{-2}$&$1.1 \times10^{-1}$&$6.3 \times10^{-2}$ &$8.8 \times10^{-3}$&$1.6 \times10^{-4}$\\
$^{17}$O &$6.7 \times10^{-4}$ &$7.3 \times 10^{-4}$&$1.1 \times 10^{-2}$& $4.4 \times 10^{-2}$&$4.5 \times 10^{-2}$ &$1.8 \times 10^{-2}$\\
$^{18}$O &$2.8 \times10^{-7}$ &$2.3 \times 10^{-7}$&$4.3 \times 10^{-6}$& $1.1 \times 10^{-4}$&$3.2 \times 10^{-4}$ &$1.2 \times 10^{-3}$\\
$^{18}$F &$5.3 \times10^{-9}$ &$1.2 \times 10^{-8}$&$2.6 \times 10^{-7}$& $7.5 \times 10^{-6}$&$2.3 \times 10^{-5}$ &$1.2 \times 10^{-4}$\\
$^{19}$F &$4.6 \times10^{-9}$ &$1.2\times 10^{-9}$&$2.4 \times 10^{-8}$& $2.4 \times 10^{-7}$&$1.4 \times 10^{-6}$ &$8.5 \times 10^{-6}$\\
$^{20}$Ne &$9.2 \times10^{-2}$ &$9.3 \times 10^{-2}$&$9.1 \times 10^{-2}$& $9.0 \times 10^{-2}$&$7.6 \times 10^{-2}$ &$1.9 \times 10^{-2}$\\
$^{21}$Ne &$1.7 \times10^{-6}$ &$3.9 \times 10^{-6}$&$1.4 \times 10^{-4}$& $5.8 \times 10^{-5}$&$4.6 \times 10^{-5}$ &$1.4\times 10^{-5}$\\
$^{22}$Ne&$1.2\times10^{-3}$ &$1.1\times 10^{-3}$&$1.2 \times 10^{-3}$& $8.6 \times 10^{-4}$&$1.1 \times 10^{-4}$ &$1.5 \times 10^{-6}$\\
$^{22}$Na &$1.1\times10^{-4}$ &$5.4 \times 10^{-5}$&$1.1 \times 10^{-4}$& $6.6 \times 10^{-4}$&$1.6 \times 10^{-3}$ &$3.9 \times 10^{-3}$\\
$^{23}$Na&$3.1 \times10^{-3}$ &$1.6 \times 10^{-4}$&$2.8 \times 10^{-3}$& $1.7 \times 10^{-3}$&$4.8 \times 10^{-3}$ &$6.5 \times 10^{-3}$\\
$^{24}$Mg &$3.2 \times10^{-5}$ &$2.4 \times 10^{-6}$&$9.7 \times 10^{-4}$& $9.2 \times 10^{-5}$&$4.5 \times 10^{-5}$ &$5.7 \times 10^{-4}$\\
$^{25}$Mg &$1.9 \times10^{-2}$ &$1.6 \times 10^{-2}$&$1.6 \times 10^{-2}$& $4.7 \times 10^{-3}$&$1.9 \times 10^{-3}$ &$1.7 \times 10^{-3}$\\
$^{26}$Mg&$2.4 \times10^{-3}$ &$1.4 \times 10^{-3}$&$1.8 \times 10^{-3}$& $4.4 \times 10^{-4}$&$1.3 \times 10^{-4}$ &$2.9 \times 10^{-4}$\\
$^{26}$Al &$3.7 \times10^{-4}$&$1.6 \times 10^{-3}$&$1.4 \times 10^{-3}$& $1.2 \times 10^{-3}$&$2.5 \times 10^{-4}$&$7.9 \times 10^{-5}$ \\
$^{27}$Al &$2.7 \times10^{-3}$&$4.8 \times 10^{-3}$&$4.2 \times 10^{-3}$& $5.3 \times 10^{-3}$&$2.0 \times 10^{-3}$&$2.0 \times 10^{-3}$ \\
$^{28}$Si &$9.5 \times10^{-4}$&$3.7 \times 10^{-3}$&$2.4 \times 10^{-3}$& $1.7 \times 10^{-2}$&$2.0 \times 10^{-2}$ &$1.4 \times 10^{-2}$\\
$^{29}$Si &$1.8 \times10^{-5}$&$1.8 \times 10^{-5}$&$1.9 \times 10^{-5}$& $4.5 \times 10^{-4}$&$1.8 \times 10^{-3}$ &$4.9 \times 10^{-3}$\\
$^{30}$Si &$1.9 \times10^{-5}$&$2.0 \times 10^{-5}$&$2.0 \times 10^{-5}$& $8.7 \times 10^{-4}$&$7.1 \times 10^{-3}$&$9.5 \times 10^{-3}$ \\
$^{31}$P &$5.2 \times10^{-6}$&$5.2 \times 10^{-6}$&$5.2 \times 10^{-6}$& $1.7 \times 10^{-4}$&$3.8 \times 10^{-3}$ &$8.9 \times 10^{-3}$\\
$^{32}$S &$2.6\times10^{-4}$&$2.6 \times 10^{-4}$&$2.6 \times 10^{-4}$& $3.1 \times 10^{-4}$&$1.3 \times 10^{-2}$&$7.3 \times 10^{-2}$ \\
$^{33}$S &$2.1\times10^{-6}$&$2.1\times 10^{-6}$&$2.1 \times 10^{-6}$& $2.0 \times 10^{-6}$&$5.9 \times 10^{-5}$&$4.8 \times 10^{-3}$ \\
$^{34}$S &$1.4\times10^{-5}$&$1.4 \times 10^{-5}$&$1.4 \times 10^{-5}$& $1.3 \times 10^{-5}$&$1.9 \times 10^{-5}$&$2.6 \times 10^{-3}$ \\
$^{35}$Cl &$3.1\times10^{-6}$&$3.1 \times 10^{-6}$&$3.1 \times 10^{-6}$& $4.6 \times 10^{-6}$&$1.7 \times 10^{-5}$ &$4.6 \times 10^{-3}$\\
$^{36}$Ar &$5.8 \times10^{-5}$&$5.8 \times 10^{-5}$&$5.8 \times 10^{-5}$& $4.8 \times 10^{-5}$&$1.0 \times 10^{-5}$&$1.1 \times 10^{-3}$ \\
$^{40}$Ca &$4.8\times10^{-5}$&$4.8 \times 10^{-5}$&$4.8 \times 10^{-5}$& $4.8 \times 10^{-5}$&$4.8 \times 10^{-5}$&$8.4 \times 10^{-5}$ \\
\enddata
\tablenotetext{a}{All abundances are Mass Fraction.}

\tablenotetext{b} {No mass was ejected so that these are the abundances for the surface zone.}

\tablenotetext{c}{The ``0.0'' in this row means that the abundance of $^7$Li is less than $10^{-15}$.}




\end{deluxetable}

\clearpage

\begin{deluxetable}{@{}lcccccc}
\tablecaption{Ejecta Abundances for 50-50 mixture assuming MDTNR in ONe White Dwarfs \tablenotemark{a} \label{ONe5050MDTNRabund}}
\tablewidth{0pt}
\tablecolumns{7}
 \tablehead{ \colhead{WD Mass (M$_\odot$): }&
\colhead{0.6\tablenotemark{}} & 
\colhead{0.8\tablenotemark{}} &
\colhead{1.0\tablenotemark{}} &
\colhead{1.15\tablenotemark{}} &
\colhead{1.25\tablenotemark{}} &
\colhead{1.35\tablenotemark{}}} 
\startdata
$^1$H&$3.4 \times10^{-1}$&$3.3 \times10^{-1}$&$3.5 \times10^{-1}$&$3.1 \times10^{-1}$&$2.8 \times10^{-1}$&$2.4 \times10^{-1}$ \\
$^{3}$He&$8.2 \times10^{-7}$  &$9.0 \times 10^{-7}$&$6.1 \times 10^{-6}$& $4.9 \times 10^{-8}$&$7.8 \times 10^{-9}$ &$4.6 \times 10^{-10}$\\
$^4$He&$1.6 \times10^{-1}$&$1.7 \times10^{-1}$&$1.4 \times10^{-1}$&$1.8 \times10^{-1}$&$2.1\times10^{-1}$&$2.2 \times10^{-1}$\\
$^{7}$Be&$2.1 \times10^{-8}$  &$5.4 \times 10^{-7}$&$3.4 \times 10^{-6}$& $3.7 \times 10^{-6}$&$1.0\times 10^{-5}$ &$1.1\times 10^{-5}$\\
$^{7}$Li  &$1.2 \times10^{-14}$&$1.8 \times 10^{-13}$&$4.1 \times 10^{-10}$& $6.6\times 10^{-13}$&$3.0 \times 10^{-13}$&$1.1 \times 10^{-13}$\\
$^{12}$C &$1.0 \times10^{-3}$&$3.9 \times10^{-3}$&$6.6 \times 10^{-4}$& $3.4 \times 10^{-3}$&$1.2 \times 10^{-2}$ &$1.3 \times 10^{-2}$\\
$^{13}$C &$4.7 \times10^{-4}$&$3.5 \times10^{-3}$&$1.7 \times10^{-3}$ &$4.3 \times 10^{-3}$&$1.4 \times 10^{-2}$&$1.3 \times 10^{-2}$\\
$^{14}$N&$3.4\times10^{-2}$ &$4.0 \times 10^{-2}$&$5.0 \times10^{-3}$&$1.5 \times10^{-2}$&$3.1 \times10^{-2}$&$2.9 \times10^{-2}$\\
$^{15}$N &$3.3 \times10^{-5}$ &$2.0 \times10^{-3}$&$4.0 \times10^{-3}$& $7.4 \times 10^{-2}$&$1.1 \times10^{-1}$&$1.3 \times 10^{-1}$ \\
$^{16}$O &$2.3 \times10^{-1}$ &$2.1 \times10^{-1}$&$2.4 \times10^{-1}$&$7.2 \times10^{-2}$ &$1.2 \times10^{-2}$&$1.4 \times10^{-3}$\\
$^{17}$O &$1.4\times10^{-3}$ &$2.8 \times 10^{-3}$&$1.7 \times 10^{-2}$& $9.8 \times 10^{-2}$&$8.0 \times 10^{-2}$ &$7.1 \times 10^{-2}$\\
$^{18}$O &$6.0 \times10^{-7}$ &$6.2 \times 10^{-7}$&$8.4 \times 10^{-6}$& $9.0 \times 10^{-5}$&$9.4 \times 10^{-5}$ &$6.8 \times 10^{-4}$\\
$^{18}$F &$2.9 \times10^{-8}$ &$3.2 \times 10^{-8}$&$4.7 \times 10^{-7}$& $7.0 \times 10^{-6}$&$8.3 \times 10^{-6}$ &$1.0 \times 10^{-4}$\\
$^{19}$F &$4.8 \times10^{-9}$ &$4.0 \times 10^{-9}$&$2.4 \times 10^{-8}$& $4.7 \times 10^{-7}$&$8.8 \times 10^{-7}$ &$9.4 \times 10^{-6}$\\
$^{20}$Ne &$1.8\times10^{-1}$ &$1.8\times 10^{-1}$&$1.7 \times 10^{-1}$& $1.7 \times 10^{-1}$&$1.6 \times 10^{-1}$ &$9.0 \times 10^{-2}$\\
$^{21}$Ne &$3.7 \times10^{-6}$ &$1.3 \times 10^{-5}$&$3.4 \times 10^{-4}$& $9.0 \times 10^{-5}$&$7.0 \times 10^{-5}$ &$4.8 \times 10^{-5}$\\
$^{22}$Ne&$2.2 \times10^{-3}$ &$2.1 \times 10^{-3}$&$2.2 \times 10^{-3}$& $8.2 \times 10^{-4}$&$1.5 \times 10^{-5}$ &$7.5 \times 10^{-7}$\\
$^{22}$Na &$1.6 \times10^{-4}$ &$9.4 \times 10^{-5}$&$3.6 \times 10^{-4}$& $6.1 \times 10^{-4}$&$5.1 \times 10^{-4}$ &$2.6 \times 10^{-3}$\\
$^{23}$Na&$7.4 \times10^{-3}$ &$4.3 \times 10^{-4}$&$9.2 \times 10^{-3}$& $2.4 \times 10^{-3}$&$2.4 \times 10^{-3}$ &$1.0 \times 10^{-2}$\\
$^{24}$Mg &$8.4 \times10^{-5}$ &$1.3 \times 10^{-5}$&$3.3 \times 10^{-3}$& $2.0 \times 10^{-5}$&$3.1 \times 10^{-5}$ &$3.6 \times 10^{-4}$\\
$^{25}$Mg &$3.8 \times10^{-2}$ &$3.1 \times 10^{-2}$&$3.2 \times 10^{-2}$& $2.0 \times 10^{-3}$&$2.6 \times 10^{-3}$ &$8.7 \times 10^{-3}$\\
$^{26}$Mg&$4.7 \times10^{-3}$ &$2.5 \times 10^{-3}$&$3.6 \times 10^{-3}$& $8.9 \times 10^{-5}$&$1.4 \times 10^{-4}$ &$6.2 \times 10^{-4}$\\
$^{26}$Al &$6.5 \times10^{-4}$&$4.5 \times 10^{-3}$&$3.7 \times 10^{-3}$& $5.7 \times 10^{-4}$&$1.1 \times 10^{-3}$&$3.4 \times 10^{-3}$ \\
$^{27}$Al &$5.3 \times10^{-3}$&$1.2 \times 10^{-2}$&$9.0 \times 10^{-3}$& $3.0 \times 10^{-3}$&$4.5 \times 10^{-3}$&$1.4 \times 10^{-2}$ \\
$^{28}$Si &$1.2 \times10^{-3}$&$6.9 \times 10^{-3}$&$2.7 \times 10^{-3}$& $5.6 \times 10^{-2}$&$4.5 \times 10^{-2}$ &$4.8 \times 10^{-2}$\\
$^{29}$Si &$1.7 \times10^{-5}$&$1.8 \times 10^{-5}$&$1.8 \times 10^{-5}$& $1.5 \times 10^{-3}$&$1.6 \times 10^{-3}$ &$3.7 \times 10^{-3}$\\
$^{30}$Si &$1.2 \times10^{-5}$&$1.3 \times 10^{-5}$&$1.3 \times 10^{-5}$& $6.0 \times 10^{-3}$&$1.9 \times 10^{-2}$&$2.3 \times 10^{-2}$ \\
$^{31}$P &$4.1 \times10^{-6}$&$4.0 \times 10^{-6}$&$4.1 \times 10^{-6}$& $1.0 \times 10^{-3}$&$8.1 \times 10^{-3}$ &$1.4 \times 10^{-2}$\\
$^{32}$S &$2.0\times10^{-4}$&$2.0 \times 10^{-4}$&$2.0 \times 10^{-4}$& $4.4 \times 10^{-4}$&$1.0 \times 10^{-2}$&$6.2 \times 10^{-2}$ \\
$^{33}$S &$1.6 \times10^{-6}$&$1.6 \times 10^{-6}$&$1.6 \times 10^{-6}$& $1.6\times 10^{-6}$&$2.7 \times 10^{-5}$&$3.7 \times 10^{-3}$ \\
$^{34}$S &$9.5\times10^{-6}$&$9.4 \times 10^{-6}$&$9.5 \times 10^{-6}$& $7.1 \times 10^{-6}$&$6.7 \times 10^{-6}$&$1.5 \times 10^{-3}$ \\
$^{35}$Cl &$1.9 \times10^{-6}$&$1.9 \times 10^{-6}$&$1.9 \times 10^{-6}$& $4.4 \times 10^{-6}$&$8.1 \times 10^{-6}$ &$1.5 \times 10^{-3}$\\
$^{36}$Ar &$3.9 \times10^{-5}$&$3.8 \times 10^{-5}$&$3.9 \times 10^{-5}$& $2.2\times 10^{-5}$&$4.8 \times 10^{-6}$&$2.9 \times 10^{-4}$ \\
$^{40}$Ca &$3.0\times10^{-5}$&$3.0 \times 10^{-5}$&$3.0 \times 10^{-5}$& $3.0 \times 10^{-5}$&$3.0 \times 10^{-5}$&$3.3 \times 10^{-5}$ \\
\enddata
\tablenotetext{a}{All abundances are Mass Fraction.}





\end{deluxetable}

\clearpage

Table \ref{ONe2575MFBabund}, Table \ref{ONe5050MFBabund}, Table \ref{solarabund}, Table \ref {ONe2575MDTNRabund}, and Table \ref{ONe5050MDTNRabund} allow us to compare the abundance predictions for all the simulations that we have discussed in the previous sections.  The columns list the isotopes along the left hand column and the prediction for each WD mass is given in the other columns.  The WD mass is given at the top and the chosen composition is given in the table caption.  Table \ref{ONe2575MFBabund} provides the abundances for the MFB study with the 25\% WD - 75\% Solar composition, Table \ref{ONe5050MFBabund} provides the same set of data for the  50\% WD - 50\% Solar MFB simulations,  Table \ref{solarabund} shows the abundance predictions for the pure Solar simulations, while Table \ref {ONe2575MDTNRabund}, and Table \ref{ONe5050MDTNRabund} tabulate the abundances for the MDTNR 25\% WD - 75\% Solar composition and the MDTNR 50\% WD - 50\% Solar composition,  respectively.

\begin{figure}[htb!]
\includegraphics[width=1.0\textwidth]{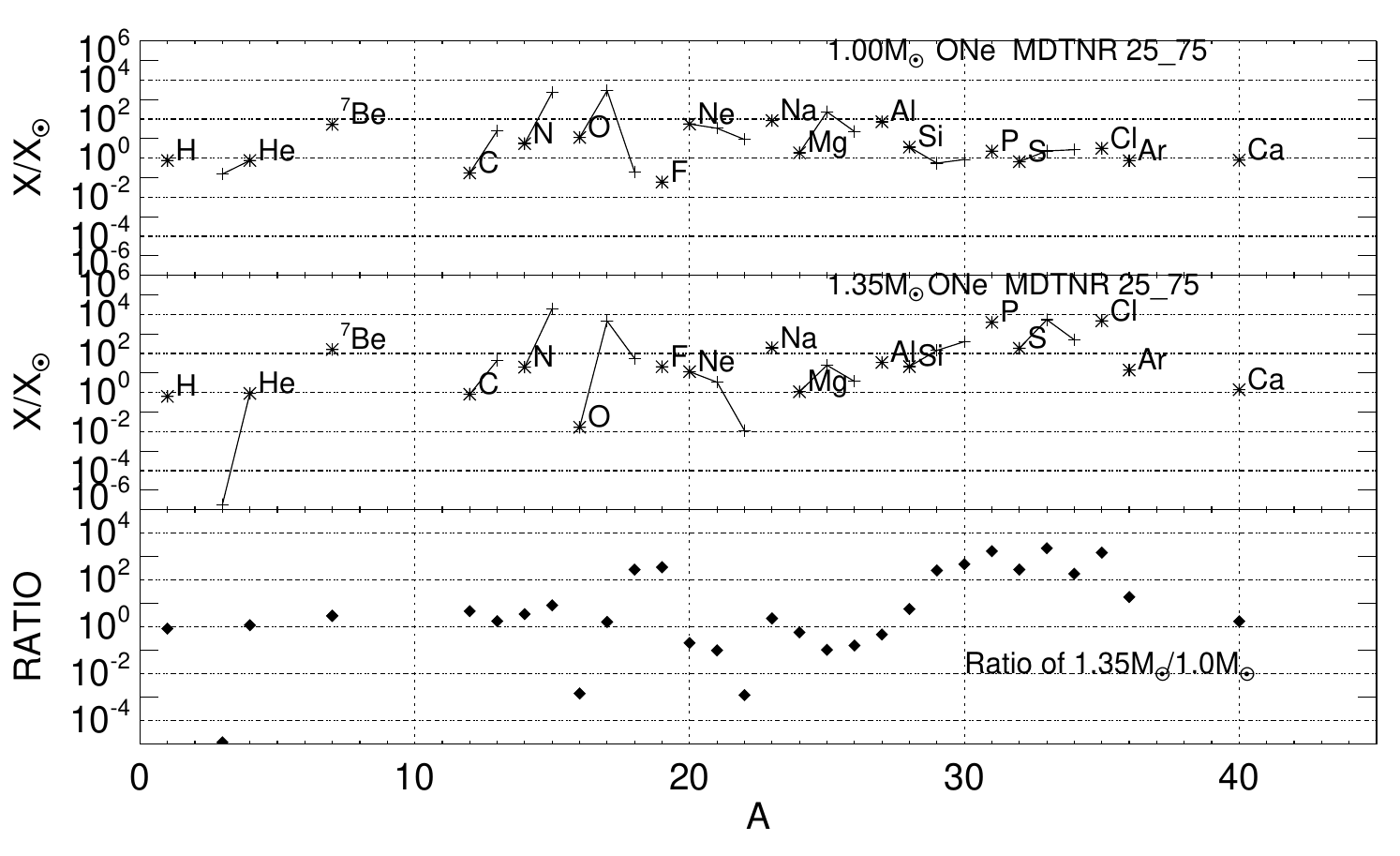}
\caption{Top panel: the abundances of the stable isotopes from hydrogen to calcium in the ejecta for the
1.0 M$_\odot$  ONe WD sequence with 25\% WD matter and 75\% Solar matter.  The $x$-axis is the atomic mass, A, and the
$y$-axis is the logarithmic ratio of the isotopic abundance divided by the
Solar abundance \citep{lodders_2003_aa}.  We also include $^7$Be in this plot, even though it is radioactive, 
because of its large overproduction and because it will decay to $^7$Li.  
The initial $^7$Li is depleted during the evolution (see the tables of isotopes).  As in \citet{timmes_1995_aa}, the most abundant isotope of a given element
is designated by ``$*$'' and all isotopes of a given element
are connected by solid lines.  Any isotope above 1.0 is
overproduced in the ejecta and a number of light, odd isotopes are
significantly enriched in the ejecta as is $^7$Be.
Middle panel: the same plot as the top panel but for the 1.35 M$_\odot$ simulation with
25\% WD matter and 75\% Solar matter.  Because of the higher
peak temperature in this simulation, phosphorus, sulfur, and chlorine are significantly enriched in addition to the light, odd isotopes of carbon, nitrogen, and oxygen. 
 {Bottom panel: The ratio of the abundances in 
the middle panel to the abundances in the top panel. We note the large depletion of $^3$He shown in the middle panel. This result was used, in part, by \cite{pepin_2011_aa} to identify ONe nova grains in anomalous interplanetary particles. Virtually all the intermediate mass elements are enriched in the 1.35 M$_\odot$ simulation while $^{16}$O and $^{20}$Ne are depleted, as expected in a TNR on an ONe WD.}}
\label{figure2575ONE}
\end{figure}

\begin{figure}[htb!]
\includegraphics[width=1.0\textwidth]{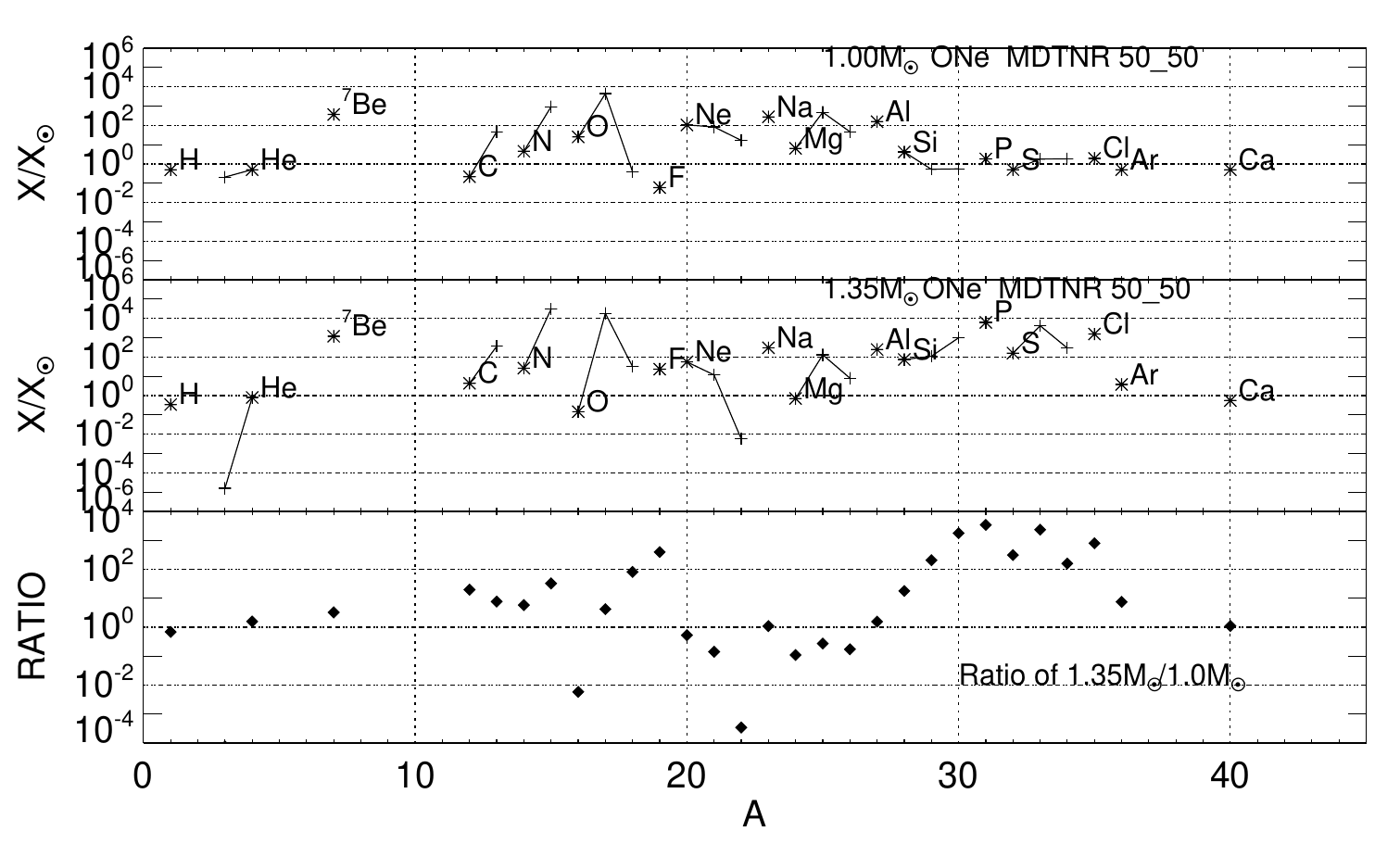}
\caption{Top panel: the same plot as in Figure \ref{figure2575ONE} but for the 1.0 M$_\odot$ 
simulation with 50\% WD matter and 50\% Solar matter. The most enriched species are
$^{13}$C, $^{15}$N, $^{17}$O, and $^7$Be. Middle panel:  the same plot as the Middle panel in Figure \ref{figure2575ONE} but for the simulation with
a mass of 1.35 M$_\odot$ and the 50\% WD and 50\% Solar composition.  Because of the higher temperatures reached in the 1.35 M$_\odot$ 
simulation, the isotopic overproduction reaches to much higher mass nuclei such as $^{31}$P, $^{32}$S, and $^{35}$Cl.  { As in Figure 
\ref{figure2575ONE}, the bottom panel shows the ratio of the middle panel to the top panel. In these simulations, the abundance of $^3$He is so depleted that the ratio lies below the bottom of the y-axis. This depletion was noted by \citet{pepin_2011_aa}.  Most of the isotopes with A greater than 3 are far more enriched in the simulation on a 1.35 M$_\odot$ WD.  The isotope, $^{31}$P, is of special interest  since \citet{banerjee_2023_ac} report
that photoionization models of infrared spectra of the ejecta of V1716 Sco (containing a massive WD) suggest that phosphorus is extremely enriched }}
\label{figure5050ONE}
\end{figure}

 In these simulations, the abundance of $^3$He is so depleted that the ratio lies below the bottom of the y-axis, as noted by Pepin et al. (2011). Most of the isotopes with A greater than 3 are far more enriched in the simulation on a 1.35 M$_\odot$ WD. Especially note $^{31}$P, which Banerjee et al. (2023) report is extremely enriched in IR observations of V1716 Sco.

\begin{figure}[htb!]
\includegraphics[width=1.0\textwidth]{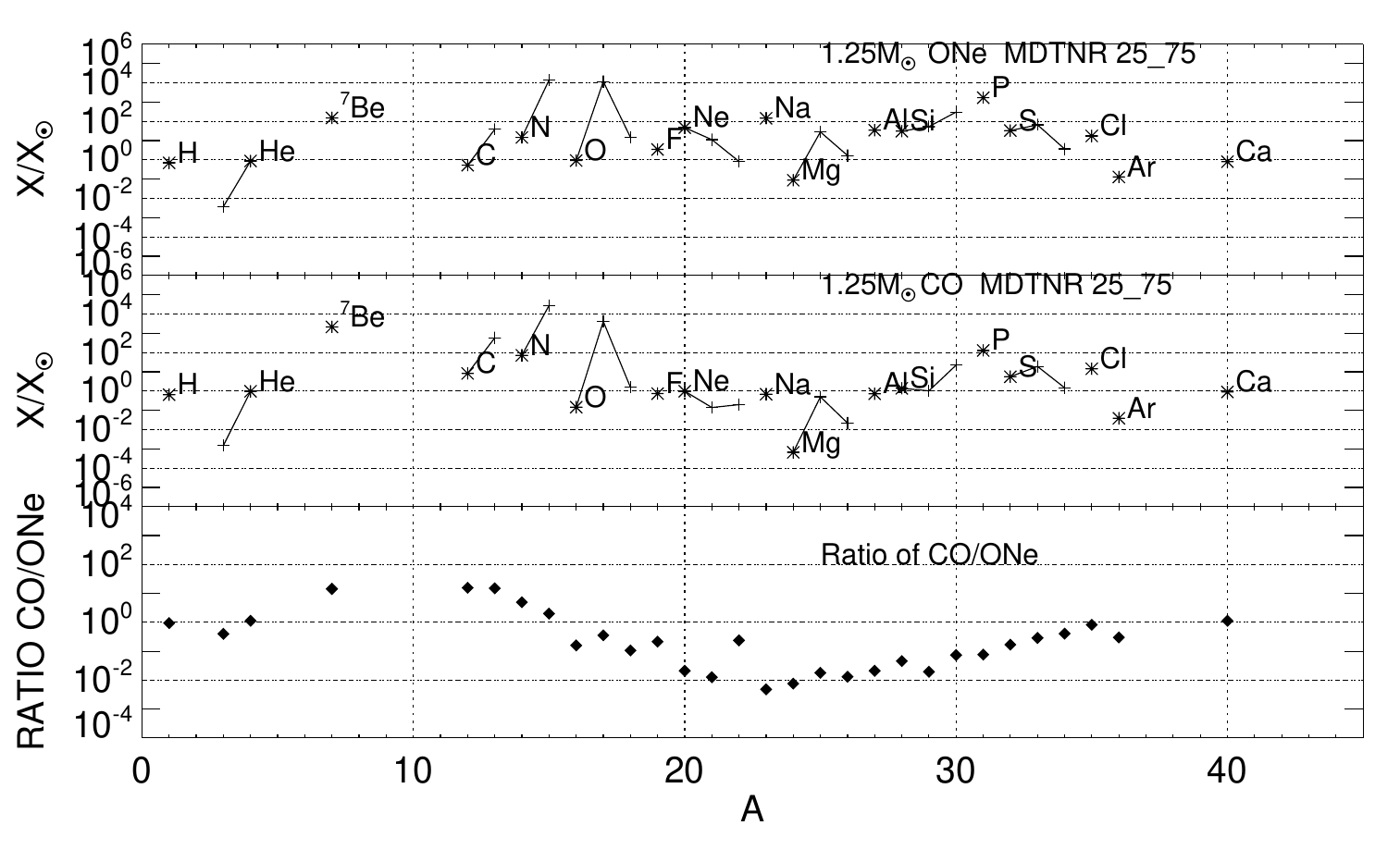}
\caption{Top panel: the same plot as in Figure \ref{figure2575ONE} but for the 1.25 M$_\odot$ ONe
simulation with 25\% WD matter and 75\% Solar matter. The most enriched species are $^7$Be, 
$^{13}$C, $^{15}$N, $^{17}$O, and $^{31}$P. Middle panel:  the same plot as the Middle panel in Figure \ref{figure2575ONE} but for the simulation with
a mass of 1.25 M$_\odot$ and the 50\% WD and 50\% Solar composition for a CO WD. These results are taken from \citet{starrfield_2020_aa} and are provided here to show the differences in the results for the two compositions.  The major difference, as expected, is that $^7$Be is much more abundant in 
the CO simulation.  
In contrast, $^{24}$Mg is considerably more depleted in the CO simulation. 
 {Bottom panel: As in Figures \ref{figure2575ONE} and \ref{figure5050ONE},
 this panel shows the ratio of the middle panel to the top panel.  We see that while the CO simulation enriches the light elements, compared to the ONe simulations, most of the intermediate mass elements are depleted except for phosphorus, sulfur, and chlorine. }}
\label{oneco2575}
\end{figure}

\begin{figure}[htb!]
\includegraphics[width=1.0\textwidth]{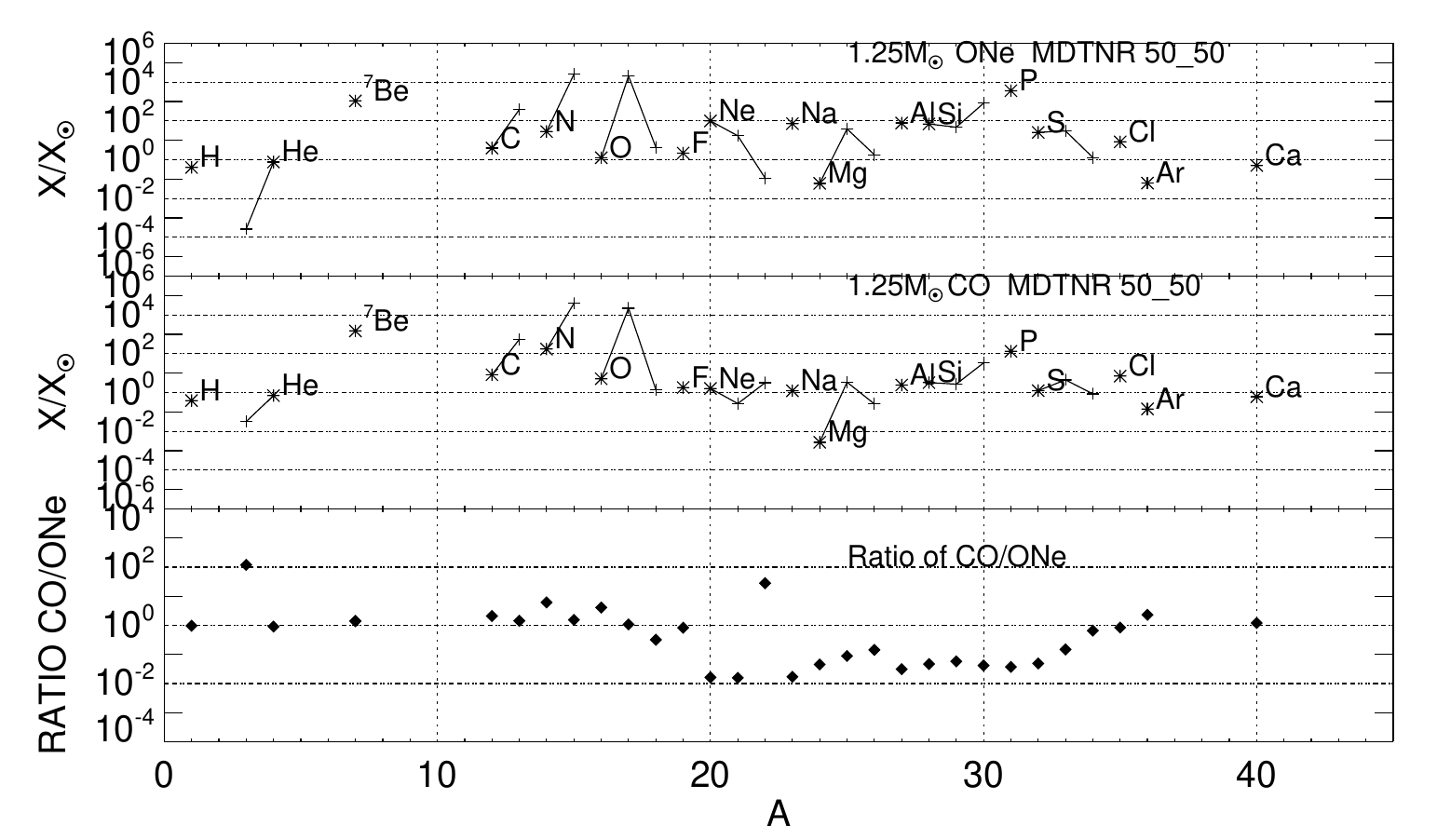}
\caption{Top panel: the same plot as in Figure \ref{figure2575ONE} but for the 1.25 M$_\odot$ ONe
simulation with 50\% WD matter and 50\% Solar matter. The most enriched species are $^7$Be,
$^{13}$C, $^{15}$N, $^{17}$O, and $^{31}$P. The depletion of $^3$He in the ONe results for both 1.25M$_\odot$ and 1.35M$_\odot$ simulations
were used (in part) by \citet{pepin_2011_aa} to identify pre-Solar grains from ONe CNe.  Middle panel:  the same plot as the Middle panel in Figure \ref{figure2575ONE} but for the CO simulation with
a mass of 1.25 M$_\odot$ and the 50\% WD and 50\% Solar composition. These results are taken from \citet{starrfield_2020_aa} and are provided here
to show the differences in the results for the two compositions.  In contrast to the results shown in Figure \ref{oneco2575},  $^7$Be has approximately the
same overabundance with respect to Solar in both compositions.  {This can be seen clearly in the bottom panel which as in the 3 previous plots shows the ratio of the abundances in the middle panel to the top panel.} }
\label{oneco5050}
\end{figure}

The production plots are Figure \ref{figure2575ONE}  (Top panel: 1.0 M$_\odot$ 25\_75 MDTNR; Middle panel: 1.35 M$_\odot$ 25\_75 MDTNR;  {Bottom panel: ratio of the abundances from the middle panel to the top panel}) and Figure \ref{figure5050ONE} (Top panel: 1.0 M$_\odot$ 50\_50 MDTNR; Middle panel: 1.35 M$_\odot$ 50\_50 MDTNR;  {Bottom panel: ratio of the abundances from the middle panel to the top panel}). 
In the next two figures, we compare our ONe results with the CO results from \citet{starrfield_2020_aa}. These are 
Figure \ref{oneco2575}  (Top panel: 1.25 M$_\odot$ 25\_75 MDTNR ONe; Middle panel: 1.25 M$_\odot$ 25\_75 MDTNR CO;  {Bottom panel: ratio of the abundances from the middle panel to the top panel}) and Figure \ref{oneco5050} (Top panel: 1.25 M$_\odot$ 50\_50 MDTNR ONe; Middle panel: 1.25 M$_\odot$ 50\_50 MDTNR CO;  {Bottom panel: ratio of the abundances from the middle panel to the top panel}). These plots  show the abundances of the stable isotopes (also including $^7$Be since it quickly decays to $^7$Li)  divided by the Lodders (2003) Solar abundances.
In these four figures, the $x$-axis is the atomic mass number and the $y$-axis is the logarithmic ratio of
the ejecta abundance divided by the Solar abundance of the same isotope.  The most abundant isotope of a given element is marked
by an asterisk and isotopes of the same element are connected by solid lines and labeled by the given element.  As in the earlier figures, 25\_75 refers to the mixture with 25\% WD and 75\% Solar matter while 50\_50 refers to the mixture with 50\% WD and 50\% Solar matter. 


\subsection{The production of the radioactive nucleus $^7$Be in ONe classical novae}
\label{lithium}

Because of the discoveries of $^7$Be, and its decay product $^7$Li, in CNe ejecta, we 
report in this subsection that our mixed  sequences are ejecting amounts of $^7$Be (which decays to $^7$Li after the simulation has ended) that are significantly enriched with respect to Solar $^7$Li.  
The observations of enriched $^7$Be are reported in \citet{tajitsu_2015_aa, tajitsu_2016_aa, izzo_2015_aa, izzo_2018_aa, molaro_2016_aa, selvelli_2018_aa,wagner_2018_aa, 
woodward_2020_aa, molaro_2020_aa, izzo_2020_aa, izzo_2022_aa, arai_2021_aa, dellavalle_2020_aa, molaro_2022_aa, molaro_2023_aa}, and qualitatively validated earlier predictions \citep{starrfield_1978_aa,
hernanz_1996_aa, jose_1998_aa, yaron_2005_aa}, and are, in part, the motivation for our new theoretical studies. 
CNe produce $^7$Li via a process originally described by \citet{cameron_1971_aa} for red giants.   
\citet{starrfield_1978_aa}, followed in more detail by \citet{hernanz_1996_aa} and \citet{jose_1998_aa}, investigated
the  formation of $^7$Be during the TNR.  They determined the amount of $^7$Be
carried to the surface by convection, and which survived before it could be destroyed by the $^7$Be(p,$\gamma$)$^8$B
reaction occurring in the nuclear burning region.  If it survives, by being transported to cooler regions, $^7$Be decays via electron-capture to $^7$Li with an $\sim$53 day half-life \citep{bahcall_1969_ab}.

Our simulations confirm that a TNR on an ONe WD also overproduces $^7$Be with
respect to Solar material and in amounts that imply that both CO and ONe CNe are responsible for a significant amount of galactic $^7$Li. Since $^6$Li is produced by spallation in the interstellar medium \citep{fields_2011_aa},  its abundance in the Solar system should not correlate with $^7$Li.  \citet{hernanz_2015_aa} gives an excellent discussion of the cosmological importance of detecting $^7$Li in nova explosions.

\begin{figure}[htb!]
\includegraphics[width=1.0\textwidth]{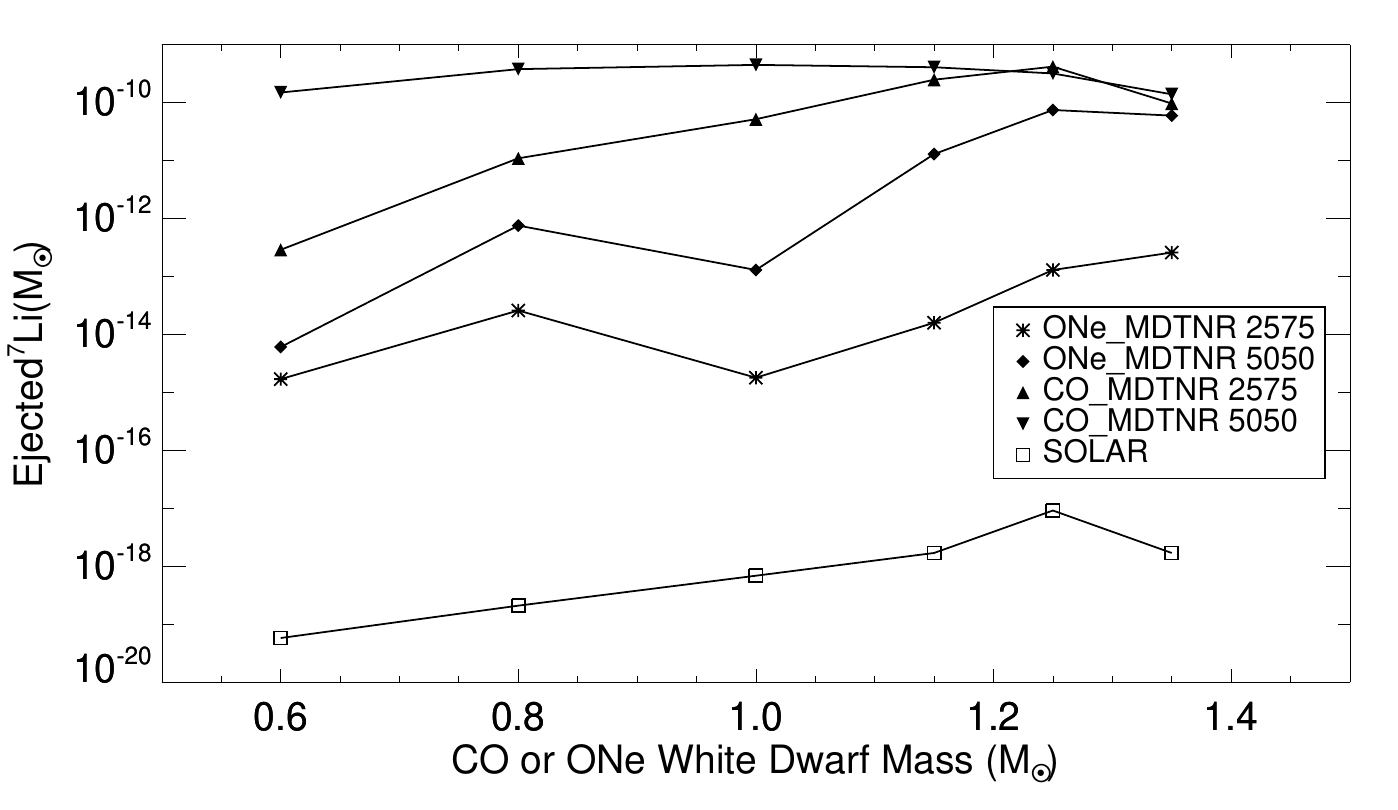}
\caption{The predicted $^7$Li abundance in the ejecta as a function of WD mass in units of Solar masses.  This plot includes both the results from this paper and those from \citet{starrfield_2020_aa}. As expected \citep{jose_1998_aa}, the CO simulations do eject more $^7$Be than the ONe simulations but the simulations on 50\% ONe WD matter and 50\% Solar are not far below for the most massive WDs. The TNRs on WDs reach sufficiently 
high temperatures to deplete the initial $^7$Li present in the accreted material.  The TNR then produces $^7$Be which is mixed to the surface by convection during the TNR and we actually plot that nucleus.  $^7$Be decays ($\sim$ 53 day half-life) after the end of the simulations. The simulations where we mix from the beginning (MFB) eject far less $^7$Li 
and are not plotted here. The simulation with Solar abundances on a 1.25 M$_\odot$ WD did not eject
any material  {so the plotted point is the $^7$Be abundance in the surface zone.}}
\label{figureli7mass}
\end{figure}

\begin{figure}[htb!]
\includegraphics[width=1.0\textwidth]{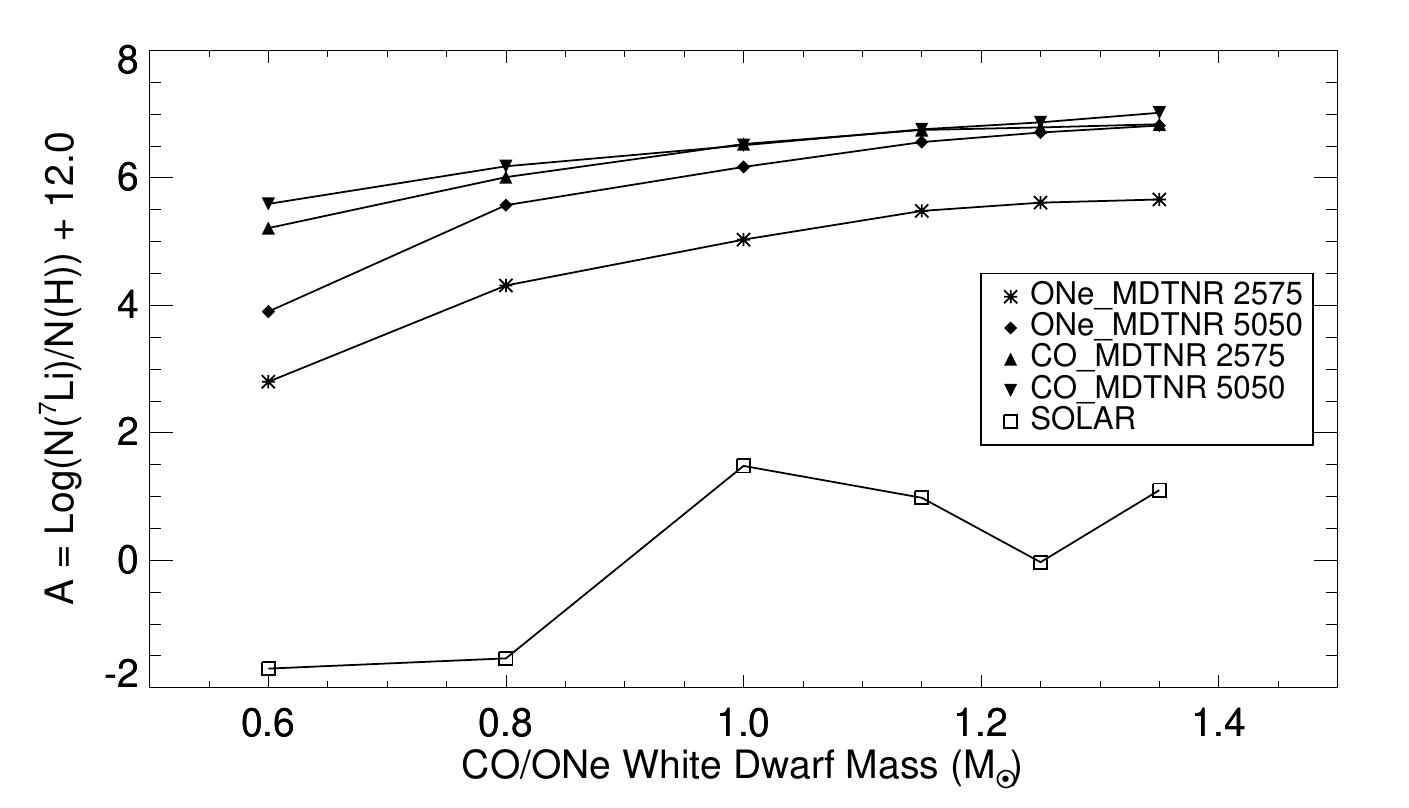}
\caption{The predicted $^7$Li abundance in the ejecta as a function of WD mass in A = Log(N($^7$Li)/N(H)) + 12.  This plot includes both the results from this paper and those from \citet{starrfield_2020_aa}. As expected \citep{jose_1998_aa}, the CO simulations do eject more $^7$Be than the ONe simulations but the simulations on 50\% ONe WD matter and 50\% Solar are not far below for the most massive WDs.  The simulation with Solar abundances on a 1.25 M$_\odot$ WD did not eject
any material  {so the plotted point is the abundance in the surface zone.}}
\label{Avalue}
\end{figure}

\begin{deluxetable}{@{}lcccccc}
\tablecaption{Comparison of  the results from our ONe simulations with previous work  \label{comparison7be}}
\tablewidth{0pt}
\tablecolumns{7}
\tablehead{ \colhead{ONe WD Mass (M$_\odot$): }&
\colhead{1.0}&
\colhead{1.15}&
\colhead{1.15}&
\colhead{1.25}&
\colhead{1.35}} 

\startdata
Core \%{\tablenotemark{a}}&50&25&50&50&50\\
\hline
{\bf $^7$Be ejecta abundance by mass}&&&&&&\\
\hline
\citet{jose_1998_aa}& $2.3 \times 10^{-7}$ & $4.6\times 10^{-8}$&$6.0 \times 10^{-7}$&
$6.9 \times 10^{-7}$&$1.3 \times 10^{-6}$\\
\citet{rukeya_2017_aa}&\nodata& $5.9 \times 10^{-8}$ & $8.6\times 10^{-7}$&$6.9\times 10^{-7}$&
$2.4 \times 10^{-6}$\\
\citet{starrfield_2009_aa}&\nodata&\nodata&\nodata&$1.4\times 10^{-8}$&$1.9\times 10^{-7}$\\
MFB (This Work)&$9.8 \times 10^{-8}$ & $2.6\times 10^{-8}$&$4.2\times 10^{-7}$&
$6.3 \times 10^{-7}$&$7.8 \times 10^{-7}$\\
MDTNR (This Work)&$3.4 \times 10^{-6}$ & $1.2\times 10^{-6}$&$3.7\times 10^{-6}$&
$1.0 \times 10^{-5}$&$1.1 \times 10^{-5}$\\
\hline
{\bf $^{22}$Na ejecta abundance by mass}&&&&&&\\
\hline
\citet{jose_1998_aa}& $4.8 \times 10^{-5}$ & $3.1\times 10^{-5}$&$5.3 \times 10^{-5}$&
$9.6 \times 10^{-5}$&$6.0 \times 10^{-4}$\\
\citet{starrfield_2009_aa}&\nodata&\nodata&\nodata&$4.5\times 10^{-3}$&$2.3\times 10^{-2}$\\
MFB (This Work)&$9.3 \times 10^{-5}$ & $5.2\times 10^{-5}$&$2.2\times 10^{-4}$&
$7.7 \times 10^{-4}$&$5.3 \times 10^{-4}$\\
MDTNR (This Work)&$3.6 \times 10^{-4}$ & $6.6\times 10^{-4}$&$6.1\times 10^{-4}$&
$5.1 \times 10^{-4}$&$2.6 \times 10^{-3}$\\
\hline
{\bf $^{26}$Al ejecta abundance by mass}&&&&&&\\
\hline
\citet{jose_1998_aa}& $2.7 \times 10^{-3}$ & $1.8\times 10^{-4}$&$9.3\times 10^{-4}$&
$5.4 \times 10^{-4}$&$7.2 \times 10^{-4}$\\
\citet{starrfield_2009_aa}&\nodata&\nodata&\nodata&$2.1\times 10^{-3}$&$3.0\times 10^{-3}$\\
MFB (This Work)&$4.3 \times 10^{-3}$ & $2.3\times 10^{-4}$&$5.9\times 10^{-4}$&
$5.6 \times 10^{-4}$&$6.4 \times 10^{-4}$\\
MDTNR (This Work)&$3.7 \times 10^{-3}$ & $1.2\times 10^{-3}$&$5.7\times 10^{-4}$&
$1.1 \times 10^{-3}$&$3.4 \times 10^{-3}$\\
\hline
{\bf Ejected Mass (M$_\odot$)}&&&&&&\\
\hline
\citet{jose_1998_aa}& $4.7 \times 10^{-5}$ & $2.3\times 10^{-5}$&$1.9 \times 10^{-5}$&
$1.4 \times 10^{-5}$&$4.4 \times 10^{-6}$\\
\citet{rukeya_2017_aa}&\nodata& $1.9 \times 10^{-5}$ & $1.2\times 10^{-5}$&$7.3\times 10^{-6}$&
$1.9\times 10^{-6}$\\
\citet{starrfield_2009_aa}&\nodata&\nodata&\nodata&$1.5\times 10^{-5}$&$1.7\times 10^{-5}$\\
MFB (This Work)&$3.4 \times 10^{-7}$ & $1.6\times 10^{-7}$&$5.1\times 10^{-7}$&
$5.6 \times 10^{-7}$&$1.4 \times 10^{-6}$\\
MDTNR (This Work)&$3.9 \times 10^{-8}$ & $1.1\times 10^{-6}$&$3.6\times 10^{-6}$&
$7.5 \times 10^{-6}$&$5.5 \times 10^{-6}$\\
\hline
{\bf Retention Efficiency:  (M$_{\rm acc}$-M$_{\rm ej}$)/M$_{\rm acc}$}&&&&&&\\
\hline
\citet{jose_1998_aa}& 0.26&0.28&0.41&0.36&0.18\\
\citet{starrfield_2009_aa}&\nodata&\nodata&\nodata&0.75&0.39\\
MFB (This Work)&0.99&0.99&0.98&0.96&1.00\\
MDTNR (This Work)&1.00&0.97&0.90&0.62&0.45\\
\enddata
\tablenotetext{a}{The numbers in this row are the percent of core material in the simulation.}
\end{deluxetable}

The ONe production plots: Figure \ref{figure2575ONE} and Figure \ref{figure5050ONE}, show that $^7$Be is enriched over Solar for both mixtures that we studied and for WD masses exceeding 1.0 M$_\odot$.  
This is in contrast to the assumption that only CO CNe produce significant amounts of $^7$Be and is in better agreement with the observations that do show both composition classes of CNe produce $^7$Be \citep{molaro_2023_aa}.
 There are other isotopes that are more enriched than $^7$Be such as $^{15}$N,  $^{17}$O, and $^{31}$P.  These results are partially confirmed by
infrared spectra of V1716 Sco which, when analyzed with CLOUDY \citep{ferland_2023_aa}, show that phosphorus is highly enriched in the ejecta \citep{banerjee_2023_ac}. We also provide Figures  \ref{oneco2575} and \ref{oneco5050} which compare the results for the ONe simulations to the CO simulations \citep{starrfield_2020_aa}.   Figure \ref{oneco2575}  compares the same 25\% WD - 75\% Solar composition but for a 1.25 M$_\odot$ WD and we see that $^7$Be is significantly more enriched in the CO simulation.  Figure \ref{oneco5050} shows the differences between the simulations for 1.25 M$_\odot$ WDs but for the 50\% WD - 50\% Solar composition.  The $^7$Be enrichment is about equal and about a thousand times Solar.

Table \ref{ONe2575MFBabund}, Table \ref{ONe5050MFBabund}, Table \ref{solarabund}, Table \ref {ONe2575MDTNRabund}, and Table \ref{ONe5050MDTNRabund} show how the abundance of $^7$Be varies as a function of the composition and WD mass. All values in the table are given in mass fraction of ejected material.  We also tabulate the ejected abundance of $^7$Li but in all cases it is significantly smaller than $^7$Be.  These tables show that the $^7$Be abundance is an increasing function of WD mass for all compositions.  

We show the amount of $^7$Li (actually produced as $^7$Be) in the ejected material in Solar masses
in Figure \ref{figureli7mass} as a function of  WD mass.  This plot provides a comparison of the results for Solar, ONe, and CO simulations with the CO data taken from \citet{starrfield_2020_aa}.  The ONe 50\% WD - 50\% Solar simulations for WD masses exceeding 1.15 M$_\odot$ produce nearly the same amount of $^7$Li as the CO simulations. The ejected amount is least abundant in the pure Solar accretion simulations and
most abundant in the 50\% WD - 50\% Solar MDTNR simulations.  The amounts of $^7$Be produced in the ONe simulations are sufficiently close to those in the CO simulations that we are confident that both compositional classes of CNe contribute to the production of $^7$Li in the galaxy.
Observations of both CO and ONe novae in outburst show large enrichments of $^7$Li  \citep{molaro_2022_aa, molaro_2023_aa}.

While the nucleus produced during the TNR is  $^7$Be, we do not follow the simulations sufficiently long for $^7$Be to decay to $^7$Li ($\sim$ 53 d).  Moreover, all the initial $^7$Li (or $^6$Li) in the accreting material is destroyed by the TNR. Both Tables \ref{evolONeMFB} and \ref{evolONeMDTNR} give the $^7$Li abundance (assuming that the $^7$Be has decayed) both as the amount of $^7$Li ejected with respect to the Solar value (N($^7$Li/H)$_{\rm ej}$/N($^7$Li/H)$_{\odot}$) and the $^7$Li ejected in solar masses.

In Table \ref{comparison7be}, we compare the values for the ejected amount of $^7$Be from both our MFB and MDTNR studies with those in \citet{hernanz_1996_aa}, \citet{ jose_1998_aa}, \citet{rukeya_2017_aa}, and \citet{starrfield_2009_aa}.  We also compare the ejecta abundances for $^{22}$Na, $^{26}$Al, the amount of mass ejected and RE in this table but defer the discussion of those results to later sections. 
\citet{rukeya_2017_aa} also provide a comparison with \citet{jose_1998_aa}.  The top row lists the WD mass and the next row gives the specific mixture, either 25\% WD matter or 50\% WD matter.  The next set of rows is the comparison of the $^7$Be results from each of the studies listed in the left column. 
The values in the first four rows all  assume MFB.

Although there are differences between the microphysics and treatment of convection in SHIVA \citep{jose_1998_aa} and NOVA (such as the opacities, equations of state, 
nuclear reaction rate library),  there is reasonably good agreement in our MFB predictions of $^7$Li  ejecta abundances, compared to \citet{jose_1998_aa}, except for the
1.35 M$_\odot$ simulation, where our MFB simulation predicts about half the amount as  \citet{jose_1998_aa}.  
The agreement is also good when comparing our results with \citet{rukeya_2017_aa} who used 
MESA \citep{paxton_2011_aa, paxton_2013_aa, paxton_2015_aa, paxton_2016_aa, paxton_2018_aa} in their study.
However, when we examine the results for the MDTNR simulations, as expected, they exceed, in some cases, the other simulations
by as much as a factor of 10.  Therefore, we again predict that both CO and ONe CNe are producing $^7$Li.

\begin{figure}[htb!]
\includegraphics[width=1.0\textwidth]{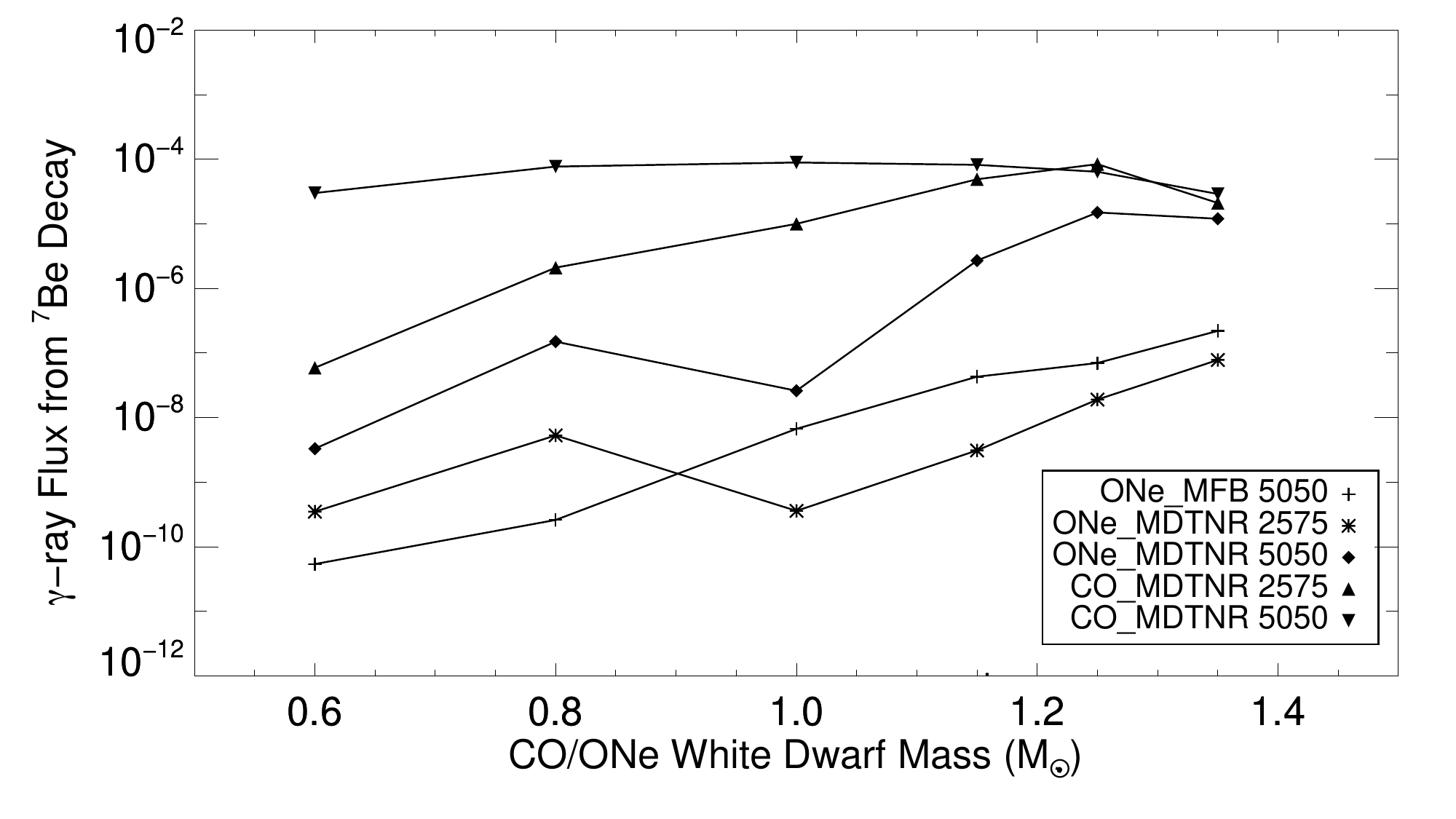}
\caption{The predicted $\gamma$-ray flux (478 keV), in  photons cm$^{-2}$ s$^{-1}$ (at 1 kpc), from
the decay of $^7$Be using \citet[][Eq. 7]{gehrz_1998_aa}.  For our CO nova results we predict $\sim 4 \times 10^{-5}$ photons cm$^{-2}$ s$^{-1}$
at a distance of 1 kpc.  However, for our ONe results we obtain values ranging from $\sim 2 \times 10^{-10}$ photons cm$^{-2}$ s$^{-1}$ to
$\sim 3 \times 10^{-6}$ photons cm$^{-2}$ s$^{-1}$ for a distance of 1 kpc.  The drop at 1.0 M$_\odot$ for the two ONe simulations is caused by a drop in the amount of mass ejected in these simulations.}
\label{be7gamma}
\end{figure}

In \citet{starrfield_2020_aa}, we reported an ejected $^7$Be mass of $ 2.0 \times 10^{-10}$ M$_\odot$ for CO novae. 
In this paper we report a peak ejected $^7$Be mass of $\sim 6 \times 10^{-11}$M$_\odot$ for ONe novae (the 50\% WD - 50\% Solar simulation on a 1.25M$_\odot$ WD).  The cause of this difference is explained in \citet{jose_1998_aa}. We can use these two values and predict the 478 keV $\gamma$-ray emission from the decay of $^7$Be using \citet[][Eq. 7]{gehrz_1998_aa}. Our results are given in Figure \ref{be7gamma}.  The values depend on the amount of mass ejected and the predicted $^7$Be abundance for each simulation.  The drop in the emission at 1.0 M$_\odot$ is caused
by that simulation not ejecting any matter.  

The values of the ejected mass of $^7$Be are taken  from the initial ejection and do not assume any decays.  Light curves for the $\gamma$-ray emission can be found in \citet{hernanz_2008_cn} and \citet{leung_2022_aa}.  For our CO nova results we predict $\sim 4 \times 10^{-5}$ photons cm$^{-2}$ s$^{-1}$  at a distance of 1 kpc.  In contrast, for our ONe results, we obtain values ranging from $\sim 2 \times 10^{-10}$ photons cm$^{-2}$ s$^{-1}$ to $\sim 3 \times 10^{-6}$ photons cm$^{-2}$ s$^{-1}$ for a distance of 1 kpc.  For CNe at larger distances , we need to scale by D$^{-2}$ where D
is the distance in kpc (the 4$\pi$ is included in the constant). As emphasized by \citet{hernanz_2008_cn},
 these values are too low for an INTEGRAL detection unless the CN is closer than 1 kpc.
 
 \citet{denissenkov_2021_aa} re-examined the production of $^7$Be in CO novae both analytically and numerically, and showed that it is the abundance of $^4$He (and not $^3$He as previously believed) that determines the final abundance of  $^7$Be. A similar result was found by \citet{starrfield_2020_aa} where the simulations with 25\% WD and 75\% Solar produced more $^7$Be than those with 50\% WD and 50\% Solar.  Since we predict that the mass of the WD is growing as a result of the CN phenomena and most of the  hydrogen has fused to helium in the material that remains on the WD, it is possible that simulations with increased $^4$He will produce a larger amount of $^7$Be decreasing the tension between current theory and observations.  Simulations with increased $^4$He, approximating the effects of previous  outbursts, are in progress.

We also show in Table \ref {comparison7be} the comparison of the amount of ejected mass and the RE.  For all cases, except that for the 50\% WD - 50\% Solar simulation on a WD mass of 1.35M$_\odot$, the sequences listed for \citet{jose_1998_aa} all eject more mass than either \citet{rukeya_2017_aa} or our MFB and MDTNR calculations.  Our MFB results are considerably smaller than either of the other two studies.  In tests done to better understand this difference, we find that the introduction of the new electron degenerate conductivities as described in \citet{cassisi_2007_aa} strongly effects the structure of the TNR and reduces the amount of ejected material.  In addition, \citet{jose_1998_aa} used fewer mass zones ($\sim$ 35) with (probably) larger masses although they only include the envelope. Interestingly, the two simulations that eject the most mass are those from \citet{starrfield_2009_aa}.  However, our MDTNR 50\% WD - 50\% Solar simulation at  a WD mass of 1.35 M$_\odot$, does predict that more mass is ejected than reported in either \citet{jose_1998_aa} or  \citet{rukeya_2017_aa}. 

We display in the last set of rows the RE for each simulation where it was possible to determine both an ejected and accreted mass.  \citet{rukeya_2017_aa} do not give both.  As has already been discussed by  \citet{starrfield_2020_aa}, \citet{jose_1998_aa}  eject more material than in our simulations.  Although the RE for their simulations is roughly flat, we find that the RE decreases with increasing WD mass in the MDTNR studies while our MFB studies never eject a significant amount of mass.  Nevertheless, in none of the studies listed in Table \ref{comparison7be} does
the RE become negative so that all the simulations predict that the WD is growing in mass.

\begin{figure}[htb!]
\includegraphics[width=1.0\textwidth]{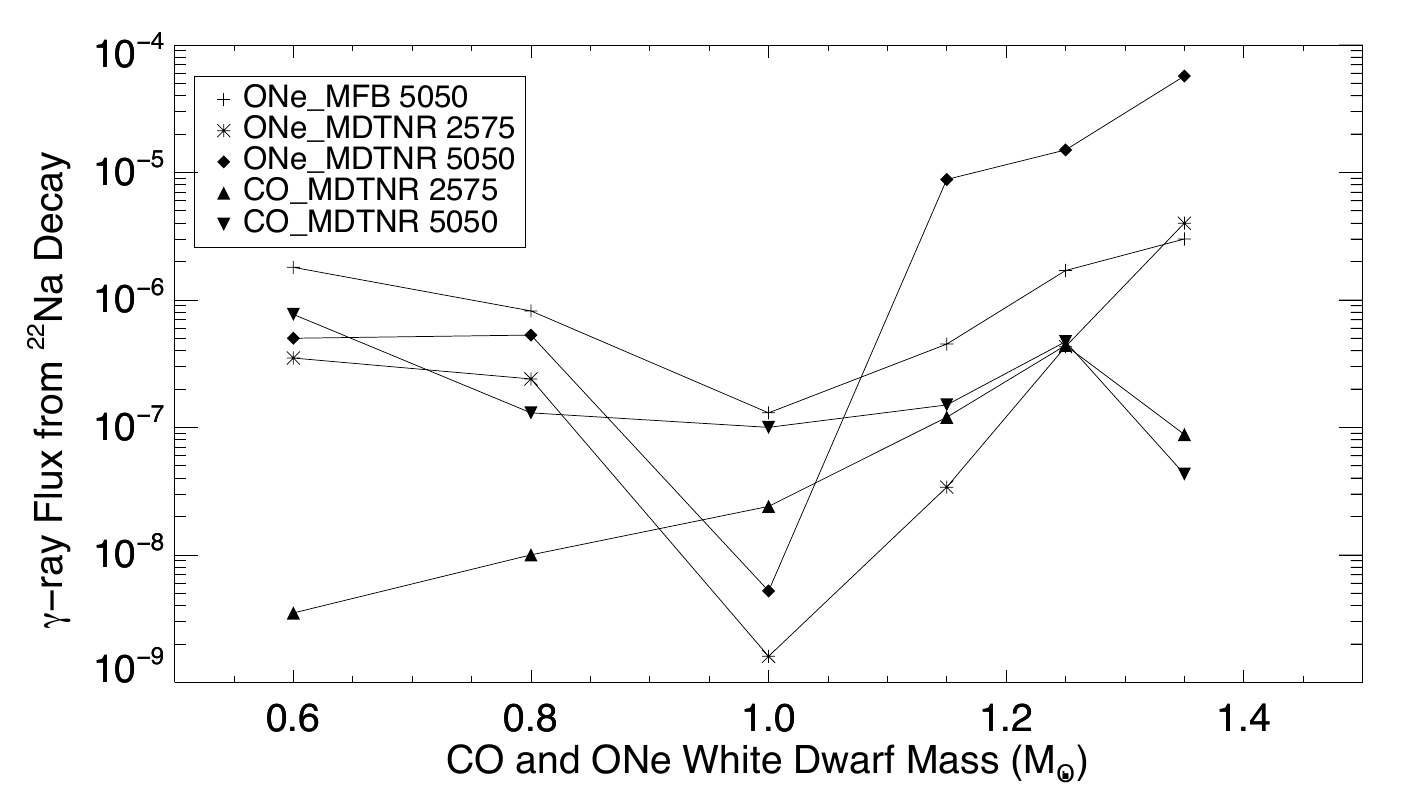}
\caption{We use \citet[][Eq. 5]{gehrz_1998_aa} to predict the $\gamma$-ray emission from both CO and ONe simulations.  For our CO nova results we predict values $< 10^{-6}$ photons cm$^{-2}$ s$^{-1}$ at a distance of 1 kpc.  However, for our ONe results, we obtain values ranging from $\sim 8\times 10^{-7}$ photons cm$^{-2}$ s$^{-1}$ to $\sim 7 \times 10^{-5}$ photons cm$^{-2}$ s$^{-1}$ for a distance of 1 kpc. We use the abundances determined from the ejecta and assume a distance of 1 kpc. The large drop in emission at a WD mass of 1.0 M$_\odot$ is caused by a large drop in ejected mass.}
\label{na22gamma} 
\end{figure}

\subsection{The production of the radioactive nuclei $^{22}$Na and $^{26}$Al  in ONe classical novae}
\label{sodium}

In addition to $^7$Be, there are two other radioactive nuclei produced in CN explosions that are of interest:
$^{22}$Na and $^{26}$Al. $^{22}$Na decays to $^{22}$Ne by a $\beta^+$-decay reaction,  $^{22}$Na $->$ $^{22}$Ne+e$^+$+$\nu$, leading to the emission of a 1.275 MeV $\gamma$-ray. The half-life is 2.6 yr and, while it has been long predicted, it has not been detected in CNe   \citep{hernanz_2008_cn, leung_2022_aa}.  Here we update the prediction of the amount of $^{22}$Na produced in ONe novae to values exceeding $10^{-8}$M$_\odot$ for the most massive WDs and a 50\% WD and 50\% Solar composition.  We then use \citet[][Eq. 5]{gehrz_1998_aa} and predict the $\gamma$-ray emission from both CO and ONe simulations 
from $^{22}$Na decay (Figure \ref{na22gamma}).  These results are for a distance of 1 kpc and use the abundances determined at the time material is ejected.  Light curves for the $\gamma$-ray emission can be found in  \citet[][]{hernanz_2008_cn} and \citet[][and references therein]{leung_2022_aa}.

For our CO nova results we predict $\sim10^{-7}$ photons cm$^{-2}$ s$^{-1}$
at a distance of 1 kpc.   For our ONe results, however, we obtain values ranging from $\sim 10^{-6}$ photons cm$^{-2}$ s$^{-1}$ to
nearly $10^{-4}$ photons cm$^{-2}$ s$^{-1}$ for a distance of 1 kpc  at ejection.  For CNe at larger distances we need to scale by D$^{-2}$ where D
is the distance in kpc.  The large drop in emission at a WD mass of 1.0 M$_\odot$ is caused by a large drop in ejected mass.  We did not vary any
initial conditions to increase the amount of ejected mass for the simulation at 1.0 M$_\odot$ but, if we could increase the ejected mass by a factor of $10^3$ (to fit with the other simulations), the
 $\gamma$-ray emission would still be below the peak value realized at a WD mass of 1.35 M$_\odot$.  As expected, ONe CNe eject more $^{22}$Na than CO novae with a concomitant
increase in $\gamma$-ray emission. 

\begin{figure}[htb!]
\includegraphics[width=1.0\textwidth]{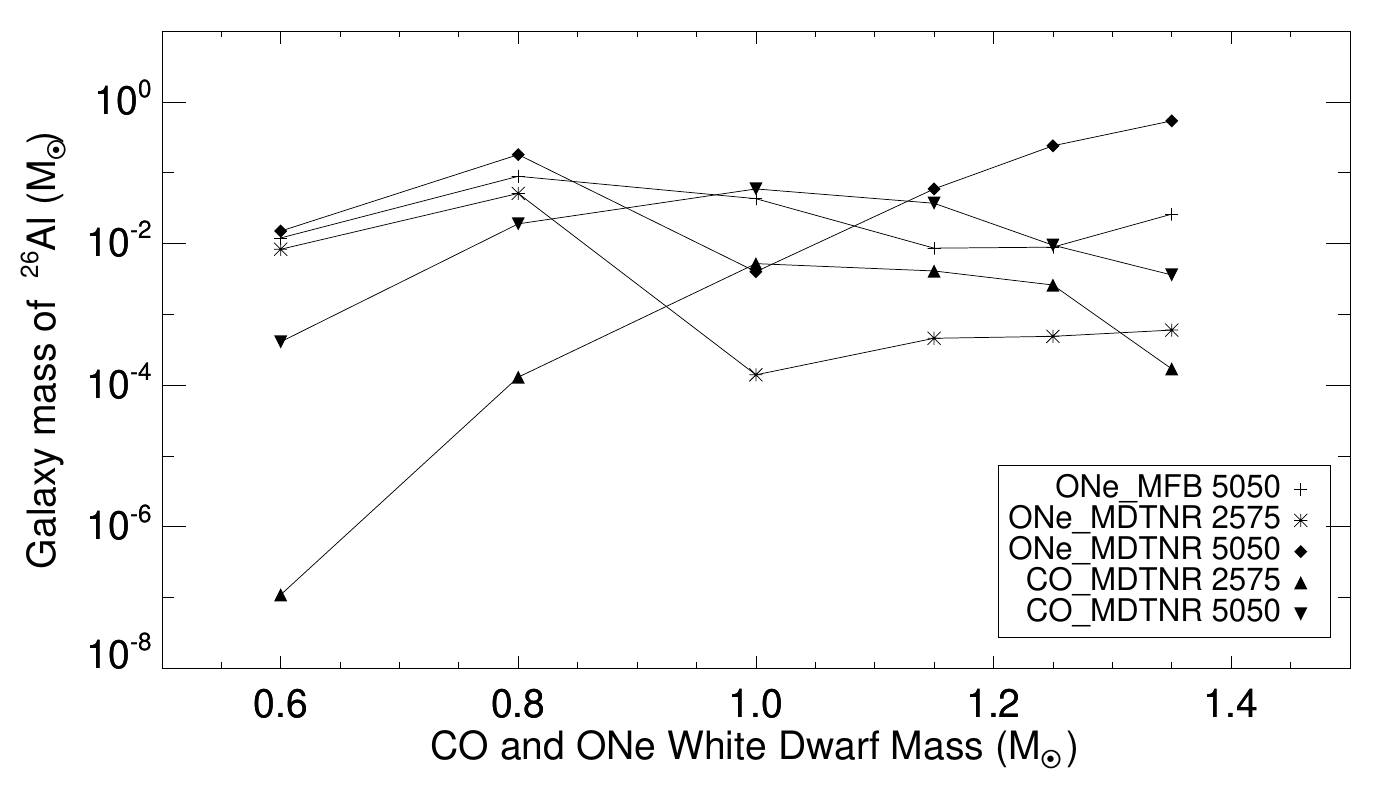}
\caption{We use \citet[][Eq. 6]{gehrz_1998_aa} to predict the amount of $^{26}$Al in the galaxy in M$_\odot$.  See text for more details. Again, the drop in the ONe MDTNR amounts at 1.0M$_\odot$ is caused by the small amount of ejected mass. Raising the value to fit with the WD masses both below and above could raise our predictions to interesting amounts at this mass.  The decline in the ejected amount as the WD mass increases (see Table \ref{evolONeMDTNR}) is also caused by a decline in the ejected mass.}
\label{al26galaxy}
\end{figure}

Another radioactive nucleus that is produced and ejected during an ONe outburst is $^{26}$Al.   \citet{urey_1955_aa} was the first to suggest that radioactive decay of $^{26}$Al was important for the heating of the small bodies in the Solar System. However, it was not until \citet{lee_1977_aa} found excesses of $^{26}$Mg in the Allende meteorite, that it was demonstrated that $^{26}$Al was present in the early Solar System. Because its half life, $7.2 \times 10^5$ yr, is short compared to the formation of the Solar System, this result implied that $^{26}$Al must have been produced by some astrophysical process shortly before the formation of the Solar System and mixed with the pre-solar nebula as it was forming. 

Its presence in the interstellar medium (ISM) was confirmed by the discovery, by HEAO-3, of the 1.809 MeV $\gamma$-ray line, which results from the decay of  $^{26}$Al  to the first excited state of $^{26}$Mg \citep{mahoney_1982_aa}.  \citet{mahoney_1984_aa} later used their detection to estimate that there was about 3 M$_\odot$ of $^{26}$Al  in the ISM and examined the possibility that CNe could produce this isotope.  More recent studies have lowered the observed amount of $^{26}$Al in the ISM to $\sim$ 2 M$_\odot$ \citep{diehl_2010_aa}.

We have argued previously that CNe could be a significant contributor to both the observed amount of $^{26}$Al in the galaxy and the amount of $^{26}$Mg (its decay product) found in the Solar System, although the major source is likely massive stars \citep{diehl_2022_aa}.  \citet{canete_2021_aa} have now redetermined the $^{25}$Al($p$,$\gamma$)$^{26}$Si
reaction rate and they predict that CNe may be responsible for up to 15\%  of the galactic abundance of $^{26}$Al.

 We have used \citet[][Eq. 6]{gehrz_1998_aa} to predict the amount, in M$_\odot$, of $^{26}$Al in the galaxy produced by both CO and ONe CNe (Figure \ref{al26galaxy}). 
  {In contrast to the expression presented in \citet{gehrz_1998_aa}, we assume a nova rate of 50 yr$^{-1}$ \citep{dellavalle_2023_aa}, the actual 
value of the ejected mass for each simulation, but use 50\% of the calculated values of the ejected mass for each simulation for either CO or ONe novae.} 
Again, the drop in the ONe MDTNR amounts at 1.0 M$_\odot$ is caused by the small amount of ejected mass.  Raising the value to fit with the WD masses both below and above could raise our predictions to interesting amounts at this mass.  The decline in the ejected amount as
the WD mass increases (see Table \ref{evolONeMDTNR}) is caused by a general decline in the ejected mass.

\subsection{Enrichment of the other nuclei in ONe novae ejecta}
\label{others}

Figure \ref{figure2575ONE} and Figure \ref{figure5050ONE} are the production plots in which we compare the ONe MDTNR simulations. 
Figure \ref{figure2575ONE} (for the 25\% WD - 75\% Solar sequences) and Figure \ref{figure5050ONE} (for the 50\% WD - 50\% Solar sequences) show the expected results when we compare two different WD masses.  Because of the higher gravity in the
more massive WD, the TNR reaches a higher peak temperature and exhibits a higher density in the nuclear burning region than found in lower mass WDs. 
Thus, $^{13}$C, $^{15}$N, $^{17}$O, $^{29}$S, $^{31}$P,
and $^{35}$Cl are more enriched in the 50\% WD - 50\% Solar sequence.  These plots also show that $^{16}$O and $^{22}$Ne are highly depleted, and $^3$He is depleted.  Similar results in  \citet{starrfield_2009_aa} were used by \citet{pepin_2011_aa} to identify CN grains in anomalous interplanetary  particles.
The 1.35 M$_\odot$ results are shown in the middle panels of Figure \ref{figure2575ONE} and Figure \ref{figure5050ONE} and show
even more extreme enrichments for the above isotopes. In the next two production plots (Figure \ref{oneco2575} and Figure \ref{oneco5050}), we compare our ONe results with the CO results from \citet{starrfield_2020_aa}. 
These plots show for  1.25M$_\odot$ WDs and both compositions 
 that both $^7$Be and $^{13}$C are about 200 times Solar and 
$^{15}$N and $^{17}$O are nearly $10^4$ times Solar.  In contrast to the CO results, neither $^{18}$O nor $^{18}$F are depleted. 
As expected, the lighter elements are more enriched in the CO results.

Table \ref{ONe2575MFBabund} provides the abundances (mass fraction) for the MFB study with the 
25\% WD - 75\% Solar composition, Table \ref{ONe5050MFBabund} provides the same set of data for the  50\% WD - 50\% Solar MFB
simulations. In Table \ref{solarabund} we provide the predictions for the pure Solar simulations, while Table \ref {ONe2575MDTNRabund},
and Table \ref{ONe5050MDTNRabund} tabulate the abundances for the MDTNR 25\% WD - 75\% Solar composition and 
the MDTNR 50\% WD - 50\% Solar composition,  respectively. Table  \ref{solarabund} (no mixing of accreted with core material, hence, a Solar mixture only) allows us to make predictions for those CNe or RNe that do not mix with WD matter.  It shows that the ejected $^{12}$C abundance increases with WD mass,  $^{14}$N is relatively constant, and the $^{16}$O abundance strongly declines with increasing WD mass.   The odd isotopes, such  as  $^{13}$C, $^{15}$N, and $^{17}$O, increase with  WD mass.  For WD masses that exceed 1.0 M$_\odot$, the $^{13}$C abundance always exceeds that of $^{12}$C and $^{17}$O always exceeds $^{16}$O.  These ratios are strong indicators of
non-equilibrium CNO burning.  

ALMA observations of the possible merger CK Vul show enriched $^{17}$O (in C$^{17}$O) indicating that
non-equilibrium CNO burning has occurred in this enigmatic object \citep{eyres_2018_aa}.  While \citet{sneden_1975_aa} reported enriched $^{13}$C and $^{15}$N in the ejecta of Nova DQ Her(1934), there are no observations of CN ejecta that suggest that the abundance of
 $^{13}$C exceeds that of $^{12}$C. We suggest that this result can be explained by mixing of the ejecta with
 material in the accretion disk that has not fallen onto the WD.  It is also possible that the ejecta in the orbital plane have
 mixed with pre-existing gas ejected from the secondary prior to the CN outburst \citep{williams_2008_aa}.  However, if these are the solutions, then we predict
 that there should be different abundances measured for material in the equatorial plane of the system as
 compared to material ejected in the polar directions.   This is an area of discovery space for JWST's integral-field-units spatial spectral
 observations of evolved CNe and RNe.
 
 The $^{15}$N abundance roughly increases with  WD mass.  In contrast, $^{18}$O, $^{26}$Al, and $^{27}$Al  
 decrease with increasing  WD mass.  The abundance  of $^4$He increases with WD mass, consistent with more hydrogen being fused to 
 helium to produce the energy observed in the outburst, since the amount of accreted mass declines with increasing WD mass.
 
Table  \ref{ONe2575MDTNRabund}, gives the ejecta abundances for the MDTNR mixture with 25\% WD matter and 75\% Solar matter.  
The abundance of $^7$Be increases with increasing WD mass to a value of $10^{-6}$ and the initial $^7$Li is destroyed by the TNR.  
Both $^{12}$C and $^{13}$C are produced in the higher mass  WDs but there is slightly  more $^{12}$C than $^{13}$C for most WD masses.  
The $^{12}$C/$^{13}$C ratio is close to one which indicates that the CNO burning is far from equilibrium, and that mixing with an accretion disk or
material that has not undergone the TNR must occur in observed systems. The abundance of $^{14}$N is roughly constant for the more massive  WDs while $^{15}$N reaches a peak abundance 
of 0.12 for a 1.35 M$_\odot$ WD and is nearly that value for the other massive WDs.  
The abundance of $^{15}$N exceeds that of $^{14}$N for WD masses from 1.15 M$_\odot$ to 1.35 M$_\odot$. 
 In contrast to the Solar abundance simulations (Table \ref{solarabund}), the abundances of
$^{17}$O, $^{18}$O, and $^{31}$P increase with WD mass.  $^{26}$Al and $^{27}$Al reach a 
maximum abundance at 1.0 M$_\odot$ and then decline with increasing WD mass as the peak temperature in the nuclear burning region increases 
during the TNR.  The ratio of their abundances is $\sim$0.1.

Table \ref{ONe5050MDTNRabund} provides a listing of the ejecta abundances for the MDTNR
 mixture with 50\% WD matter and 50\% Solar matter.   The $^7$Be
abundance reaches a maximum at 1.35 M$_\odot$ and the $^7$Li abundance decreases with increasing WD mass. $^7$Li is destroyed during the TNR so that  all the $^7$Li listed in the tables and that observed in CNe ejecta must arise from the decay of $^7$Be produced in the outburst.   

The ejecta abundances of $^{12}$C and $^{13}$C increase with WD mass.   The abundance of $^{14}$N increases with WD mass to about 3\% while $^{15}$N reaches a maximum at 1.35 M$_\odot$.
The odd isotopes are so abundant that molecular studies of CN ejecta should discover large amounts of $^{12}$C$^{15}$N,  $^{13}$C$^{14}$N, and in some cases
$^{13}$C$^{15}$N.  The detection of these molecular species would provide strong observational support for the results of these simulations, and possibly could
be used to determine the composition of the underlying WD.

The abundance of $^{16}$O declines with increasing WD mass
while that of $^{17}$O increases but then reaches a maximum value at 1.15 M$_\odot$. In contrast, that of $^{18}$O increases as the WD mass increases.  The abundance of  $^{26}$Al reaches
a maximum at 1.35 M$_\odot$ while that of $^{27}$Al increases up to 1.35 M$_\odot$.  The ratio of their abundances varies from about 0.3 down to about 0.1, values which are smaller than
found in the 25\% WD - 75\% Solar MDTNR studies.  The abundances of $^{22}$Na, $^{31}$P and $^{35}$Cl also increase with  WD mass.  

\section{ Discussion}
\label{discuss}

The results reported here are guided by the recent multi-D studies of convection in the accreted layers of WDs driving mixing of the core matter with the accreted matter after the TNR is ongoing  \citep[][and references therein]{casanova_2010_aa,
 casanova_2010_ab, casanova_2011_aa, casanova_2011_ab, casanova_2016_aa, casanova_2018_aa, jose_2014_aa, jose_2020_aa}.  This technique  provides a range of model outcomes that are compatible with observed CNe physical parameters 
reported in the literature.  The difference between \citet{starrfield_2020_aa} and this paper,
is that the enriched nucleus is $^{16}$O instead of $^{12}$C.  As a direct result,
peak temperatures and energy generation rates are lower in the simulations reported here than in
those described in \citet{starrfield_2020_aa} for the same WD mass. As a consequence, less mass is ejected for the same
WD mass and \.M, which increases the RE. For some of the simulations described in
earlier sections, no mass is ejected and {\it the WD is growing in mass at the mass accretion rate}.  

While the primary argument against the secular evolution of the WDs in CNe to near the Chandrasekhar Limit is the belief that the consequences of the CN outburst is a net loss in mass of the WD,  the major result of this paper, along with \citet[][and references therein]{starrfield_2020_aa}, is
that the WDs are growing in mass as a consequence of accretion onto the WD and the resulting CN outburst.
The growth in mass of the WD does not require the existence of steady burning which was originally proposed, using semi-analytic calculations, by   \citet{nomoto_1982_aa} and \citet{ fujimoto_1982_aa, fujimoto_1982_ab}.  In any case, their calculations are invalidated by the \citet{schwarzschild_1965_aa} thin shell instability using full stellar evolution calculations.

\citet{starrfield_2020_aa} discussed in some detail the observational and theoretical arguments
suggesting that CO novae can evolve to the Chandrasekahr Limit and, thereby, should be considered as
SN Ia progenitors.  We do not repeat that discussion here except to point out that the same arguments imply
that an ONe nova can evolve to Accretion Induced Collapse (AIC). Descriptions of the formation of neutron stars
from AIC of WDs can be found in  \citet[][and references therein]{ruiter_2019_aa} and \citet[][and references therein]{wangb_2022_aa}. 

However, NOVA  is only able to follow one outburst and reaching to nearly  the Chandrasekhar Limit requires many such cycles of accretion-TNR-ejection - accretion.  While this has yet to be done with either CO or ONe enriched material (this may have been done in the study of
 \citet{rukeya_2017_aa} but they only reported their ejected mass not the accreted mass); multi-cycle evolution and the growth in mass of the
 WD has been done with Solar accretion studies \citep{epelstain_2007_aa, starrfield_2014_aa, hillman_2015_aa, hillman_2015_ac, hillman_2016_aa, starrfield_2017_aa}.  The multi-cycle studies reported in \citet{newsham_2014_aa, starrfield_2014_aa} and \citet{starrfield_2017_aa} were done with 
 MESA \citep[][and references therein]{paxton_2011_aa, paxton_2013_aa, paxton_2018_aa} while those described by
 \citet{epelstain_2007_aa, hillman_2015_aa} and \citet{hillman_2016_aa} were done with the code of \citet[][and references therein]{kovetz_2009_aa}. 
 The effects of recurrent outbursts, assuming compositions enriched with WD matter, both on the outburst properties and the WD mass, are a topic ripe for investigation.

In addition, some RNe are repeating sufficiently often that their WDs must have grown in mass so that they are now close to the Chandrasekhar Limit.  One such system is the M31 RN in M31 (M31N 2008-12a) which is outbursting about once per year and
has opened up a large cavity in the ISM surrounding the system \citep{darnley_2016_aa, darnley_2017_aa, darnley_2017_ab, darnley_2019_aa, henze_2015_aa, henze_2018_aa}.  In contrast to speculation \citep{gilfanov_2010_aa, kuuttila_2019_aa},
it is neither X-ray nor UV luminous between outbursts.  If the continued TNRs on this WD have built up a thick helium layer, then it seems likely that
at some time in the future, it will result in a SN Ia event similar to that of SN2020eyj which exploded into a helium rich circumstellar medium \citep{kool_2023_aa}.

Of great importance, some of the ejected isotope abundances in these simulations also fit the isotopic ratios of
Ne and He measured in anomalous Interplanetary Dust Particles (IDPs) and are unique compared to solar system compositions \citep{pepin_2011_aa}. Their similarities to isotope ratios reported in \citet{starrfield_2009_aa}, and confirmed in this paper, are evidence that dust produced in classical nova explosions found its way into solar system pre-solar grains \citep{pepin_2011_aa, bose_2019_aa}. 
While \citet{bose_2019_aa} compared the compositions of 30 pre-solar SiC grains with the ejected isotopic abundances in 
our CO nova simulations, the grains studied by \citet{pepin_2011_aa} analyzed material 
captured on stratospheric collectors flown to sample dust from comets 26P/Grigg-Skjellerup and 55P/Tempel-Tuttle. The
isotopic ratios used in \citet{pepin_2011_aa}  are virtually unchanged in this paper (in spite of the changes in
microphysics, number of zones, and initial composition) attesting to the robustness of these predictions.  More
details on the CO predictions can be found in \citet{bose_2019_aa}.

Another indicator of a system with a massive WD is the time to decline 2 magnitudes from maximum (t$_2$ time) or 3 magnitudes from maximum (t$_3$ time).  
This rate of decline is determined by the amount of ejected mass and the ejecta velocities.  
The less ejecta mass, the earlier in the evolution the expanding gases become optically thin, and the peak emission transitions from the optical into the ultraviolet and then shorter wavelengths.  
Since the more massive WDs accrete less material before the TNR is initiated, unless they mix with a great deal of underlying WD matter, they will eject much less material.  In addition the faster the ejecta velocities, the more rapidly the expanding gases reaches radii of $\sim 10^{12}$ cm and begin to become optically thin.  
The ejection velocities depend on the total energy of the outburst, and our simulations show that the more massive WDs are the most energetic with the highest velocities.  
While the RNe such as U Sco have extremely fast decline times, three CNe (V1500 Cyg, V838 Her, and V1674 Her) have had t$_2$ times less than 2 days.  
They also exhibited large ejection velocities \citep[][and references therein]{woodward_2022_ab}.  We predict that as the WD grows in mass and its mass approaches the Chandrasekhar Limit, and if accretion results in matter that is enriched in $^{12}$C (for example), then the t$_2$ for the explosion will become shorter and shorter.  It is likely, that there could be insufficient ejecta for a visible outburst and at ``peak light'' the emission peak would be in the UV, EUV,  or even shorter wavelengths.  These are targets for an all-sky UV satellite such as ULTRASAT
(\url{https://apd440.gsfc.nasa.gov/ultrasat/}).

\subsection{The Amount of $^7$Li in the Galaxy produced by Classical Novae}
\label{lithiumgalaxy}

Our results confirm that ONe novae are also overproducing $^7$Be, which decays to $^7$Li
long after we end our simulations.  Figure \ref{figureli7mass} shows that both CO and ONe explosions on WDs more massive than 0.8 M$_\odot$ produce reasonable amounts of
$^7$Li.  In \citet{starrfield_2020_aa}, we reported an ejected $^7$Be mass of $ \sim 2.0 \times 10^{-10}$ M$_\odot$ for CO novae. 
In this paper we report an ejected $^7$Be mass of $\sim 6 \times 10^{-11}$M$_\odot$ for ONe novae.  The CO results are roughly
constant and exceed $10^{-10}$M$_\odot$ for WD masses greater than 0.8 M$_\odot$.  In contrast, only the simulations on the
most massive ONe WDs reach $\sim 6 \times 10^{-11}$ M$_\odot$ and for the 50\% WD matter plus 50\% solar matter composition. 
Nevertheless, it is likely that the WDs in ONe CNe are more massive than 1.0 M$_\odot$.   {We also cite the observational studies that show
that both ONe and CO CNe are contributing to the $^7$Li abundance in the galaxy (see Figure 9 in \citet{molaro_2022_aa} and
Figure 13 in \citet{molaro_2023_aa}).}

The amount of $^7$Be we predict from our simulations, and those of others, in combination with the observations,
allow us to assert that both CO and ONe CNe are responsible for a significant fraction of the $^7$Li in the galaxy.  While our CO and ONe
simulations imply that CNe have provided about 100 M$_\odot$ of $^7$Li during the lifetime of the galaxy, the observations of $^7$Be and $^7$Li 
\citep{tajitsu_2015_aa, tajitsu_2016_aa, izzo_2015_aa, izzo_2018_aa, molaro_2016_aa, selvelli_2018_aa,wagner_2018_aa, woodward_2020_aa, molaro_2020_aa, molaro_2022_aa, molaro_2023_aa} report much higher values than we predict.  In fact, at least 10 times higher than previously predicted \citep{starrfield_1978_aa, hernanz_1996_aa, jose_1998_aa} implying CNe may be responsible for all the $^7$Li in the galaxy \citep[][and references therein]{molaro_2023_aa}.
 
What is the total amount of $^7$Li in the galaxy?  As reported in \citet{starrfield_2020_aa}, \citet{lodders_2009_ab} give a value of $2.0 \times 10^{-9}$ for the solar system meteoritic abundance  of $^7$Li/H by number.  We convert to mass fraction by multiplying by 7 and obtain $1.4 \times 10^{-8}$ for X($^7$Li)/X(H).   We assume that the total mass of the stars in the  galaxy is $\sim 10^{11}$ M$_\odot$ and the mass fraction of hydrogen is 0.71 \citep{lodders_2009_aa,lodders_2009_ab}. Therefore,
 the total mass of $^7$Li in the galaxy should be $0.71 \times 10^{11}$$ \times$$ 1.4 \times 10^{-8}$ or $\sim$1000 M$_\odot$. 
  {The total mass of primordial lithium is $\sim$300 M$_\odot$, halo stars contribute $\sim$100 M$_\odot$, and spallation contributes
 about 100 M$_\odot$ leaving about 500 M$_\odot$ to come from other sources such as CNe and RNe (P. Molaro 2023 private
 communication).}
 
The most recent discussion of the importance of CNe for $^7$Li in the Galaxy is that of
\citet[][and references therein] {molaro_2023_aa} who address the discoveries of $^7$Li and $^7$Be in CNe
 ejecta, both ONe and CO.  {It is beyond the scope of this paper to include either population synthesis or Galactic
 Chemical Evolution (GCE) studies of the contributions of CNe and RNe to the abundance of $^7$Li in the Galaxy.  
 Fortunately, however, those have already been done in detail by others \citep{cescutti_2019_aa, grisoni_2019_aa, matteucci_2020_aa, kemp_2021_ab, kemp_2022_ab} and we see no need to repeat those studies.  We agree with their conclusions which are that the theoretical
 studies of CNe and RNe produce insufficient $^7$Li to make up the difference between known contributors and
 unknown contributors.  However, sufficient $^7$Li is observed in CN and RN outbursts so that they
 should be considered the ``unknown'' contributors. }

\section{Conclusions}
\label{conclude}
\begin{enumerate}

\item Multi-dimensional studies show that there is sufficient mixing, once the TNR has evolved to where the convective region nearly reaches the surface,
 to agree with observations of the ejecta abundances  \citep[][and references therein]{casanova_2018_aa, jose_2020_aa}.  
 This mixing occurs via convective entrainment (dredge-up of WD outer layers into the accreted material)
 during the TNR and does not affect the total amount of accreted material since it occurs after the accretion phase of the outburst.

\item By accreting solar material (rather than assuming mixing occurs during accretion), more matter is accreted before the TNR.  Reducing the metallicity to
values in agreement with the Magellanic Clouds, or even lower, further reduces the initial $^{16}$O abundance, allowing even more material to be accreted before the TNR is initiated \citep{starrfield_1999_ac, jose_2007_aa}. 

\item As a result, the amount of accreted material is an inverse function of the initial abundance of $^{12}$C for CO novae and the initial
$^{16}$O abundance for ONe novae.

\item Both a lack of mixing with the ONe WD (solar accretion) or mixing with the ONe WD (MFB) during accretion, results in an outburst that is less violent with little material (accreted plus WD) ejected during the outburst.  Thus, explosions on ONe WDs result in the WD growing in mass.  
We measure this by tabulating the Retention Efficiency (M$_{\rm acc}$ -  M$_{\rm ej}$)/M$_{\rm acc}$) (see Figure \ref{retention}). 
This has been shown both by following one outburst with NOVA and repeated outbursts with MESA \citep{starrfield_2016_aa}. 
The growth in WD mass implies that ONe systems can ultimately reach a mass where they undergo Accretion Induced Collapse and
become neutron stars \citep{ruiter_2019_aa}.

\item Our simulations with 50\% ONe WD and 50\% Solar matter, mixed after the TNR is underway, ejected a larger fraction of accreted material but not as much as was accreted.  They also reached higher peak temperatures and ejected more material moving at higher velocities 
then those with only 25\% WD and 75\% solar matter.

\item Our simulations confirm that both ONe and CO novae are overproducing $^7$Be, which decays to $^7$Li after we have ended our simulations.  
This result is in agreement with the observations of enriched $^7$Be in CN explosions, although the observed values exceed our predictions and those of others.

\item Observations of CNe in outburst imply that far more mass is ejected and more $^7$Li is produced then predicted by our simulations.  If this also holds true for $^{26}$Al 
such that it is more enriched than we predict, then CNe could be responsible for most of the $^{26}$Al in the galaxy.

\item Although our simulations using a composition of 25\% WD matter and 75\% solar matter (MDTNR) on low mass WDs, either
CO or ONe,  do not appear to resemble the outbursts of typical CNe, they are sufficiently slow to resemble the recent outburst of Nova Velorum 2022 \citep{aydi_2023_aa} and we predict that 
the WD in this system has a very low mass.

\item The predictions of an early X-ray flash, now detected by eRosita \citep{konig_2022_aa}, are supported by the simulations reported in this paper.

\item  \citet{bose_2019_aa} and \citet{iliadis_2018_aa} have identified CO grains from novae in meteorites while \citet{pepin_2011_aa} have identified grains from ONe
novae in anomalous interplanetary particles from comets.  These results confirm that material ejected from CNe explosions penetrated the material
that formed the Solar System.

 \item  { Finally, as we report in this paper and reported in \citet{starrfield_2020_aa}, our simulations imply that WDs are growing in mass.  Observational studies of the masses of the WDs in CN systems suggest a broad range in WD mass.  For example, \citet{sion_2019_aa} report a WD mass for the RN CI Aql of 0.98 M$_\odot$ and \citet{shara_2018_ab} report that the mean WD mass for 82 Galactic CNe is 1.13 M$_\odot$, and for 10 RNe is 1.31 M$_\odot$. \citet{selvelli_2019_aa} analyzed 18 CNe, using data from both IUE and Gaia, and report that many WDs in CNe have masses above the canonical value for single WDs of 0.6 M$_\odot$. }

\end{enumerate}


We thank the anonymous referee for their comments which greatly improved the plots and discussion in this manuscript.
We acknowledge useful discussion and encouragement from D. Banerjee, A. Evans, R. Gehrz, K.Hensley, M. Darnley, E. Aydi, J. Jos\'e,  M. Hernanz, S. Kafka,  L. Izzo, P. Molaro, I. Perron, M. della Valle, A. Shafter, L. Takeda and the attendees at EWASS18, COSPAR 2018, and HEAD 2019 for  their comments.  We espcially thank P. Molaro for his comments on an earlier draft of this paper which have been incorporated into the manuscript..
This work was supported in part by the U.S. DOE under Contract No. DE-FG02- 97ER41041. SS and MB acknowledge partial support from a NASA Emerging Worlds grant to ASU (80NSSC22K0361) as well as support to SS from his ASU Regents' Professorship, WRH is supported by the U.S. Department of Energy, Office of Nuclear Physics, and CEW acknowledges support from NASA Grant 80NSSC19K0868.   

{\bf Data Availability}
The data underlying this publication will be shared on reasonable request to the first author.

\bibliography{references_iliadis,starrfield_master}

\end{document}